\documentclass[a4paper]{scrartcl}
\usepackage[ansinew]{inputenc}
\usepackage{amsmath,amssymb,amstext,color,bm,verbatim,graphicx,float}
\usepackage[margin=0.9in,bottom=1in,top=1in]{geometry}
\usepackage{fancyhdr}
\usepackage{lastpage}
\usepackage{cancel}
\usepackage{booktabs}
\usepackage{enumitem}

\begin{document}
\begin{center}
\huge{\textbf{Odd dynamics of living chiral crystals}} 
\end{center}
\vspace{0.2cm}

\normalsize
Tzer Han Tan$^{1,2,3,5}$, Alexander Mietke$^{4,5}$, Junang Li$^{1,6}$, Yuchao~Chen$^{1,6}$, Hugh Higinbotham$^1$, \linebreak Peter~J.~Foster$^1$, Shreyas~Gokhale$^1$, J\"{o}rn~Dunkel$^4$, Nikta~Fakhri$^{1,7}$
\vspace{-0.25cm}

\begin{itemize}[leftmargin=0.15in]\small
\item[$^1$] Department of Physics, Massachusetts Institute of Technology, Cambridge, MA, USA \vspace{-0.3cm}
\item[$^2$] Quantitative Biology Initiative, Harvard University, Cambridge, MA, USA \vspace{-0.3cm}
\item[$^3$] Center for Systems Biology Dresden, Dresden, Germany \vspace{-0.3cm}
\item[$^4$] Department of Mathematics, Massachusetts Institute of Technology, Cambridge, MA, USA \vspace{-0.3cm}
\item[$^5$] These authors contributed equally and are joint first authors. \vspace{-0.3cm}
\item[$^6$] These authors contributed equally. \vspace{-0.3cm}
\item[$^7$] Corresponding author: fakhri@mit.edu
\end{itemize}

\vspace{1cm}
\normalsize
\section*{Abstract}

Active crystals are highly ordered structures that emerge from the self-organization of motile objects, and have been widely studied in synthetic~\cite{Palacci:2013eu,bililign2021chiral} and bacterial~\cite{Petroff2015,Petroff:2018js} active matter. Whether collective crystallization phenomena can occur in groups of autonomously developing multicellular organisms is currently unknown. Here, we show that swimming starfish embryos spontaneously assemble into chiral crystals that span thousands of spinning organisms and persist for tens of hours. Combining experiments, theory, and simulations, we demonstrate that the formation, dynamics, and dissolution of these living crystals are controlled by the hydrodynamic properties and natural development of embryos. Remarkably, living chiral crystals exhibit self-sustained chiral oscillations as well as various unconventional deformation response behaviors recently predicted for odd elastic materials~\cite{Scheibner:2020gm,braverman2020topological}. Our results provide direct experimental evidence for how  nonreciprocal interactions between autonomous  multicellular components may  facilitate novel nonequilibrium phases of chiral active matter.

\newpage

Symmetry breaking~\cite{Anderson393,ashcroft1976solid} is a hallmark of living~\cite{2010LiBowerman} and synthetic~\cite{Palacci:2013eu,Bricard:2013jq,2000Grzybowski,2010Lee} active matter. From the asymmetric growth of multicellular  organisms~\cite{2010LiBowerman,Naganathan:2014fc,2019Smith} to the coherent motions of swimming cells~\cite{riedel2005self,PhysRevLett.98.158102} and self-propelled colloids~\cite{Palacci:2013eu,Bricard:2013jq,2019Shen,2020Bechinger}, active systems form self-organized structures~\cite{RevModPhys.66.1481,wang2021emergent,2021Omar} with unusual material properties \cite{avron1998odd,soni2019odd,banerjee2021active} that can only emerge far from thermal equilibrium. In spite of major experimental~\cite{Palacci:2013eu,Bricard:2013jq,Petroff2015,PhysRevLett.98.158102,soni2019odd,2020Bechinger}  and theoretical~\cite{Marchetti:2013bp,shankar2020topological,2015CatesTailleur} progress over the last decade, we are only beginning to understand how complex collective behaviors of multicellular~\cite{Bi:2015hh,hartmann2018emergence,Qin71,2021Collinet} and multiorganismal~\cite{Rosenthal4690,Bialek4786} systems arise from the broken symmetries and nonequilibrium dynamics of their individual constituents.
\\

A particularly interesting class of nonequilibrium symmetry-breaking phenomena comprises the active crystallization processes recently discovered in colloidal~\cite{Palacci:2013eu} and bacterial~\cite{Petroff2015} systems. Unlike conventional passive crystals, which form upon lowering temperature and often require attractive forces, active crystallization arises from the particles' self-propulsion and can occur even in purely repulsive dilute systems~\cite{Palacci:2013eu}. A long-standing related, unanswered question is whether groups of multicellular organisms can self-organize into states of crystalline order and, if so, what emergent material properties they might exhibit. 
\\

Here, we report the discovery of spontaneous crystallization in large assemblies of developing starfish \textit{Patiria miniata} embryos (Fig.~1a). Our experimental observations show how, over the course of their natural development, thousands of swimming embryos come together to form living chiral crystal (LCC) structures that persist for many hours. In contrast to externally actuated colloidal systems,  the self-assembly, dynamics, and dissolution of these LCCs are controlled entirely by the embryos' internal developmental program (Fig.~1a,b). A quantitative theoretical analysis reveals that LCC formation arises from the complex hydrodynamic interactions~\cite{Drescher2009,Lauga2020} between the  starfish embryos. Once formed, these LCCs exhibit striking collective dynamics, consistent with predictions from a recently proposed theory of odd elasticity~\cite{Scheibner:2020gm}.

\subsection*{Self-assembly, growth, and dissolution of living chiral crystals}

\begin{figure}
\centering
\small
\includegraphics[width=1\textwidth]{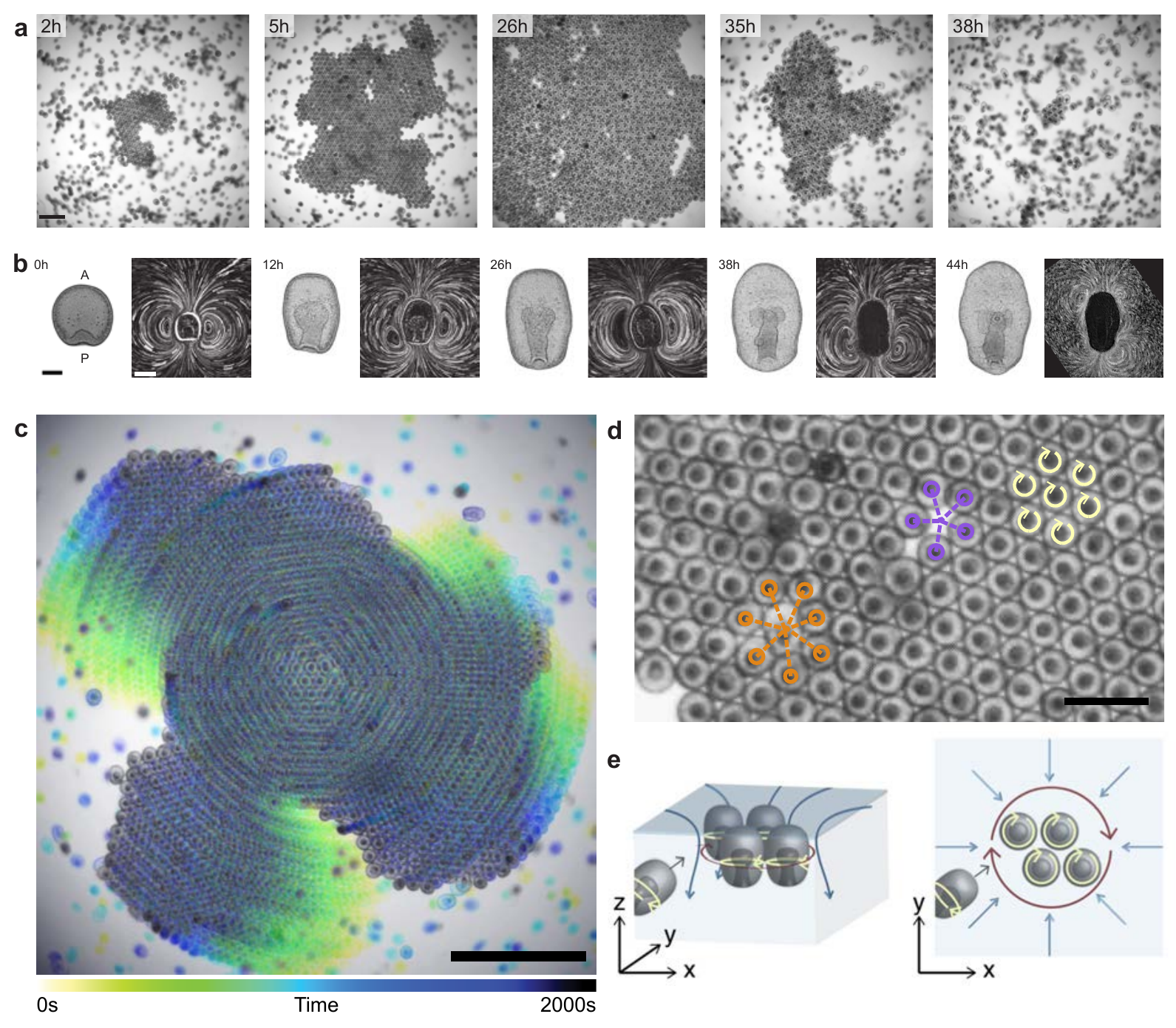}
\caption{\textbf{Developing starfish embryos self-organize into living chiral crystals.} \textbf{a,}~Time sequence of still images showing crystal assembly and dissolution. Scale bar, 1\,mm. $t=0$ hours corresponds to the onset of clustering. \textbf{b,}~Embryo morphology and flow fields change with developmental time. Shape scale bar, 100\,$\mu$m. Flow field scale bar, 200\,$\mu$m. See Extended Data Figure 1 for uncropped morphology images. \textbf{c,} Embryos assembled in a crystal perform a global collective rotation. Scale bar, 2\,mm. \textbf{d,}~Spinning embryos (yellow arrows) in the crystal form a hexagonal lattice, containing 5-fold (purple) and 7-fold (orange) defects. Scale bar, 0.5\,mm. \textbf{e,}~Schematic of embryo dynamics and fluid flows. Crystals of spinning embryos form near the air-water interface. Self-generated hydrodynamic flows lead to an effective attraction between surface-bound embryos.}
\end{figure}

During early development, starfish embryos exhibit substantial morphological changes. From the onset of gastrulation (Fig.~1b, 0\,h), embryos elongate along their anterior-posterior (AP) axis (0-44\,h) while progressively developing folds that further break shape symmetry. In parallel, the self-generated fluid flow near the embryo's surface changes~(Fig.~1b), reflecting spatial reconfigurations of cilia during growth~\cite{Gilpin:2017ef} similar to other ciliated organisms~\cite{wan2020reorganization}. Remarkably, when embryos come close to the fluid surface, they can attain a stable bound state in which their AP axes are oriented perpendicular to the fluid-air interface. Groups of surface-bound embryos can spontaneously self-organize into two-dimensional hexagonal clusters (Fig.~1a, 2-5\,h). Over time, these clusters grow into larger crystals, reaching sizes of hundreds to thousands of embryos (Fig.~1a, 26\,h) and persisting for tens of hours. As embryos develop further (Fig.~1b, 38--44\,h) crystals begin to disassemble (Fig.~1a, 35\,h) and eventually dissolve completely (Fig.~1a, 38\,h). 
\\

Viewed from above, both small and large crystals rotate clockwise (Fig.~1c), consistent with the chiral spinning motions of individual embryos about their AP axis (Fig.~2a). Large LCCs typically exhibit a high degree of hexagonal order, while also harboring lattice defects (Fig.~1d). The assembly, rotational dynamics and dissolution of LCCs can be rationalized by a hydrodynamic analysis that accounts for the flow fields generated by individual embryos (Figs.~1e and 2a-c).

\subsection*{From single embryo properties to crystal formation, chiral rotation, and dissolution}

\begin{figure}
\centering
\small
\includegraphics[width=1\textwidth]{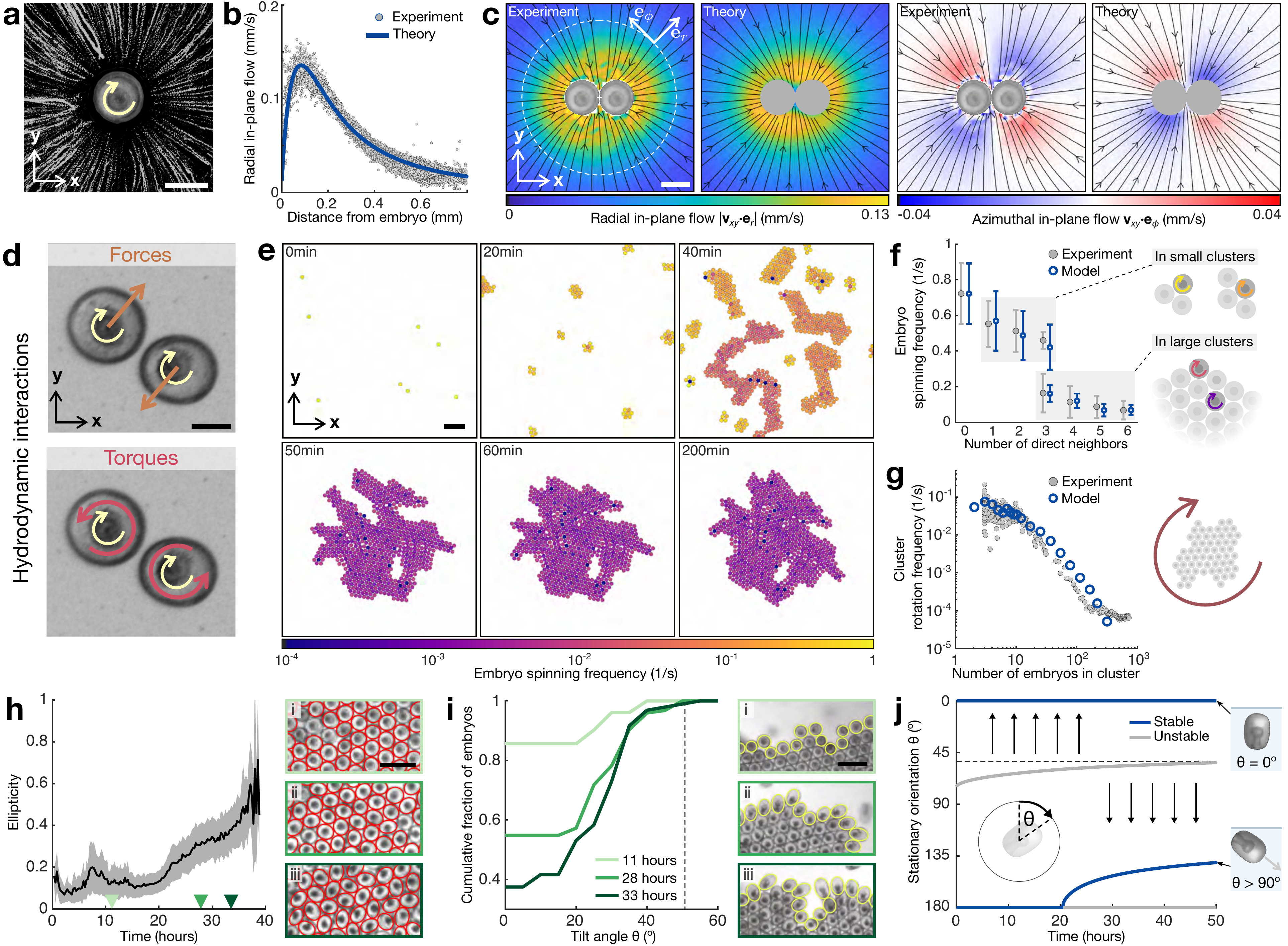}
\caption{\textbf{Single embryo properties facilitate formation, rotations and dissolution of clusters.} \textbf{a,}~Single embryo top view. Yellow arrow indicates spinning direction, gray dotted lines visualize streamlines~(SI~Sec.~1.2). \textbf{b,}~Measured radial in-flow velocities (gray dots) are well-described by Stokeslet flow below a free surface (blue line)~(SI~Sec.~2.1.6). \textbf{c,}~Flow fields surrounding bound pairs (Experiment, SI~Sec.~3.2.1) fitted by a solution of the Stokes equation~(Theory, SI~Sec.~3.2.2) that takes into account hydrodynamic interactions (\textbf{d}). \textbf{d,}~Hydrodynamic interactions cause nearby embryos to orbit around each and reduce individual spinning frequencies. \textbf{e,}~Stokeslet-mediated attraction~(\textbf{a},\textbf{b}) and hydrodynamic near-field interactions of spinning particles (\textbf{d}) in an experimentally constrained minimal model~(SI~Sec.~2.2)  reproduce crystal formation and rotation dynamics seen in the experiments. 
\textbf{f,}~Single embryo spinning frequencies in small clusters ($\le4$ embryos) and in larger clusters ($\approx100$ embryos). Error bars denote standard deviations of measurements (Experiment) and from simulations (Model) (SI~Sec.~2.2.3).
\textbf{g,}~Assuming a cluster-size dependent reduction of the individual embryo's spinning activity (SI Sec.~2.2.2) leads to good agreement with measured whole-cluster rotation frequencies. \textbf{h,}~Ellipticity of embryo shapes (right: top-view outlines in red, SI~Sec.~3.5) increases during development, leading to increasingly noisy steric interactions among spinning embryos in clusters. Gray band depicts standard deviation. \textbf{i,}~Embryos at cluster boundaries exhibit increasing AP axis tilt angles as development progresses (right: projection outlines in yellow, SI~Sec.~3.5). Dashed line: critical angle at which bound states of late embryos become unstable. \textbf{j,}~Stationary orientations and stability of a microswimmer with hydrodynamic properties akin to developing embryos~(SI Secs.~2.1.1--2.1.5). A decreasing critical angle (gray line) and the increase in effective noise (\textbf{h},\textbf{i}) increase the rate of embryos leaving cluster boundary and fluid surface, ultimately driving the dissolution of clusters. Scale bar, 200\,$\mu$m (\textbf{a},\textbf{c}), 100\,$\mu$m~(\textbf{d}), 1\,mm~(\textbf{e}), 500\,$\mu$m~(\textbf{h},\textbf{i}).}
\label{fig:Model}
\end{figure}

To understand the hydrodynamic interactions underlying the cluster dynamics, we first analyzed the fluid flow around individual embryos bound below the air-water interface~(Fig.~\ref{fig:Model}a,b). Observed along the AP axis, fluid moves radially inward towards the embryo, reaches maximum speed 0.1--0.2\,mm/s lateral to the embryo surface~(Fig.~\ref{fig:Model}b), and eventually moves toward the bottom of the well~(Fig.~1e). The radial in-flow generated by isolated embryos can be described as a Stokeslet flow~(Fig.~\ref{fig:Model}b, blue curve), a solution of the Stokes equation that describes the generic fluid flow around an external force~(SI~Sec.~3.2.2). This force is related to the negative buoyancy of embryos. Indeed, the buoyant weight force $F_g=1.7\pm0.4\,$nN estimated from sedimentation speeds of immobilized embryos~(SI~Sec.~1.4) is close to the Stokeslet strength $F_{\text{st}}=2.6\pm0.3\,$nN obtained from fitting radial in-plane flow fields (Fig.~1b, SI~Sec.~2.1.6). 
\\

The self-generated Stokeslet flow stabilizes the upright AP-axis orientation of embryos below the fluid surface~(SI~Sec.~2.1.6). In addition, it induces an effective long-ranged hydrodynamic attraction between embryos, facilitating the assembly of clusters. Similar effects have been observed previously for bacterial and algal microswimmers near rigid surfaces~\cite{Petroff2015,Drescher2009}. Once two embryos are close together, their intrinsic spinning motions lead to an additional exchange of hydrodynamic forces and torques~(Fig.~\ref{fig:Model}d). Similar to pairs of \emph{Volvox} colonies near a rigid surface \cite{Drescher2009,Ishikawa2020}, nearby starfish embryos orbit each other, and their spinning frequency decreases compared to that of a freely spinning embryo. The excess cilia-generated torque from slower rotating embryos~\cite{Drescher2009} manifests itself in systematic azimuthal flow contributions~(Fig.~\ref{fig:Model}c). To confirm our understanding of these hydrodynamic interactions, we complemented the Stokeslet flow of each embryo with additional contributions that reflect the effects of hydrodynamic interactions  (SI~Sec.~3.2.2, SI~Fig.~S5). Flow fields fitted via this approach show good quantitative agreement with experimental measurements~(Fig.~\ref{fig:Model}c, SI~Fig.~S6). 
\\

Based on these insights, we experimentally constrained a minimal model in which upright spinning embryos are represented by rigid disks interacting through hydrodynamic Stokeslet-mediated pairwise attraction, and through pairwise transverse force and torque exchanges~(SI~Sec.~2.2). Using the Stokeslet strength determined from fits as in Fig.~\ref{fig:Model}b, and a parameterization of transverse interactions based on rotation frequency measurements of bound pairs and triplets~(SI~Sec.~2.2.2), this minimal model predicts the self-organized formation of rotating clusters similar to those seen in the experiments~(Fig.~\ref{fig:Model}e). Assuming a cluster-size dependent reduction of the individual embryo's spinning activity to match whole-cluster rotation rates (SI Sec.~2.2.2), the model quantitatively captures the experimentally observed reduction of individual embryo rotation frequencies in both small and large clusters~(Fig.~\ref{fig:Model}f), as well as their collective translation into global cluster rotation rates~(Fig.~\ref{fig:Model}g).
\\

To investigate how developmental changes of embryos contribute to the dissolution of a cluster, we followed the time-dependent morphology and hydrodynamics of embryos. Body shape anisotropies perpendicular to the AP axis increase almost five-fold over the course of experiments~(Fig.~2h). Such anisotropies cause neighboring embryos to 'bump' into each other when closely packed and spinning within a cluster, introducing an effective source of noise in the LCC lattice. The increased interaction noise is particularly visible at cluster boundaries, where embryos become more and more tilted as their morphological development progresses~(Fig.~2i), increasing their tendency to leave or to be scattered off a cluster. Using additional flow field measurements of single embryos at different time points~(SI~Sec.~3.3), we parameterized an orientational stability diagram that reveals a bistable nature of bound state orientations~(Fig.~2j, SI~Sec.~2.1): In addition to a stable upright orientation ($\theta=0^\circ$), downwards oriented stable orientations ($\theta>90^\circ$) exist for which embryos are expected to swim away from the surface. These two orientations are separated by an unstable critical angle (Fig.~2j, gray). The increase of effective noise as characterized in Fig.~2j,h contributes to an increased rate at which embryos tilt beyond this critical angle and therefore represents a key factor in the eventual dissolution of clusters.

\begin{figure}
\centering
\small
\includegraphics[width=1\textwidth]{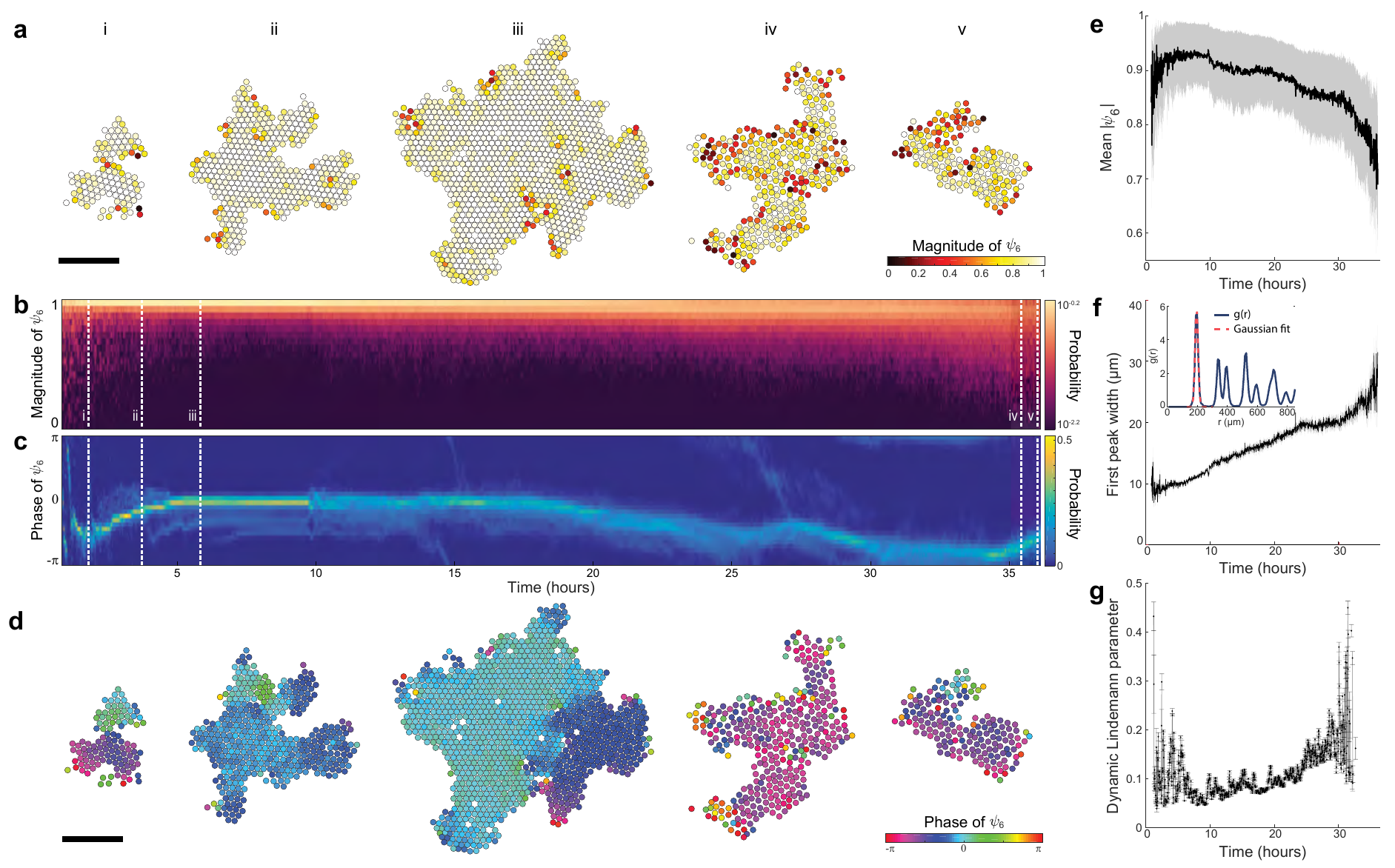}
\caption{\textbf{Crystalline order first increases and then decreases as embryos develop.} 
\textbf{a,}~Embryo centroids at different time points color-coded by the magnitude of the orientational order parameter~$|\psi_6|$ (SI~Sec.~3.6).  Scale bar,~2mm.
\textbf{b}\,--\,\textbf{c,}~The measured probability distribution of $|\psi_6|$ spreads to smaller values after about 20 hours, indicating a loss of bond-orientational order (\textbf{b}). The ensuing drift of the mean phase angle (\textbf{c}) signals dynamical restructuring of the crystal. 
\textbf{d,}~Embryo centroids at different times color-coded by the phase of $\psi_6$. Scale bar,~2mm.  Time slices corresponding to snapshots (i-v) in \textbf{a} and \textbf{d} are indicated by white dotted lines. 
\textbf{e,}~Average magnitude of $\psi_6$ decays after $\sim 10$ hours, confirming a decrease in orientational order. Error bars indicate standard deviation. 
\textbf{f,}~The widening of the first peak of the radial pair distribution function $g(r)$ indicates increased variation in the distance between nearest neighbors. Error bars indicate 95\% confidence interval from Gaussian fit. Inset: Example pair distribution function, $g(r)$, and Gaussian fit to the first peak (SI~Sec.~3.7).
\textbf{g,}~The increase of the dynamic Lindemann parameter  with developmental time signals a progressive destabilization of the crystal lattice. Error bars indicate standard deviation of 20 consecutive time points (SI~Sec.~3.8).}
\end{figure}

\subsection*{Increase and decrease of crystalline order with development}

A striking feature of the LCCs is that they nucleate, grow and dissolve naturally as embryos progressively develop (Fig.~1a). To quantify the evolution of crystalline order, we calculated the local order parameter $\psi_6(\mathbf{r}_i)= |\psi_6|_ie^{i\phi_i}$ at embryo positions $\mathbf{r}_i$ in the co-rotating frame of the cluster (SI~Secs.~3.1~and~3.6). Measurements of $\psi_6(\mathbf{r}_i)$ determine the local phase $\phi_i$ representing the crystal orientation, as well as the magnitude of hexagonal order $|\psi_6|_i$~\cite{nelson1979dislocation}. Initially, small clusters merge together along different crystal axes, resulting in grain boundaries and broad distributions of $|\psi_6|_i$ and $\phi_i$ (Fig.~3a-d, (i)). Within 5~hours of crystal formation, LCCs undergo rapid internal restructuring during which subdomains align. This results in large, nearly defect-free crystals with a high degree of hexagonal order ($\langle |\psi_6|_i \rangle \approx0.9$) and a narrow distribution of local bond orientation (Fig.~3a-d, (ii-iii)). This highly ordered state persists for several hours.
\\

As development progresses, changes in morphology and surrounding flow fields (Figs.~1b and 2h) lead to a decreased crystalline order. Specifically, the probability density of $|\psi_6|_i$ spreads to smaller values (Fig.~3b,e \hbox{$t>20$\,hours}), quantitatively indicating a loss of orientational order. A similar spread is observed in the average phase angle $\phi_i$, indicating the loss of a well-defined, global crystal orientation~(Fig.~3c, $t>20$\,hours). After about 30~hours, disorder dominates and the crystal dissolves over a period of 10~hours (Fig.~3a-d, (iv-v)). 
\\

Furthermore, we identified a progressive loss of translational order prior to dissolution as quantified by the radial pair distribution function $g(r)$ (inset Fig.~3f, SI~Sec.~3.7). Specifically, the first peak width of $g(r)$ -- representing the variability of nearest neighbor distances -- was found to increase with development (Fig.~3f). Consequently, deviations from an ideal hexagonal lattice become more frequent and translational order is reduced as embryos develop.
\\

To examine whether the evolution of orientational and translational order is also reflected in dynamic crystal properties, we determined the dynamic Lindemann parameter  (SI~Sec.~3.8), which characterizes the strength of fluctuations in the crystal lattice~\cite{zahn2000dynamic}. In the crystalline phase (5-25h), the dynamic Lindemann parameter increases with time (Fig.~3g) and indicates a progressive destabilization of the crystal lattice, consistent with the observed loss of orientational~(Fig.~3e) and translational order~(Fig.~3f), and with the increased interaction noise due to changes in the embryo morphology~(Fig.~2h,i). Large fluctuations of the dynamic Lindemann parameter at early and late times are due to the small crystal sizes and the highly dynamic nature of growing and dissolving clusters.
\\

Taken together, the systematic decay of  orientational, translational and dynamic order with developmental time shows how morphological changes at the single-embryo level (Figs.~1b, 2h) can autonomously drive LCCs through a dissolution transition reminiscent of solid-gas phase transitions.

\begin{figure}
\centering
\small
\includegraphics[width=1\textwidth]{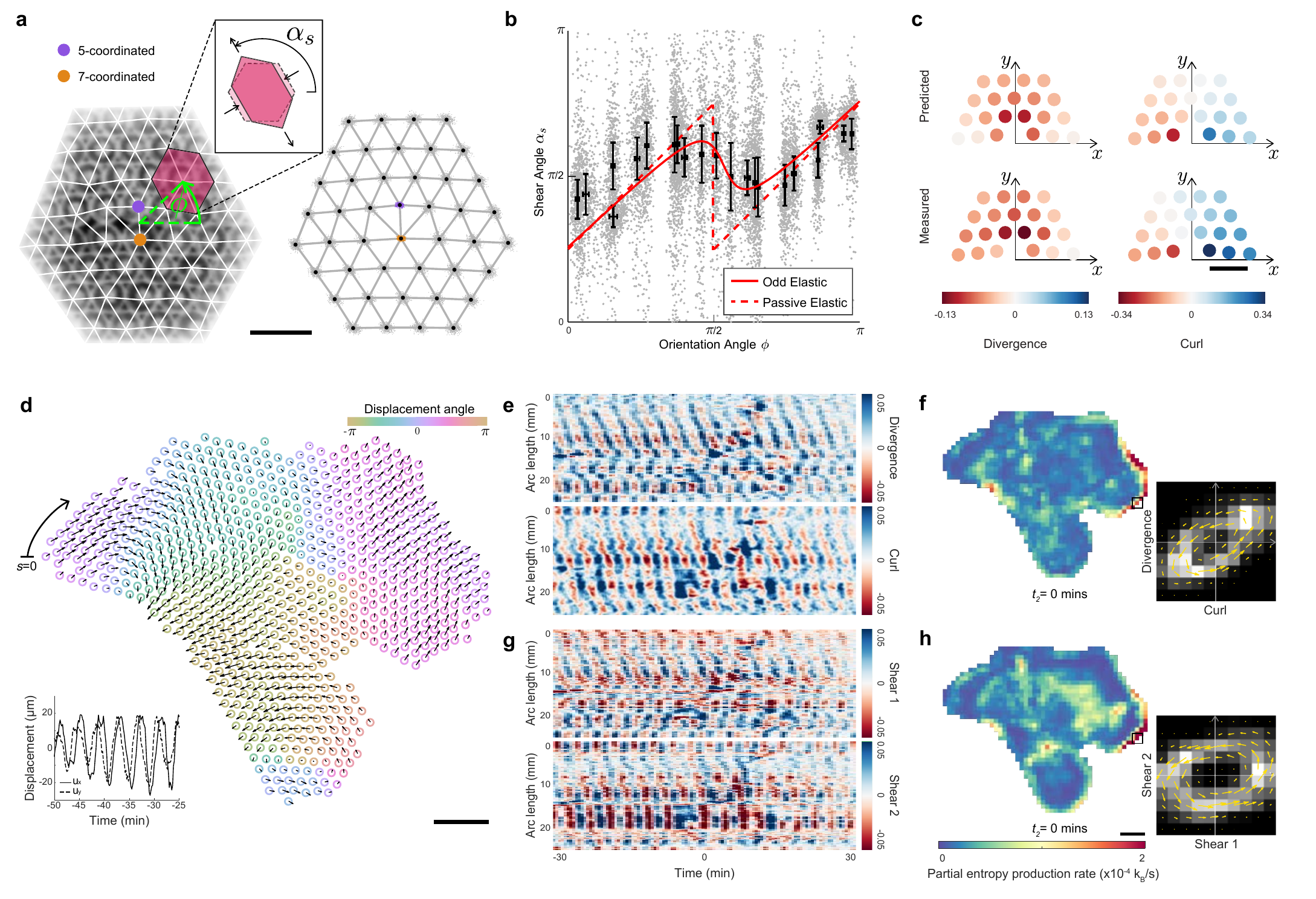}
\caption{\textbf{Defect strains and displacement waves exhibit signatures of odd elasticity.} \textbf{a,}~Shear strain angles $\alpha_s$ near a defect -- defined by a pair of embryos with 5 (purple) and~7 (orange) neighbors -- encode information about effective material properties. Left:~Delaunay triangulation overlaid with crystal. Right:~Embryo centroid positions (gray dots) collected over 80\,min, black dots depict average positions identified as lattice sites (SI Sec.~3.11.1). Scale bar,~500\,$\mu$m. \textbf{b,}~Measured shear angle $\alpha_s$ seen along~$\phi$ for data in \textbf{a} (gray dots) and averages at lattice sites (black symbols, error bars depict standard deviation). Dashed line:~$\alpha_s(\phi)$ predicted for passive elastic solids (no fit parameter). Solid line:~Best fit including contributions from an isotropic odd elastic solid~\cite{braverman2020topological} (SI~Sec.~3.11.4). \textbf{c,}~Odd elastic moduli obtained from fits of spatial shear strain profiles predict measured divergence and curl strain components with good quantitative agreement~(SI~Sec.~3.11.5). Scale bar,~0.4\,mm. \textbf{d,}~Snapshot of embryo displacements during cluster oscillations~(SI~Sec.~3.9). Inset: $x$- and $y$- displacement components of a representative embryo indicate robust oscillations with frequency $\approx 0.26\,$min$^{-1}$. Scale bar,~1\,mm. \textbf{e,}~Space-time kymographs of the strain components divergence and curl along the boundary (SI Sec.~3.9). Oscillation with similar amplitude are also present in the bulk~(SI Fig.~S10). \textbf{f,}~Spatial map of the partial entropy production rate computed in the strain component space of curl and divergence (SI Sec.~3.12.1). Scale bar,~1\,mm. Inset: Probability density current in the curl-divergence strain component space computed at the location of the black box shows a representative counter-clockwise strain cycle (SI~Sec.~3.12.2).
\textbf{g},\textbf{h,}~An analogous analysis in the shear strain component space yields similar results as in~\textbf{e} and~\textbf{f}.}
\end{figure}

\subsection*{Signatures of odd elasticity and emergence of chiral displacement waves}
Starfish embryos are inherently chiral and spin about their AP axis in a left-handed manner (Figs.~1d, 2a). This chiral spinning motion leads to distance-dependent, transverse lubrication interactions between pairs of embryos (Fig.~2d). A coarse-graining of our minimal model~(SI Sec.~2.3) suggests that these interactions could lead to effective material properties of LCCs that emulate an odd elastic material~\cite{Scheibner:2020gm,braverman2020topological}. Odd elasticity theory complements the conventional elastic response of passive isotropic solids to compression and shear -- a bulk modulus $B$ and a shear modulus $\mu$ -- by odd bulk and shear moduli $A$ and $K^o$, respectively. Odd elasticity can emerge in active isotropic solids that are chiral~\cite{Scheibner:2020gm}. 
\\

To identify signatures of odd elasticity in our experimental data, we use the fact that LCCs typically harbor lattice defects~(Figs.~1d, 4a). The defects locally deform LCCs, with the deformation field encoding  information about the effective material properties. We quantify deformations in the frame co-rotating with the cluster by measuring the displacement $\mathbf{u}(\mathbf{r}_i,t)=(u_x,u_y)^\top$ of embryos at position~$\mathbf{r}_i$ and time $t$ from a regular lattice (SI~Sec.~3.9). By computing the displacement gradient tensor $u_{ij}=\partial_iu_j$ ($i=x,y$) (SI~Secs.~3.9, 3.11), relative deformations can be expressed in terms of four strain components~\cite{Scheibner:2020gm}: divergence ($u^0=u_{xx}+u_{yy}$), curl/rotations ($u^1=u_{yx}-u_{xy}$), and shear strain components ``shear~1" ($u^2=u_{xx}-u_{yy}$) and ``shear~2" ($u^3=u_{yx}+u_{xy}$). We first analyzed profiles of the local shear elongation angle $\alpha_s(\phi)=\text{arg}(u^2+iu^3)/2$ (Fig.~4a, inset) measured at different lattice sites (Fig.~4a) surrounding a defect pair. In a general isotropic linearly elastic solid, $\alpha(\phi)$ is independent of the distance from the defect~\cite{braverman2020topological} and in a conventional passive solid, where all moduli except $B$ and $\mu$ vanish, $\alpha_s$ is parameter-free~\cite{braverman2020topological} (dashed line, Fig.~4b). In contrast, our measured values of $\alpha_s$ (Fig.~4b, gray dots) averaged at lattice sites (black symbols) show a more complex pattern that can only be explained by allowing for a more exotic material response that may include nonvanishing odd moduli~\cite{braverman2020topological} (solid line, Fig.~4b, SI~Sec.~3.11.4).  We then fit in a second step the full spatial profiles of the shear strain components $u^2$ and~$u^3$ (SI~Sec.~3.11.5), which yields results consistent with the shear angle analysis (Fig.~4b), and in addition provides the relative odd moduli estimates $A/\mu\approx8$ and $K^o/\mu\approx7$. Finally, we validate these fit results by predicting the remaining strain components $u^0$ and $u^1$ (Fig.~4c, top) that had not been used so far and find very good quantitative agreement with experiments~(Fig.~4c, bottom).\\

The presence of odd moduli raises the possibility that LCCs can support self-sustained chiral waves and strain cycles, similar to those recently predicted in odd elastic materials~\cite{Scheibner:2020gm}. In the displacement fields $\mathbf{u}(\mathbf{r}_i,t)$ introduced above, we indeed observe the propagation of chiral displacement waves~(Fig.~4d, SI~Secs.~3.9, 3.10) that persist for more than an hour. The existence of such waves in an overdamped LCC is a direct manifestation of its nonequilibrium nature. The frequency of the dominant, chiral modes~(SI~Fig.~S10), 0.28\,min$^{-1}$, is close to the spinning frequency of individual embryos within the cluster, 0.33\,min$^{-1}$, (SI~Sec.~3.4), suggesting that these modes are directly linked to the spinning motion of embryos.
\\ 

A generic feature of systems with nonreciprocal interactions is that mechanical work can be extracted from quasistatic cyclic processes. Specifically, in materials with an odd elastic response, work can be extracted from cyclic deformations (strain cycles)~\cite{Scheibner:2020gm}~(SI~Sec.~2.2.1). To investigate whether strain cycles exist in an LCC, we determined the displacement gradient tensor, $u_{ij}=\partial_iu_j$ with $i,j\in\{x,y\}$. As evident from kymographs measured along the boundary (Fig.~4e,g) and in the bulk (SI~Sec.~3.9) of the LCC, all strain components exhibit long-lived oscillation that span the whole cluster. Moreover, in the space of suitable strain component pairs (insets Fig.~4f,h), strain cycles are found that have the same handedness almost everywhere in the cluster~(SI~Sec.~3.12.2). Such strain cycles are theoretically predicted as part of the chiral waves that odd elastic solids can support~\cite{Scheibner:2020gm}. Together with the signs of the measured odd moduli, $A,\,K^o>0$, we conclude that oscillating LCCs are effectively doing work on the surrounding fluid~(SI~Sec.~2.3.1).
\\

Strain waves in materials with finite odd elastic moduli can give rise to work and dissipation cycles~\cite{Scheibner:2020gm}. To quantify the lower bounds of the associated entropy production rates, we estimated the statistical irreversibility of strain cycles using recently developed frameworks of stochastic thermodynamics~\cite{battle2016broken,Li:2019dy}. By calculating the local phase-space currents in strain space~(SI~Sec.~3.12), we constructed spatial maps of the local entropy production rates arising in the relevant strain component spaces (Fig.~4f,h). These maps reveal spatio-temporal variations of the entropy production rates, with higher rates appearing mostly in the vicinity of vacancy defects and boundary regions. Spatially integrated entropy production rates exhibit in both spaces temporal maxima during the period of most active wave propagation (SI~Fig.~S15). 
 
\subsection*{Discussion}

Our combined experimental and theoretical results demonstrate how morphological changes in developing multicellular organisms can lead to the self-assembly and dissolution of living crystals with broken chiral symmetry. By observing starfish embryos over two days post gastrulation, we have identified hydrodynamic and morphological single-embryo properties that facilitate these self-organized processes. Over the course of several hours, thousands of embryos can come together to form a macroscopic non-equilibrium  material that carries signatures of odd elasticity. Driven by the embryos' inherent activity, these living crystal structures   support self-sustained chiral waves that exemplify  upward energy transport from the individual microscopic constituents to the macro-scale. More broadly, such living chiral crystals can serve as a paradigmatic active matter system to elucidate  principles of collective self-organization, nonequilibrium  thermodynamics, and exotic material properties that emerge from nonreciprocal interactions.

\newpage
\subsection*{Acknowledgments}
We thank Colin Scheibner, William Irvine, Ned Wingreen, Jinghui Liu and Yu-Chen Chao for helpful discussions. This research was supported by the Sloan Foundation Grant (G-2021-16758) to N.F. and J.D., and the National Science Foundation CAREER Award to N.F.. T.H.T. acknowledges support from the NSF-Simons Center for Mathematical and Statistical Analysis of Biology at Harvard (award number 1764269) and Harvard Quantitative Biology Initiative as NSF-Simons Postdoctoral Fellow. T.H.T. acknowledges support from the Center for Systems Biology Dresden as ELBE Postdoctoral Fellow. A.M. acknowledges support from a Longterm Fellowship from the European Molecular Biology Organization (ALTF~528-2019) and a Postdoctoral Research Fellowship from the Deutsche Forschungsgemeinschaft (Project~431144836). Y.C. acknowledges support from MIT Department of Physics Curtis Marble Fellowship. P.J.F. and S.G. acknowledge support from the Gordon and Betty Moore Foundation as Physics of Living Systems Fellows through grant no. GBMF4513. J.D. was supported by the Robert E. Collins Distinguished Scholarship fund. N.F., J.D. and S.G. are grateful to KITP program \lq ACTIVE20: Symmetry, Thermodynamics and Topology in Active Matter\rq, supported in part by the National Science Foundation under grant no. NSF PHY-1748958. We thank the MIT SuperCloud~\cite{Reuther2018} for providing access to its HPC resources.
\\

\bibliographystyle{unsrt}
\bibliography{references}

\end{document}


\begin{center}
\Huge{\textbf{Supplementary Information}}
\end{center}
\begin{center}
\LARGE{\textbf{Odd dynamics of living chiral crystals}} 
\end{center}

\normalsize
Tzer Han Tan$^{1,2,3,5}$, Alexander Mietke$^{4,5}$, Junang Li$^{1,6}$, Yuchao~Chen$^{1,6}$, Hugh Higinbotham$^1$, \linebreak Peter~J.~Foster$^1$, Shreyas~Gokhale$^1$, J\"{o}rn~Dunkel$^4$, Nikta~Fakhri$^{1,7}$
\vspace{-0.25cm}

\begin{itemize}[leftmargin=0.15in]\small
\item[$^1$] Department of Physics, Massachusetts Institute of Technology, Cambridge, MA, USA \vspace{-0.3cm}
\item[$^2$] Quantitative Biology Initiative, Harvard University, Cambridge, MA, USA \vspace{-0.3cm}
\item[$^3$] Center for Systems Biology Dresden, Dresden, Germany \vspace{-0.3cm}
\item[$^4$] Department of Mathematics, Massachusetts Institute of Technology, Cambridge, MA, USA \vspace{-0.3cm}
\item[$^5$] These authors contributed equally and are joint first authors. \vspace{-0.3cm}
\item[$^6$] These authors contributed equally. \vspace{-0.3cm}
\item[$^7$] Corresponding author: fakhri@mit.edu
\end{itemize}

\vspace{1cm}
\normalsize
\renewcommand\contentsname{} 
\tableofcontents

\newpage
\section{Experiment}

\subsection{Preparation of starfish embryos}\label{sec:Prep}
Starfish \emph{Patiria Miniata} were procured from South Coast Bio-Marine LLC. The animals were kept in salt water fish tanks maintained at 15$\,^\circ$C. To fertilize  embryos, we first extracted oocytes and sperm separately. Ovaries were extracted through a small incision made at the bottom of the starfish and were then carefully fragmented using a pair of scissors to release the oocytes. Extracted oocytes were washed twice with calcium free seawater to prevent maturation and incubated in artificial filtered seawater (FSW) at 15$\,^\circ$C. The testes were extracted similarly and kept in a 1\,mL Eppendorf tube at 4$\,^\circ$C. To fertilize embryos, we first matured the oocytes by adding the hormone 1-methyladenine. After one hour, sperm extract was added to the culture at a 1:10000 dilution. Fertilized embryos were cultured in FSW at 15$\,^\circ$C for the first 24~hours before being moved to 20$\,^\circ$C. 

\subsection{Clustering and flow field experiments}\label{sec:PIVexp}
All experiments have been performed at 20$\,^\circ$C. For clustering experiments, the appropriate number of embryos was transferred to a well of a 24-well plate (VWR sterilized tissue culture plates, Catalog Number 10861-558, single well diameter: $15.7\,$mm). The total water level height was, in all well-experiments, approximately 14\,mm. Images were taken in 10\,s intervals using a dissection scope (Nikon SMZ745) with a high-speed CMOS digital camera (Amscope MU500) attached at the eyepiece. 

To measure flow fields around embryos, 2\,$\mu$m sized polystyrene beads (Bangs Lab, Catalog Number PS05001) were added to the medium. ``Top-view" cross-sectional flow field experiments (main text Fig.~2a--c, Fig.~\ref{fig:PIVfittingtrip}) were performed with embryos oriented vertically near the fluid-air interface. Maximum projections of the bead dynamics are used to visualize stream lines (main text Fig.~1a). The corresponding data analysis is detailed in Sec.~\ref{sec:FFanalsis}. ``Side-view" flow field experiments (main text Fig~1b, Fig.~\ref{fig:SideViewFitResults}) were performed by confining embryos and beads in a flow cell, where double-sided tape was used as a spacer to generate a flat channels with a height of $\approx$ 100\,$\mu$m. The flow cell was sealed with Valap, and flows were imaged using bright-field illumination on a Leica microscope (DMIL LED) at a 50\,ms interval with either a 10x/0.25 (N PLAN CY) or a 4x/0.10 (HI PLAN) objective. The corresponding data analysis is detailed in Sec.~\ref{sec:FFanalsisSide}. Embryo morphology images (main text Fig.~1b and Fig.~\ref{fig:Morphology}) at different time points were obtained using the same experimental setup.

\subsection{Verification of proper development after cluster formation}\label{sec:Morp}
\begin{figure}[!t]
\centering
\includegraphics[width=0.97\textwidth]{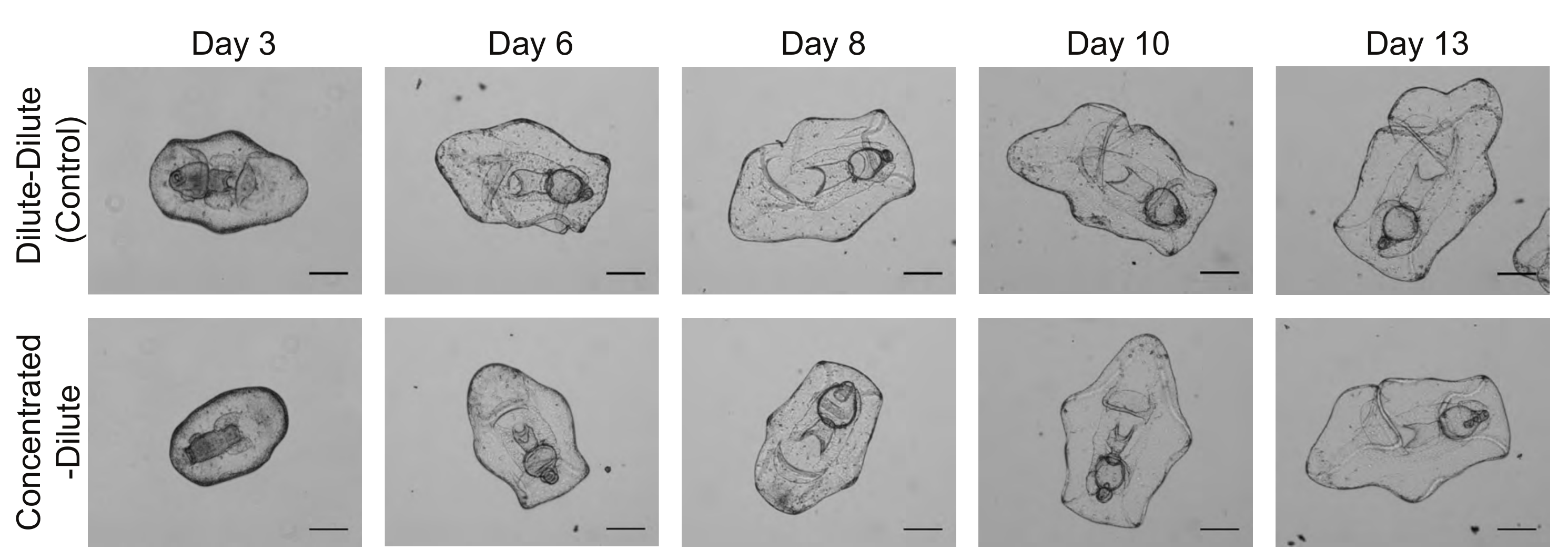}
\caption{\textbf{Embryo development after cluster formation.} Cluster formation and dissolution takes place during the first 3 days post fertilization~(dpf).
Top row:~Morphologies of embryos that were right after fertilization cultured at low density ($<10$ embryos/ml). Time points indicate dpf. Bottom row:~Embryos that underwent cluster formation were transferred at 3\,dpf to a dilute well for further culturing and imaging up to 13\,dpf. Morphologies of embryos that formed living chiral crystals show no notable differences when compared to the control (top-row). For experimental details see Sec.~\ref{sec:Morp}. Scale bar,~100\,$\mu$m.}
 \label{fig:Morphology}
\end{figure}

To verify that the embryos develop properly after they formed a living chiral crystal, we performed a standard clustering experiment (see Sec.~\ref{sec:PIVexp}) until about 72 hours post fertilization (hpf) when clusters have dissolved. The embryo solution was then diluted down to a low concentration of less than $10$ embryos/ml and randomly selected embryos were regularly imaged for up to 10 days following cluster dissolution~(Fig.~\ref{fig:Morphology}, bottom row). At this point, embryos had developed well into the bipinnaria stage~($\approx3$ days post fertilization). As a control, we cultured embryos right after fertilization at a low density ($<10$ embryos/ml). The morphology of these control embryos (Fig.~\ref{fig:Morphology}, top row), is comparable to the morphology of embryos that underwent cluster formation, indicating that the formation of living chiral crystal does not interfere with the development into more advanced larval stages. We note that, even when embryo suspensions were kept at high density up to day 13 post fertilization, no additional collective phenomena comparable to the cluster formation during the first 2 to 3 days post fertilization were observed.

\subsection{Sedimentation of inactive embryos}\label{sec:SedExps}
To estimate the embryo's negatively buoyant weight force $F_g$, we performed a sedimentation experiment. Embryos were immobilized by treatment with 1\,mM of the metabolic inhibitor~\cite{ishii2014mitochondrial} Sodium Azide $NaN_3$. They were then transferred to a long Pasteur pipette (diameter 7\,mm) and released from the top position. The process of immobilized embryo sedimentation was video-recorded and embryos were tracked to determine sedimentation velocities $v_s$. Using the measured velocities, $v_s=(0.6\pm0.1)\,$mm/s (mean\,$\pm$\,standard deviation, $n=7$), an effective embryo size of $L=(\ell_{\text{min}}+\ell_{\text{maj}})/2\approx(150\pm30)\,\mu$m  during mid to later stages of experiments (see Fig.~\ref{fig:SideViewFitResultsfortheo}c,d), we find $F_g/\eta=(1.7\pm0.4)\,$mm$^2$/s from equating the approximate Stokes drag $6\pi\eta Lv_s$ with the weight force $F_g$. Using viscosity of water ($\eta=1\,$mPa$\cdot$s at 20$\,^\circ$C) the negatively buoyant weight force is therefore of the order $F_g\approx1-2\,$nN.

\section{Theory}
The length and time scales, $L\sim 150\,\mu$m (typical embryo size, see Fig.~\ref{fig:SideViewFitResultsfortheo}c,d) and $v\sim 0.1\,$mm/s (typical flow speed), relevant for the embryo dynamics correspond to a Reynolds number of $\text{Re}\sim1.5\times10^{-2}$ and therefore allow for a theoretical description of our system in terms of overdamped Stokesian hydrodynamics. Within this framework, we provide in this section a quantitative description of the hydrodynamics and orientational stability of surface-bound embryos~(Sec.~\ref{sec:Hydbd}), introduce an experimentally parametrized minimal model of cluster formation~(Sec.~\ref{sec:HydrModel}) and discuss effective macroscopic material properties expected to arise from the microscopic interactions among embryos~(Sec.~\ref{sec:MappingDetails}).

\subsection{Hydrodynamics of single embryos and their bound states}\label{sec:Hydbd}
The fluid flows surrounding embryos are generated by the beating of cilia that cover the embryo's body. Such an interactions with the fluid can lead to the propulsion and rotation of embryos, and it affects how embryos interact with each other. For low-Reynolds number microswimmers it is well-established that their dynamics and hydrodynamic interactions can be modulated by nearby fluid-substrate interfaces~\cite{Dufresne2000,Squires2001,Drescher2009,Spagnolie2012} or free fluid surfaces~\cite{Lauga2020}. Building on the corresponding ideas, we aim to develop in this section a quantitative and mechanistic understanding of how starfish embryos form stable surface bound states and how interactions between them lead to the observed phenomenology of cluster formation and dissolution. While the analysis presented here focuses exclusively on hydrodynamic effects, we note that also nonhydrodynamic effects, such as bottom- or top-heaviness, can in principle affect the orientational stability of microswimmers~\cite{Drescher2009}.

The theoretical analysis of hydrodynamic embryo properties and bound states proceeds in several steps: First, using Stokes flow singularities, we introduce a generic description of the flow fields surrounding a microswimmer below a free fluid surface as a function of its body axis orientation (Secs.~\ref{sec:FFms} and \ref{sec:Image}). Second, using Faxen's law, we derive a criterion for the orientational stability of such a generic microswimmer as function of the singularity amplitudes~(Sec.~\ref{sec:StabAnalFull}). In the last step, we explain how the singularity amplitudes can be parametrized by experiments as a function of developmental time~(Sec.~\ref{sec:DevModePara}) and evaluate the derived stability criterion~(\ref{sec:stabtimefinal}). Finally, we focus on the dominant Stokeslet singularity and show that it is essential for both the upright bound state stability and for generating an effective hydrodynamic attraction that facilitates the formation of clusters~(\ref{sec:Stokeslet}). 

\subsubsection{Far-field description of fluid flow surrounding microswimmers}\label{sec:FFms}
To understand the hydrodynamic effects that contribute to bound state stabilization and destabilization near a fluid-air interface, we require a tractable parametrization of the fluid flows that surround embryos. Here, we use a systematic expansion in terms of flow singularities that are constructed from the Green's function of the Stokes equation $\eta\nabla^2\mathbf{v}-\nabla p=0$, where $\eta$ is the viscosity and the pressure $p$ is determined by the incompressibility condition $\nabla\cdot\mathbf{v}=0$. Up to order $1/r^3$, the relevant axisymmetric flow singularities are given by~\cite{Spagnolie2012,Kim2013}
\begin{subequations}\label{eq:singall}
\begin{align}
\mathbf{v}_{\text{st}}(\mathbf{r};\mathbf{p})&=a\frac{R_0}{r }\left[\mathbf{p}+(\mathbf{p}\cdot\mathbf{e}_r)\mathbf{e}_r\right]\hspace{6.5cm}\text{(Stokeslet)}\label{eq:Stokeslet_uf}\\
\mathbf{v}_{\text{fd}}(\mathbf{r};\mathbf{p})&=b\frac{R_0^2}{r^2 }\left[3(\mathbf{p}\cdot\mathbf{e}_r)^2-1\right]\mathbf{e}_r\hspace{5.7cm}\text{(Force-dipole)}\label{eq:Fd}\\
\mathbf{v}_{\text{fq}}(\mathbf{r};\mathbf{p})&=c\frac{R_0^3}{r^3 }\left[\mathbf{p}-3(\mathbf{p}\cdot\mathbf{e}_r)^2\mathbf{p}-9(\mathbf{p}\cdot\mathbf{e}_r)\mathbf{e}_r+15(\mathbf{p}\cdot\mathbf{e}_r)^3\mathbf{e}_r\right]\hspace{0.55cm}\text{(Force-quadrupole)}\label{eq:Fq}\\
\mathbf{v}_{\text{sd}}(\mathbf{r};\mathbf{p})&=d\frac{R_0^3}{r^3 }\left[3(\mathbf{p}\cdot\mathbf{e}_r)^2\mathbf{e}_r-\mathbf{p}\right]\hspace{5.5cm}\text{(Source-dipole)}.\label{eq:Sd}
\end{align}
\end{subequations}
Here, $\mathbf{r}=(x,y,z)^\top$, $r=|\mathbf{r}|$, $\mathbf{e}_r=\mathbf{r}/r$, and~$\mathbf{p}$ is a unit vector ($|\mathbf{p}|=1$) that will be identified below with the embryo's major body axis. The length scale $R_0$ is in general related to the size of the microswimmer. However, its concrete value depends on the specific context in which the singularities Eqs.~(\ref{eq:singall}) are used and may differ from the length $L=(\ell_{\text{min}}+\ell_{\text{maj}})/2\approx150\,\mu$m (see Fig.~\ref{fig:SideViewFitResultsfortheo}c,d) that we use mostly for basic scale arguments. Due to the absence of systematic rotary flows surrounding single embryos, rotlet contributions with $|\mathbf{v}|\sim1/r^3$ are not included in this analysis and only become relevant in the case of embryo interactions (Sec.~\ref{sec:PIVfit}). 
The coefficients $a,b,c$ and $d$ in Eqs.~(\ref{eq:singall}) are determined by the details of how the microswimmer interacts with the fluid, in particular by the flows generated near the surface of the microswimmer's body. To make this connection explicit, we consider a nondeforming spherical surface of radius~$R_0$ at rest, $v_r(r=R_0)=0$ and $\mathbf{v}(r\rightarrow\infty)=0$, that is immersed in an unbound fluid and generates a tangential surface flow
\begin{equation}\label{eq:vtheta}
v_{\theta}(r=R_0)=\alpha\sin\theta+\beta\sin2\theta+\gamma\sin3\theta.
\end{equation}
Here, $(r,\theta,\phi)$ define spherical coordinates that are related to the Cartesian coordinates above by $x=r\sin\theta\cos\phi,\,y=r\sin\theta\sin\phi,\,z=r\cos\theta$, and $\alpha$, $\beta$ and $\gamma$ denotes parameters that will later be determined from experiments (see Secs.~\ref{sec:DevModePara} and \ref{sec:FFanalsisSide}). Using analytic solutions of the Stokes equation~\cite{Mietke2019}, implementing the boundary condition Eq.~(\ref{eq:vtheta}) and comparing the solution with the singularities in Eqs.~(\ref{eq:singall}) for $\mathbf{p}=\mathbf{e}_z$, one finds far field flows of the form 
\begin{equation}\label{eq:SingExp}
\mathbf{v}=\mathbf{v}_{\text{st}}+\mathbf{v}_{\text{fd}}+\mathbf{v}_{\text{fq}}+\mathbf{v}_{\text{sd}}+\mathcal{O}(R_0^4/r^4)    
\end{equation}
with coefficients given by
\begin{equation}\label{eq:SingCoeffs}
a=\frac{\gamma}{10}-\frac{\alpha}{2}\hspace{1cm}b=-\beta\hspace{1cm}c=-\frac{2\gamma}{3}\hspace{1cm}d=\frac{\alpha}{2}-\frac{11\gamma}{30}.
\end{equation}
In Sec.~\ref{sec:DevModePara}, we describe how flow field measurements can be used to parametrize the coefficients $\alpha$, $\beta$ and $\gamma$ in Eq.~(\ref{eq:vtheta}) over the relevant developmental time frame.

The orientation $\mathbf{p}=\mathbf{e}_z$ used to derive the coefficients in Eq.~(\ref{eq:SingCoeffs}) reflects the azimuthal symmetry axis of the boundary condition Eq.~({\ref{eq:vtheta}}) and therefore fully determines the symmetry axis of this solution. Using the same coefficients $a,b,c$ and~$d$ with another orientation $\mathbf{p}=\mathbf{p}_0$ in the flow singularities Eqs.~(\ref{eq:singall}) therefore yields a far field flow that solves an analog boundary value problem to order $\mathcal{O}(R_0^4/r^4)$ in which the boundary condition Eq.~({\ref{eq:vtheta}}) is rotated such that the azimuthal symmetry axis is given by $\mathbf{p}_0$. Hence, the singularity expansion Eq.~(\ref{eq:SingExp}) with coefficients given in Eq.~(\ref{eq:SingCoeffs}) provides a tractable parametrization of flows surrounding a microswimmer with arbitrary orientation~$\mathbf{p}$.

\subsubsection{Image construction near free fluid surfaces}\label{sec:Image}
In the next step, we describe how the far-field flows derived in the previous section are modified by the nearby fluid-air interfaces. We place the origin of the coordinate system on the fluid surface and align the $z$-axis with the surface normal pointing away from the fluid. For simplicity, we do not take into account shape changes of the fluid surface due to the presence of embryos and treat it as a nondeforming and shear stress-free interface. This corresponds to nondeforming, stress-free boundary conditions $v_z|_{z=0}=0$ and $E_{xz}|_{z=0}=E_{yz}|_{z=0}=0$, where \hbox{$\mathbf{E}=[\nabla\mathbf{v}+(\nabla\mathbf{v})^\top]/2$} is the strain rate tensor. Consider then any of the flow singularities in Eqs.~(\ref{eq:singall}) at a position $\mathbf{r}_0=(x_0,y_0,-h)^{\top}$, where $h>0$ is the distance below the surface. Flow fields $\mathbf{v}=\bar{\mathbf{v}}$ that obey nondeforming, free boundary conditions at $(x,y,0)$ can be constructed by complementing the singularity with a suitable image~\cite{Lauga2020} above the surface at $\mathbf{r}'_0=(x_0,y_0,h)^{\top}$. For example, the flow field
\begin{equation}
\bar{\mathbf{v}}_{\text{st}}(\mathbf{r};\mathbf{p},\mathbf{r}_0)=\mathbf{v}_{\text{\text{st}}}(\mathbf{r}-\mathbf{r}_0;\mathbf{p})+\mathbf{v}_{\text{\text{st}}}(\mathbf{r}-\mathbf{r}_0';\mathbf{p}')\label{eq:Stokeslet}
\end{equation}
with $\mathbf{p}'=(p_x,p_y,-p_z)^{\top}$ corresponds to a solution of the Stokes equation that describes fluid flows originating from a Stokeslet at $\mathbf{r}_0=(x_0,y_0,-h)^{\top}$ below a nondeforming, stress-free fluid surface at $z=0$. The same construction as Eq.~(\ref{eq:Stokeslet}) also yields force-dipole, force-quadrupole and source-dipole flows that solve the Stokes equation below a nondeforming, stress-free fluid surface.

\subsubsection{Stationary orientations and stability near free surfaces}\label{sec:StabAnalFull}
Using the flow parametrization derived in Secs.~\ref{sec:FFms} and~\ref{sec:Image}, we can now study how hydrodynamic effects influence the orientation of microswimmers near a free surface. To this end, we identify the unit vector $\mathbf{p}$ as the embryo's major body axis pointing from posterior (P) to anterior (A) and use the generalized Faxen's law for a prolate ellipsoidal body~\cite{Kim2013,Spagnolie2012}: 
\begin{align}
\frac{d\mathbf{p}}{dt}&=\frac{1}{2}\left[\nabla\times\mathbf{v}'\right]\times\mathbf{p}+\chi[\mathbf{p}\times\left(\mathbf{E}'\cdot\mathbf{p}\right)]\times\mathbf{p}.\label{eq:Faxen}
\end{align}
Equation~(\ref{eq:Faxen}) describes the dynamics of the body axis orientation $\mathbf{p}$ in response to external flows~$\mathbf{v}'$, including image flow contributions. The corresponding strain rate tensor is given by $\mathbf{E}'=\left[\nabla\mathbf{v}'+(\nabla\mathbf{v}')^\top\right]/2$. The geometric parameter $\chi=(1-e^2)/(1+e^2)$, with $e=\ell_{\text{min}}/\ell_{\text{maj}}$ being the semi-minor~($\ell_{\text{min}}$) to semi-major axis~($\ell_{\text{maj}}$) ratio, captures shape anisotropies of the immersed body with $\chi=0$ for a perfect sphere and $\chi>0$ for any elongated, ellipsoidal body. For a single microswimmer below the fluid surface, $\mathbf{v}'$ in Eq.~(\ref{eq:Faxen}) is given by the microswimmer's image flow. For the image flow $\mathbf{v}'=\mathbf{v}(\mathbf{r}-\mathbf{r}_0';\mathbf{p}')$ with $\mathbf{v}$ given in Eq.~(\ref{eq:SingExp}) and singularities listed in Eqs.~(\ref{eq:singall}), we find from Eq.~(\ref{eq:Faxen}) the director dynamics
\begin{equation}\label{eq:Dynpcart}
\frac{dp_x}{dt}=\frac{3}{16}f(p_z)p_zp_x\hspace{1cm}\frac{dp_y}{dt}=\frac{3}{16}f(p_z)p_zp_y\hspace{1cm}\frac{dp_z}{dt}=\frac{3}{16}f(p_z)(p_z^2-1),    
\end{equation}
where 
\begin{equation}\label{eq:fp}
f(p_z)=\frac{2bR_0^2}{h^3}p_z+\frac{cR_0^3}{h^4}(5p_z^2-1)+\chi\left(\frac{4aR_0}{h^2}p_z+\frac{2bR_0^2}{h^3}p_z^3+\frac{cR_0^3}{h^4}(p_z^2-1)(3p_z^2+2)+\frac{dR_0^3}{h^4}(1+p_z^2)\right).    
\end{equation}
Here, $a,b,c$ and $d$ denote amplitudes of the different singularities listed in Eqs.~(\ref{eq:singall}) and $h$ is the distance to the fluid surface. Equations~(\ref{eq:Dynpcart}) imply $\frac{d}{dt}(p_xp_y^{-1})=0$, indicating that the dynamics of $\mathbf{p}$ is at all times confined to a single plane that contains the $z$-axis and is fixed by initial conditions. Without loss of generality, this plane is chosen in the following to be the $xz$-plane. Together with the normalization of the director $|\mathbf{p}|=1$, only one dynamic degree of freedom remains. Parameterizing the unit director as $p_x=\sin\theta_{\text{p}}$ and $p_z=\cos\theta_{\text{p}}$, Eqs.~(\ref{eq:Dynpcart}) finally yield
\begin{equation}\label{eq:OrientDyn}
\frac{d\theta_{\text{p}}}{dt}=\frac{3}{16}f(p_z)\sin\theta_{\text{p}}.
\end{equation}

\textit{Stationary orientations and linear stability: }The dynamics Eq.~(\ref{eq:OrientDyn}) has stationary orientations $\theta_{\text{p}}^*=0$ and $\theta_{\text{p}}^*=\pi$, corresponding to $p_z^*=\cos\theta_{\text{p}}^*=\pm1$ and therefore rationalizes the perfectly upright oriented state $p_z^*=1$ seen in experiments. Additionally, skewed stationary orientations with $\sin\theta_{\text{p}}^*\ne0$ exist that can be found numerically from~$f(p_z^*)=0$. Stationary orientations are stable if 
\begin{equation}\label{eq:StabCrit}
-\sin^2\theta_{\text{p}}^*\left.\frac{df}{dp_z}\right|_{p_z^*}+\cos\theta_{\text{p}}f(p_z^*)<0
\end{equation}
and unstable if the left-hand side of Eq.~(\ref{eq:StabCrit}) becomes larger than zero. The stability properties of the stationary upright orientation $\theta_{\text{p}}^*=0$ ($p_z^*=1$) can according to Eq.~(\ref{eq:StabCrit}) directly be read off the sign of $f(1)$ for $f(p_z)$ given in Eq.~(\ref{eq:fp}), once the singularity amplitudes $a,b,c$ and $d$ have been specified. The latter are parametrized through Eq.~(\ref{eq:SingCoeffs}) by developmental changes of the flows generated near embryo surfaces~(Sec.~\ref{sec:DevModePara}), leading to different sationary orientations and stability properties as embryo development progresses (Sec.~\ref{sec:stabtimefinal}).

\begin{figure}[!t]
\centering
\includegraphics[width=0.99\textwidth]{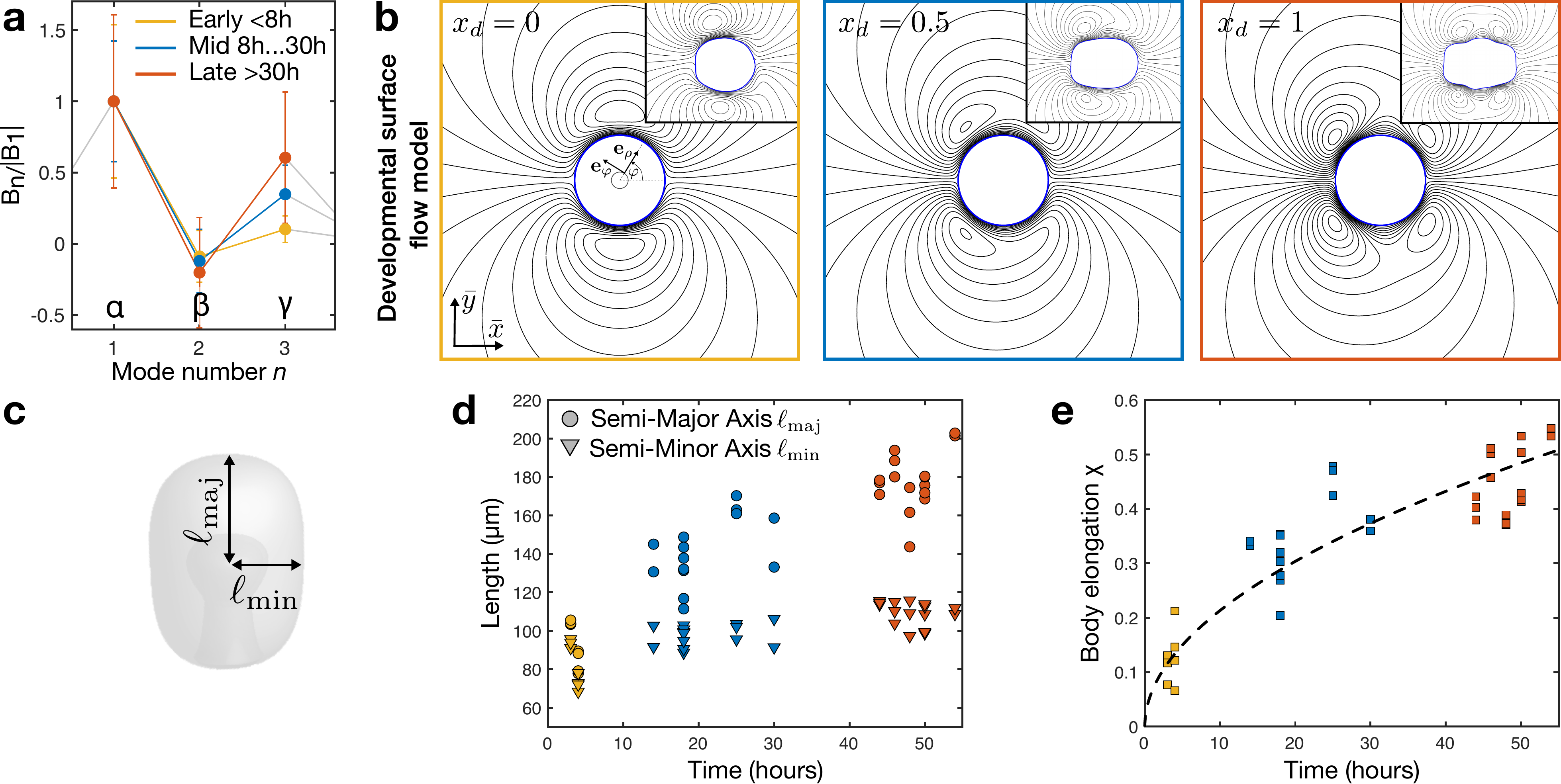}
\caption{\textbf{Developmental parametrization of lateral embryo surface flows and body aspect ratios.} \textbf{a,}~Normalized mode coefficients over development (mean and standard deviation) used to parametrize flows generated on the embryo surface via Eq.~(\ref{eq:vtheta}). Values were extracted from experimentally measured flows surrounding confined embryos (see Sec.~\ref{sec:PIVfitSide}, Fig.~\ref{fig:SideViewFitResults}) and correspond to generalized squirmer parameters~\cite{Lauga2020}. \textbf{b,}~Validating the minimal parametrization of embryo surface flows: Solutions of the Hele-Shaw flow Eq.~(\ref{eq:HS}) using boundary condition Eq.~(\ref{eq:bcHS}) with values from $\mathbf{a}$ show good agreement with the characteristic streamline features seen in experiments (insets, reproduced from Fig.~\ref{fig:SideViewFitResults}b). \textbf{c,}~Illustration of semi-minor length~$\ell_{\text{min}}$ and semi-major length~$\ell_{\text{maj}}$ used to characterize embryo size. \textbf{d,}~Measurements of $\ell_{\text{min}}$ and $\ell_{\text{maj}}$ over development. For basic scale arguments, we use a characteristic embryo size during mid to later stages $L=(\ell_{\text{min}}+\ell_{\text{maj}})/2\approx150\,\mu$m. \textbf{e,}~The embryo's body elongation is characterized by the geometric parameter $\chi=(1-e^2)/(1+e^2)$ with $e=\ell_{\text{min}}/\ell_{\text{maj}}$. Heuristic linear and power-law interpolations [Eqs.~(\ref{eq:DevParas})] of data in \textbf{a} and \textbf{e} (dashed line), respectively, are used to analyze orientational stability over development of embryos below a free fluid surface (Secs.~\ref{sec:FFms}--\ref{sec:stabtimefinal}, main text Fig.~2j).}
 \label{fig:SideViewFitResultsfortheo}
\end{figure}

\subsubsection{Mode parametrization of embryo surface flows}\label{sec:DevModePara}
The analysis from the previous section allows us to determine stationary body axis orientations and their linear stability properties and connects them through Eqs.~(\ref{eq:singall})--(\ref{eq:SingCoeffs}) directly to the fluid flows generated near the microswimmer's body surface. In the next step, we want to experimentally infer these near-surface embryo flows in a form that can be integrated via Eq.~(\ref{eq:vtheta}) into the stability analysis. To this end, we have measured fluid flows in a ``side-view" plane along the AP body axis of confined embryos at different developmental stages~(Fig.~\ref{fig:SideViewFitResults}a) and extracted a low-dimensional representation in terms of a mode expansion of Hele-Shaw flow solutions~(Sec.~\ref{sec:PIVfitSide}, Fig.~\ref{fig:SideViewFitResults}b). Despite the embryo's developmental body shape changes from a close-to-spherical to more elongated shapes over the course of experiments ($\sim0-50$\,hours), fluid flows in this plane retain a high level of symmetry and, as a result, are captured by a small number of modes that can be specified in cylindrical coordinates (Eq.~(\ref{eq:HSsol}), Fig.~\ref{fig:SideViewFitResults}c). The following two observations allow us to further simplify the parametrization of embryo surface flows~(Fig.~\ref{fig:SideViewFitResultsfortheo}a): $i)$~The modes~$B_n$, in particular $B_1,\,B_2$ and $B_3$ represent the dominant contributions to the fits, and $ii)$~due to the linearity of Eqs.~(\ref{eq:OrientDyn}) in fluid flows, stationary orientations and their stability will only depend on relative values of mode amplitudes. Akin to the definition of a squirmer parameter~\cite{Lauga2020}, we therefore focus in the following on the modes $B_1,\,B_2$ and $B_3$ and normalize each of the pooled data sets (Early, Mid, Late) by the corresponding mode amplitude~$|B_1|$. Note, for this normalization $B_2/|B_1|$ corresponds to a classical squirmer parameter~\cite{Lauga2020} and Fig.~\ref{fig:SideViewFitResultsfortheo}a suggests starfish embryos are weak pushers ($|B_2|/|B_1|\ll1$ with $B_2/|B_1|<0$). 

Finally, we verify that such a simplified parametrization does recapitulate the flow profiles seen in side-view~PIV measurements of confined embryos (Fig.~\ref{fig:SideViewFitResults}). Indeed, when solving the Hele-Shaw Eq.~(\ref{eq:HS}) around a circular unit disk with boundary flow profile
\begin{equation}\label{eq:bcHS}
V_{\varphi}=\alpha\sin\varphi+\beta\sin2\varphi+\gamma\sin3\varphi    
\end{equation}
using $\alpha=1$ and the measured values of $\beta$ and $\gamma$ indicated in Fig.~\ref{fig:SideViewFitResultsfortheo}a, we find flow solutions that exhibit the same evolution of features as seen in experiments (compare Fig.~\ref{fig:SideViewFitResults}a,b with Fig.~\ref{fig:SideViewFitResultsfortheo}b): Initially, an almost centered pair of vortices (closed streamlines) is present at the lateral sides~(Early) and moves over time towards the posterior end of embryos~(Mid), which is a consequence of the increasing pusher amplitude $|\beta|$. At later stages an increasing contribution from the quadrupole mode $\sim\gamma$ leads to the appearance of a second vortex pair near the embryo's anterior~(Late).

\subsubsection{Stationary orientations and stability as function of developmental time}\label{sec:stabtimefinal}
To integrate the two-dimensional surface flow parametrization from the previous section through the boundary condition Eq.~(\ref{eq:vtheta}) and corresponding singularity amplitudes Eqs.~(\ref{eq:SingCoeffs}) into the three-dimensional stability analysis, we assume lateral embryo surface flows are axisymmetric with respect to the AP body axis. In this case, we can identify the azimuthal angle $\varphi$ used in Eq.~(\ref{eq:bcHS}) and defined in Fig.~\ref{fig:SideViewFitResultsfortheo}b with the polar angle used in the stability analysis Secs.~\ref{sec:FFms}--\ref{sec:StabAnalFull}, i.e. $\theta\simeq\varphi$ for $\varphi\in[0,\pi)$ and $\theta\simeq2\pi-\varphi$ for $\varphi\in[\pi,2\pi)$, and may directly use the the coefficients $\alpha=1$, $\beta$ and $\gamma$ (Fig.~\ref{fig:SideViewFitResultsfortheo}a) in Eqs.~(\ref{eq:SingCoeffs}) to determine the different singularity amplitudes. Combining this with measurements of minor and major embryo body axis lengths that define the aspect ratio~$\chi=(1-e^2)/(1+e^2)$, with $e=\ell_{\text{min}}/\ell_{\text{maj}}$ (Fig.~\ref{fig:SideViewFitResultsfortheo}c,d), we can define an interpolated developmental parametrization of the embryo's properties in the form
\begin{equation}\label{eq:DevParas}
\beta(x_d)=-0.2x_d\hspace{1.5cm}\gamma(x_d)=0.6x_d\hspace{1.5cm}\chi(x_d)=0.5x_d^{1/2},
\end{equation}
where $x_d\in[0,1]$ represents a developmental coordinate that maps to experimental time points as $t_d=x_d\times50$\,h. The functions $\beta(x_d)$ and $\gamma(x_d)$ interpolate linearly towards the largest mode amplitudes of $\min(B_2/|B_1|)\approx-0.2$ and $\max(B_3/|B_1|)\approx0.6$~(Fig.~\ref{fig:SideViewFitResultsfortheo}a). The function $\chi(x_d)$ represents the best power-law fit to the data shown in Fig.~\ref{fig:SideViewFitResultsfortheo}e (black dashed line) and heuristically parametrizes developmental changes of the embryo body shape. Using the parametrization Eqs.~(\ref{eq:DevParas}) in Eqs.~(\ref{eq:OrientDyn}) and (\ref{eq:StabCrit}) with $h/R_0=1$ in Eq.~(\ref{eq:fp}) then determines stationary orientations and their linear stability as a function of developmental time~$t_d$, as depicted and discussed in~Fig.~2j of the main text.

\subsubsection{Stokeslet hydrodynamics of bound embryos}\label{sec:Stokeslet}
The analysis of the previous sections was based on fluid field measurements surrounding confined embryos within an effectively two-dimensional geometry. We finally want to verify that the basic orientational stability established by this analysis is compatible with flow fields surrounding surface-bound and freely spinning embryos in the experimental setting used to study cluster formation. In this case, we only have direct experimental access to ``top-view" flow field information recorded for bound embryos from above the fluid surface (main text Fig.~2a,b). From the flow field measurements around confined embryos used in the analysis above, we expect relative contributions to dipole flows $\sim1/r^2$ set by~$\beta$ in Fig.~\ref{fig:SideViewFitResultsfortheo}a are small. The next potentially relevant singularity contribution is short-ranged \hbox{($\sim1/r^3$)}, such that the Stokeslet flow ($\sim1/r$) is expected to be the dominant contribution to ``top-view" flow fields. We therefore focus here on direct measurements of the embryo's Stokeslet strength and its independent validation via sedimentation experiments. Additionally, we discuss in this section the impact of a curved fluid-air interface.\\

{\sffamily\bfseries Stokeslet strength and negative bouyancy}

\begin{figure}[!t]
\centering
\includegraphics[width=0.97\textwidth]{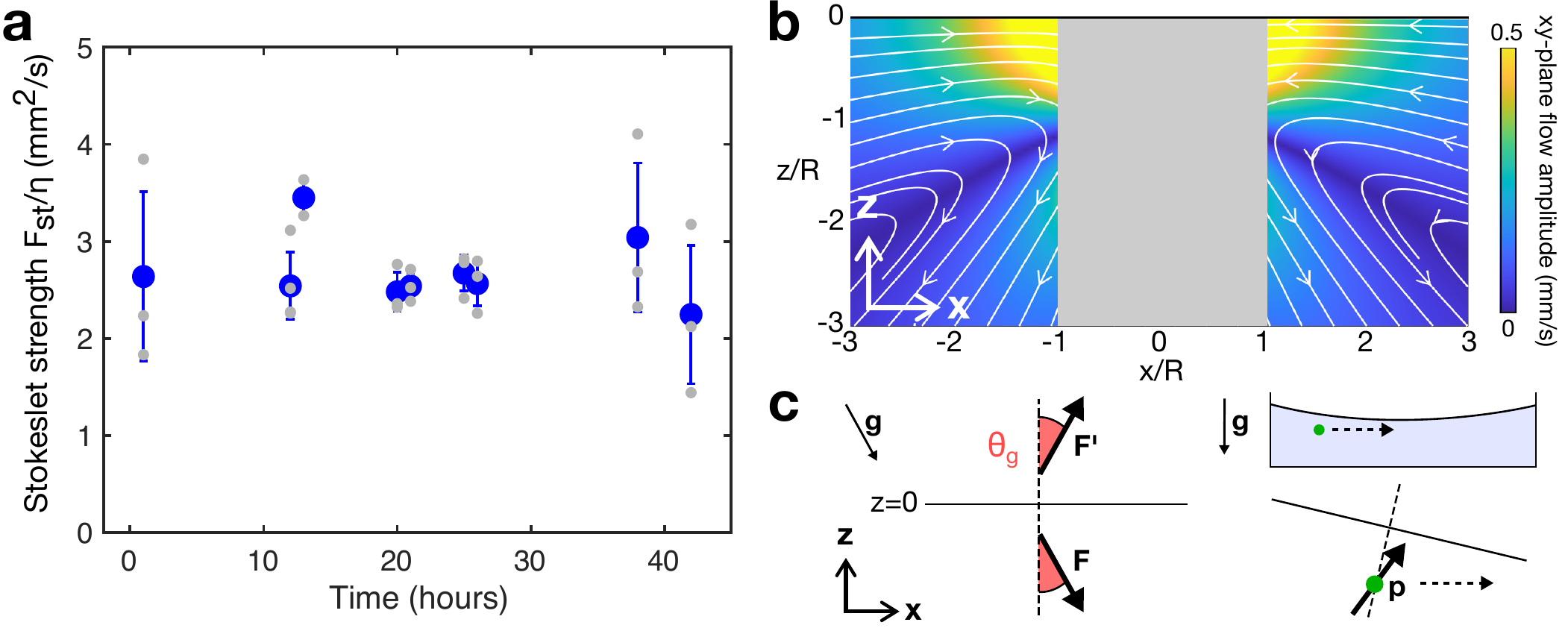}
\caption{\textbf{Stokeslet description of single embryos bound below a free fluid surface.}
\textbf{a,}~Stokeslet strength $F_{\text{st}}$ normalized by fluid viscosity $\eta$ from fits of in-plane flow fields (Sec.~\ref{sec:PIVfit}) over time course of experiments. Gray dots depict individual fit values, blue dots and error bars represent mean and standard deviation, respectively. \textbf{b,}~Side-view of Stokeslet flow below a free surface (Eq.~(\ref{eq:Stokeslet}) with $\mathbf{r}_0=(0,0,-R)^{\top}$, $F_{\text{st}}/\eta=2.6\,$mm$^2$/s, $R=110\,\mu$m). White lines depict stream lines in the $xz$-plane and colors indicate the amplitude of flows parallel to the $xy$-plane. The dominantly lateral in-flow leads to an effective attraction between embryos. \textbf{c,}~Image Stokeslet construction below a free surface ($z=0$) that allows for an angle $\theta_g$ between surface normal and gravity $\mathbf{g}$ (left) to represent embryos below a curved fluid surface present in small wells (right). Under a surface that is locally not orthogonal to gravity, a tilted stationary orientation enhances the accumulation of embryos near the well center (Sec.~\ref{sec:Stokeslet}).}
 \label{fig:StokesletHydrod}
\end{figure}

From sedimentation experiments (Sec.~\ref{sec:SedExps}), we know that starfish embryos are negatively buoyant. When bound and stationary below the surface, the weight force of embryos must then be encoded in the flow fields surrounding the embryo. To test this, we have fitted experimental measurements of top-view flow fields in the $xy$-plane using the image construction Eq.~(\ref{eq:Stokeslet}) with Stokeslet singularity Eq.~(\ref{eq:Stokeslet_uf}) and $\mathbf{p}=\mathbf{e}_z$ (see Sec.~\ref{sec:PIVfit} for details of the fitting procedure). The singularity parameter $a$ in Eq.~(\ref{eq:Stokeslet_uf}) is related to the Stokeslet force $F_{\text{st}}$ by
\begin{equation}\label{eq:Stoka}
a=-\frac{F_{\text{st}}}{8\pi R_0\eta}, 
\end{equation}
where~$\eta$ denotes the fluid viscosity. Consistent with our expectation that the Stokeslet represents the dominant contribution to ``top-view" flow fields, we find that the latter are indeed captured well by the Stokeslet flow Eq.~(\ref{eq:Stokeslet}) below a fluid surface~(Fig.~2b, main text). The fitted Stokeslet strength $F_{\text{st}}/\eta=(2.6\pm0.3)\,$mm$^2$/s is comparable to the negatively buoyant weight force $F_g/\eta=(1.7\pm0.4)\,$mm$^2$/s estimated from independent sedimentation experiments~(Sec.~\ref{sec:SedExps}) and does not change substantially during the experimental time window~(Fig.~\ref{fig:StokesletHydrod}a). Assuming a viscosity of water ($\eta=1\,$mPa$\cdot$s at 20$\,^\circ C$), we therefore find $F_{\text{st}}\approx2-3\,$nN. The fact that $F_{\text{st}}>F_g$ suggests the presence of a small counter-force $F_{\Delta}=F_{\text{st}}-F_g\approx1\,$nN. With the surface tension of a water-air interface ($\sigma_{\text{H}_2\text{O}}\approx70\times10^{-3}\,$N/m at 20$^\circ$C) and for an embryo area of approximately $\ell_{\text{min}}^2=100^2\,\mu$m$^2$ (see Fig.~\ref{fig:SideViewFitResultsfortheo}c,d) pushing against the surface, any interface curvature radius smaller than $\ell_{\text{min}}^2\sigma_{\text{H}_2\text{O}}/F_{\Delta}\approx70\,$cm is sufficient to provide such a counter-force. This value is much larger than the well diameter of $1.6\,$cm, such that normal forces resulting from surface tension are indeed a plausible source of the counter-force~$F_{\Delta}$, even if the fluid-air interface is only weakly curved near the center of the well.

The flow seen in the $xz$-plane as described by these fits is shown in~Fig.~\ref{fig:StokesletHydrod}b. As in the case of Stokeslet flow below a rigid surface~\cite{Drescher2009}, a strong lateral inwards flow will draw in nearby embryos and thereby generates an effective attraction between them. A Stokeslet of strength $F_{\text{st}}/\eta$ captures the spatial profile and strength of this attraction~\cite{Squires2001}. Below, we will use this experimental parametrization of hydrodynamic attraction to describe the corresponding embryo interactions in a minimal hydrodynamic model~(Sec.~\ref{sec:HydrModel}).

Finally, we note that the embryo's spinning motion around its own AP axis as such has no systematic effect on the surrounding fluid flow. Interestingly, a torque-free spherical body that generates an azimuthally symmetric, rotary surface flow also rotates in the lab frame, but does not drive any flow in its environment~\cite{Lauga2020}. This solution remains valid in the vicinity of a nondeforming shear stress-free interface and therefore provides a scenario that is consistent with experimental observations.\\

{\sffamily\bfseries Bound state orientations near curved fluid surfaces}

The fluid-air interfaces in finite-sized wells used in experiments are weakly curved. In the following, we show that this can lead to slightly tilted, but still linearly stable embryo orientations with an essentially upright AP~axis. For simplicity, we describe fluid flows surrounding an embryo as a pure Stokeslet flow, but note that qualitatively similar results can be found when higher order singularities as described in Secs.~\ref{sec:FFms}--\ref{sec:StabAnalFull} are included.

For a curved fluid surface, the surface normal along which image flow singularities are reflected is in general not parallel to gravity, but may enclose a finite angle $\theta_g$. We chose the plane of this tilt without loss of generality to be parallel to the $xz$-plane (Fig.~\ref{fig:StokesletHydrod}c, left) and then study the orientation dynamics in a frame where the surface normal is parallel to the $z$-axis. Specifically, we consider Eq.~(\ref{eq:Faxen}) for the image flow $\mathbf{v}'$ of a Stokeslet with fixed total force $\mathbf{F}=F_{\text{st}}(\sin\theta_g,0,-\cos\theta_g)^\top$ with $F_{\text{st}}>0$ (Fig.~\ref{fig:StokesletHydrod}c) -- such an image flow is given by Eq.~(\ref{eq:Stokeslet_uf}) with $\mathbf{p}=-(\sin\theta_g,0,\cos\theta_g)^{\top}$ and $a$ as in Eq.~(\ref{eq:Stoka}) -- which yields 
\begin{subequations}
\begin{align}
\frac{dp_x}{dt}&=\frac{F_{\text{st}}}{32\pi\eta h^2}\left(-p_z\sin\theta_g+3\chi p_z^2p_x\cos\theta_g\right)\label{eq:DynPx}\\
\frac{dp_y}{dt}&=-\frac{F_{\text{st}}}{32\pi\eta h^2}3\chi p_z^2p_y\cos\theta_g\label{eq:DynPy} \\
\frac{dp_z}{dt}&=\frac{F_{\text{st}}}{32\pi\eta h^2}\left[-p_x\sin\theta_g+3\chi \left(1-p_z^2\right)p_z\cos\theta_g\right].\label{eq:DynPz}    
\end{align} \label{eq:DynPi}
\end{subequations}
Equations~(\ref{eq:DynPi}) are for $\theta_g=0$ equivalent to the Stokeslet contribution $\sim a$ in Eqs.~(\ref{eq:Dynpcart})--(\ref{eq:fp}). The stationary orientation of interest ($\mathbf{p}$ mostly oriented along the $z$-axis) described by Eqs.~(\ref{eq:DynPi}) is given by $p^*_y=0$ and  
    \begin{subequations}
    \begin{align}
    p^*_x&=\frac{\tan\theta_g}{3\chi p^*_z}\label{eq:Px}\\
    p^*_z&=\sqrt{\frac{1}{2}+\sqrt{\frac{1}{4}-\left(\frac{\tan\theta_g}{3\chi}\right)^2}},\label{eq:Pz} 
    \end{align} \label{eq:Pfull}%
    \end{subequations}
which shows that the body axis $\mathbf{p}$ is at steady state oriented in the plane formed by the surface normal and the direction of gravity. For this solution, two important scenarios can be discussed:

\begin{itemize}
    \item\textit{Gravity aligned with the surface normal $(\theta_g=0)$: Orientational stability of bound states}
    
    In this case, Eqs.~(\ref{eq:Pfull}) imply $\mathbf{p}^*=\mathbf{e}_z$ if $\chi>0$ and a linearization of Eqs.~(\ref{eq:DynPi}) shows this state is stable against orientational perturbations. Hence, below a free surface, an anisotropic body shape $\chi>0$ (Fig.~\ref{fig:SideViewFitResultsfortheo}e) and the generation of a Stokeslet flow (Fig.~\ref{fig:StokesletHydrod}a) are in principle sufficient ingredients to stabilize the upright body axis orientation of a microswimmer, consistent with the analysis in the previous sections and with experimental observations.
    
    \item\textit{Gravity tilted relative to the surface normal $(\theta_g\ne0)$: Hydrodynamic focusing in wells}
    
    Equation~(\ref{eq:Pz}) implies in this case that a stationary solution only exist for anisotropic surface shapes with
    \begin{equation}
        \chi>\frac{3}{2}|\tan\theta_g|.\label{eq:Gammacrit}
    \end{equation}
    Furthermore, assuming Eq.~(\ref{eq:Gammacrit}) holds, we see from Eqs.~(\ref{eq:Pfull}) that the stationary orientation vector is itself tilted relative to the fluid surface normal ($p^*_x\ne0$). A linearization of Eqs.~(\ref{eq:DynPi}) around the stationary state Eqs.~(\ref{eq:Pfull}) for arbitrary~$\theta_g$ shows that such a tilted state is still linearly stable. This has important consequences for swimmers that are bound below a potentially curved fluid surface: Whenever embryos tilt relative to the surface above them, part of the thrust that is generated by ciliary beating is converted into translational motion parallel to the surface. The direction of this translation is set by the direction of the embryo tilt, which in turn is set by the orientation of gravity relative to the surface normal. From the sign of \smash{$p_x^*$} given in Eq.~(\ref{eq:Px}) it follows that the body axis tilt relative to the surface is always oriented such that, for a convexly curved fluid surface in a well, bound embryos tend to translate towards the center of the well~(Fig.~\ref{fig:StokesletHydrod}c, right). Complementing the Stokeslet-mediated attraction, this hydrodynamic focusing additionally contributes to the accumulation of embryos and thereby supports the formation of even larger clusters.
\end{itemize}

\subsection{Minimal model of cluster formation and rotation}\label{sec:HydrModel}
In this section, we describe a minimal model of interacting chiral disks that faithfully recapitulates the phenomenology of cluster formation and quantitatively accounts for experimentally observed single embryo spinning and whole cluster rotation frequencies~(Fig.~2e--g, main text). Notably, over the course of each experiment, system properties spread over almost three orders of magnitude in frequencies and in the number of embryos contained in clusters. For simplicity, our model aims at a description of the effectively two-dimensional dynamics in the $xy$-plane of a fixed number of embryos that are bound below the surface. In this case, each embryo can be represented as a disk and lateral interactions can be described by effective force and torque balance equations. 

\subsubsection{Force and torque balance}
Before formulating effective force and torque balance equations for a given embryo, we qualitatively describe the minimal set of interactions expected in our living chiral crystal system. Besides a steric repulsion between nearby embryos, the essential lateral interactions are given~by:
\begin{itemize}
    \item \textit{Hydrodynamic attraction}\vspace{0.05cm}
    
    The dominant long-ranged radial in-flow $|\mathbf{v}|\sim1/r$ from the Stokeslet generated by each embryo (main text Fig.~2a,b, see Sec.~\ref{sec:Stokeslet}), entrains surrounding embryos and thereby leads to an effective attraction between them. An analogous effect near rigid no-slip interfaces has been studied in various theoretical and experimental settings~\cite{Dufresne2000,Squires2001,Drescher2009,Petroff2015}. Lateral hydrodynamic interactions from higher order flow singularities are subdominant as they decay at least with $1/r^2$ in distance and are therefore neglected in the minimal model that is presented below.
    \item \textit{Transverse hydrodynamic force and torque exchange}\vspace{0.05cm}
    
    Two nearby rotating embryo surfaces will experience an additional exchange of forces and torques due to hydrodynamic near-field interactions~\cite{Kim2013,Drescher2009,Petroff2015}~(see Fig.~2d, main text). In particular, transverse forces within the $xy$-plane due to the neighboring embryo's spinning effectively make a pair of embryos ``roll" on each other. Together with the Stokeslet attraction, this leads to an orbiting motion of groups of embryos. Similarly, each embryo is subject to torques due to the spinning of nearby embryos, which leads to a slow-down of individual embryo spinning frequencies~(Fig.~2f, main text). The forces and torques contributing to these interactions are expected to affect the fluid flow around rotating groups of embryos in a particular fashion, a prediction that was independently verified by fitting suitable sets of singularities to measured flow fields~(Sec.~\ref{sec:PIVfit}, main text Fig.~2c, Figs.~\ref{fig:PIVfitting} and~\ref{fig:PIVfittingtrip})
\end{itemize}

Representing an embryo~$i$ by a disk with centroid position \hbox{$\mathbf{r}_i=(x_i,y_i,-h)$}, where $h$ is the distance between centroid and fluid surface, the above interactions translate into an effective force balance of the form
\begin{equation}
\frac{d\mathbf{r}_i}{dt}=\sum_{j\ne i}\left[\bar{\mathbf{v}}_{\text{st}}(\mathbf{r}_i;\mathbf{e}_z,\mathbf{r}_j)+\frac{1}{\eta R}\mathbf{F}_{\text{rep}}(|\mathbf{r}_i-\mathbf{r}_j|)+R(\omega_i+\omega_j)F_{\text{nf}}(|\mathbf{r}_i-\mathbf{r}_j|)\,\hat{\mathbf{r}}_{ij}^\perp\right],\label{eq:FB}
\end{equation}
which describes an overdamped dynamics of the in-plane centroid coordinates, $dx_i/dt$ and~$dy_i/dt$. The vector $\hat{\mathbf{r}}_{ij}^\perp=[(\hat{\mathbf{r}}_{ij})_y,-(\hat{\mathbf{r}}_{ij})_x]^\top$ is orthogonal to the unit vector $\hat{\mathbf{r}}_{ij}=(\mathbf{r}_i-\mathbf{r}_j)/|\mathbf{r}_i-\mathbf{r}_j|$ that points from the center of disk $j$ to the center of disk~$i$. The first term in Eq.~(\ref{eq:FB}) captures the Stokeslet-mediated attraction through the flow $\bar{\mathbf{v}}_{\text{st}}$ [Eqs.~(\ref{eq:Stokeslet_uf})~and~(\ref{eq:Stokeslet}) with $a=-F_{\text{st}}/(8\pi R\eta)$] generated by embryos at positions~$\mathbf{r}_j$. The second term in Eq.~(\ref{eq:FB}) implements a steric repulsion between embryos of the form  \hbox{$\mathbf{F}_{\text{rep}}(r)=-dV(r)/d\mathbf{r}$}, where $r=|\mathbf{r}|$ and 
\begin{equation}\label{eq:RepPot}
V(r)=f_{\text{rep}}\frac{R^{13}}{r^{12}}.   
\end{equation}
Here, $R$ is the average apparent radius of bound embryos when viewed along the $z$-axis from above the fluid surface and the effective repulsion force~$f_{\text{rep}}$ was inferred from experiments (Sec.~\ref{sec:ModPara}). The last term in Eq.~(\ref{eq:FB}) introduces the transverse hydrodynamic near-field forces. This transverse force is proportional to the relative velocity \hbox{$R(\omega_i+\omega_j)$} of two nearby embryo surfaces, where $\omega_i$ denotes the angular spinning frequency of embryo~$i$. $\omega_i>0$ ($\omega_i<0$) corresponds to clockwise (counter-clockwise) single embryo spinning when viewed from above the fluid surface. The amplitude of the transverse force depends on the distance \hbox{$d_{ij}=|\mathbf{r}_i-\mathbf{r}_j|-2R$} between embryo surfaces and takes the form
\begin{equation}
F_{\text{nf}}(|\mathbf{r}_i-\mathbf{r}_j|)=
\left\{
\begin{matrix}
f_0\ln\frac{d_c}{d_{ij}}\hspace{0.5cm}(d_{ij}<d_c)\\
0 \hspace{1.6cm}(d_{ij}\ge d_c)
\end{matrix}\right.,\label{eq:Finter}
\end{equation}
where the logarithmic distance dependence is a known result from lubrication theory~\cite{Kim2013,Drescher2009}. While the prefactor in Eq.~(\ref{eq:Finter}) for purely hydrodynamic interactions in the lubrication limit is also known~\cite{Kim2013}, we introduce here a phenomenological dimensionless parameter $f_0$ that is instead determined from experiments to account for potential effects of flagella-flagella interactions or of the nonspherical embryo shape. The parameter $f_0$ then characterizes the effective overall strength of the transverse force exchange~(see also Sec.~\ref{sec:ModPara}).

Anticipating the formation of clusters in experiments, where hydrodynamic attraction gets screened by the presence of neighboring embryos, we distinguish two populations of disks depending on the size of clusters they are part of at a given moment in time: A disk $i$ that is isolated or part of a small cluster of at most 3 disks interacts with all other disks $j$ via the Stokeslet flow $\bar{\mathbf{v}}_{\text{st}}$ in Eq.~(\ref{eq:FB}). A disk $i$ that is part of a cluster with more than 3 disks only experiences a Stokeslet-mediated attraction with disks $j$ with $|\mathbf{r}_i-\mathbf{r}_j|\le3.8R$, such that hydrodynamic attraction in larger clusters is restricted to nearest and second nearest neighbor interactions -- corresponding to the set of neighboring disks with which a direct lateral line-of-sight exists. 

To mimic the effect of hydrodynamic focusing of embryos towards the well center (see Sec.~\ref{sec:Stokeslet} and Fig.~\ref{fig:StokesletHydrod}c) and facilitate the formation of a single large cluster in simulations within finite time, the contribution $\mathbf{v}_w=-\frac{|\mathbf{r}_i|}{\tau_w}\hat{\mathbf{r}}_i$ with $\hat{\mathbf{r}}_i=\mathbf{r}_i/|\mathbf{r}_i|$ and $\tau_w=15\,$min was added to Eq.~(\ref{eq:FB}), effectively placing the center of the well at the in-plane coordinate origin $x=0$ and $y=0$.\\

This model is closed by a torque balance that describes the evolution of each embryo's spinning frequency. We consider a fully overdamped scenario, in which the torque balance requires for each embryo that the sum of torques from rotational drag, from the flagellar beating and from lubrication interactions with nearby embryos vanish. For embryos interacting isotropically in the $xy$-plane, this leads to an algebraic condition that determines instantaneous spinning frequencies of each embryo as
\begin{equation}\label{eq:TB}
    \omega_i=\omega_{0}-\sum_{j\ne i}(\omega_i+\omega_j)T_{\text{nf}}(|\mathbf{r}_i-\mathbf{r}_j|).
\end{equation}
Single bound embryos are spinning with angular frequency $\omega_0$ and nearby embryos slow down each other's spinning. The distance-dependence of the latter interaction again borrows from known results in lubrication theory~\cite{Kim2013,Drescher2009}, which suggests the form
\begin{equation}\label{eq:Tinter}
T_{\text{nf}}(|\mathbf{r}_i-\mathbf{r}_j|)=
\left\{
\begin{matrix}
\tau_0\ln\frac{d_c}{d_{ij}}\hspace{0.5cm}(d_{ij}<d_c)\\
0 \hspace{1.6cm}(d_{ij}\ge d_c)
\end{matrix}\right.,
\end{equation}
for the near-field contribution in Eq.~(\ref{eq:TB}). Similar to Eq.~(\ref{eq:Finter}),  Eq.~(\ref{eq:Tinter}) uses a phenomenological dimensionless parameter $\tau_0$ that is directly determined by experiments and characterizes the strength of the torques that slow down the spinning of nearby embryo. 

 \begin{figure}[!t]
\centering
\includegraphics[width=0.94\textwidth]{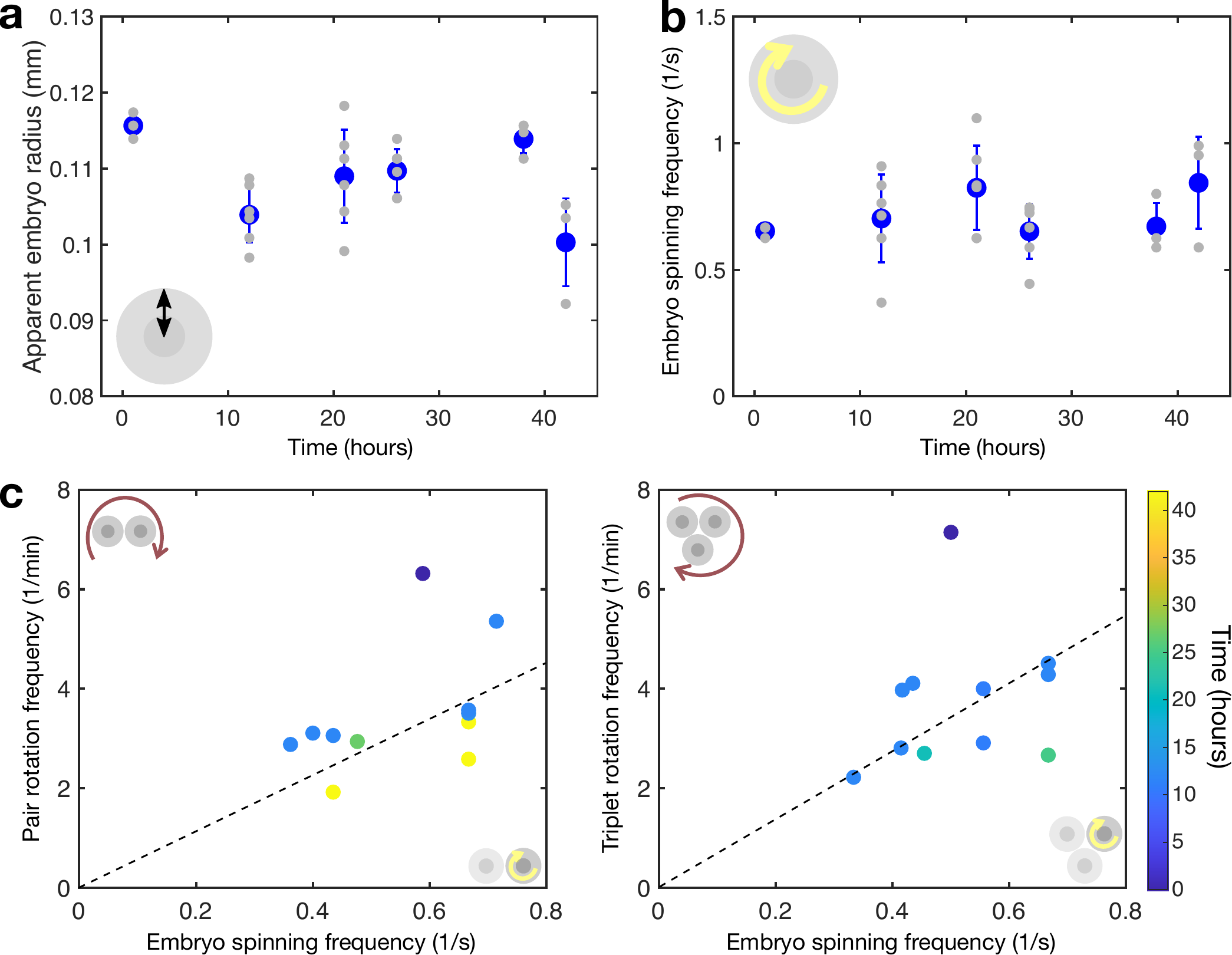}
\caption{\textbf{Measurements of embryo properties.}~\textbf{a,}~Apparent radii~$R$ of embryos bound below the fluid surface. \textbf{b,}~Spinning frequencies $\omega_0/(2\pi)$ of isolated bound embryos. In \textbf{a} and \textbf{b}, each gray dot depicts value for a different embryo (same set of embryos as in Fig.~\ref{fig:StokesletHydrod}a), blue dots and error bars represent mean and standard deviation, respectively. \textbf{c,}~Rotation frequencies of bound pairs (left) and triplets (right) increase linearly with the spinning frequency of embryos within these groups, as previously observed for pairs of $Volvox$ colonies~\cite{Drescher2009}. Qualitatively, a given embryo spinning frequency translates over time into gradually smaller group rotation frequencies, suggesting a weakening of effective hydrodynamic interactions as development progresses. The variability of single embryo spinning frequencies (\textbf{b}) is consistent with the variability of spinning frequencies within pairs and triplets and consequently leads to variability in rotation frequencies of pairs and triplets. Black dashed lines depict calibration of average embryo properties in the minimal model (see Sec.~\ref{sec:HydrModel}).} 
 \label{fig:EmbryoProps}
\end{figure}

\subsubsection{Determining model parameters from experimental data}\label{sec:ModPara}
The parameters contained in this model can be systematically determined from experimental measurements and suitable fits to the dynamics of single embryos, rotating pairs and rotating triplets. Similar to the Stokeslet strength $F_{\text{st}}$~(Fig.~\ref{fig:StokesletHydrod}a), apparent average radii $R$ and spinning frequencies $\omega_0/(2\pi)$ of isolated bound embryos are determined from direct single embryo measurements~(Fig.~\ref{fig:EmbryoProps}a,b). While we noted a slight decrease of in-plane flow speeds around single embryos towards the very end of the experimental time window, none of these parameters showed substantial changes over about~40\,hours of experiments. For given size $R$ and attraction characterized by~$F_{\text{st}}$, the repulsive force strength $f_{\text{rep}}$ in Eq.~(\ref{eq:RepPot}) sets the surface distance $d_{ij}$ between equilibrated pairs of disks in the minimal model. This distance was measured in experiments ($\approx20\,\mu$m) and $f_{\text{rep}}$ was set accordingly [see also Eq.~(\ref{eq:r0eq})]. Finally, rotation frequencies of pairs and triplets~(Fig.~\ref{fig:EmbryoProps}c) in experiments can be used to set the transverse force strength~$f_0$, while reduced embryo spinning frequencies within pairs and triplets~(Fig.~2f, main text) determine the lateral torque strength~$\tau_0$. The final parameter values used in simulations of the emergent cluster formation are listed in~Tab.~\ref{Tab:ModelParams}.\\

\begin{table}[H]
\begin{tabular}{c|c|c|c|c|c|c|c}
    Parameter & $R$\ ($\mu$m) & $\frac{\omega_0}{2\pi}$\ (Hz) & $F_{\text{st}}/\eta$\ (mm$^2$/s) & $f_{\text{rep}}/\eta$\ (mm$^2$/s)& $f_0$ & $\tau_0$ & $d_c/R$ \\
    \hline
    Value & $110$ & $0.72\pm0.17$ & $2.6\pm0.3$ & $38\pm6$ & $0.06$ & $0.12$ &  $0.5$
\end{tabular}
\caption{Parameters used in the minimal hydrodynamic model of cluster formation. Mean values of apparent radii $R$, free spinning frequencies $\omega_0/2\pi$ and Stokselet strength $F_{\text{st}}/\eta$ were set to averages of corresponding experimental measurements (Figs.~\ref{fig:StokesletHydrod}a and \ref{fig:EmbryoProps}a,b). The variability of apparent radii is represented by a commensurate variability in the repulsive force amplitude $f_{\text{rep}}$. In simulations the parameters $\omega_0$, $F_{\text{st}}$ and $f_{\text{rep}}$ have been drawn from random distributions as described in the text below to reflect the natural variability of embryo properties. \label{Tab:ModelParams}} 
\end{table}

\textit{Introducing variability of single embryo parameters in the minimal model:} Experimental measurements of bound single embryo properties (Figs.~\ref{fig:StokesletHydrod}a and \ref{fig:EmbryoProps}a,b) provide insights into the biological variability of microscopic parameters. Qualitatively, this parameter variability can be interpreted as a form of noise that is present in the system. Indeed, parameter variability in the minimal model increases the likelihood of more comprehensive neighbor rearrangements when clusters merge. In the absence of microscopic variability on the other hand, even small clusters formed in the minimal model are rather static in their shape and nearest-neighbor topology, which is in contrast to experimental observations. To include parameter variability into the model, we proceeded as follows. Stokeslet strength $F_{\text{st}}$ (Fig.~\ref{fig:StokesletHydrod}a) and isolated bound embryo spinning frequencies $\omega_0/(2\pi)$ (Fig.~\ref{fig:EmbryoProps}b)  were sampled for each disk from a normal distribution with mean and standard deviation as determined from experiments (Tab.~\ref{Tab:ModelParams}). To mimic a finite apparent size variability in a minimal fashion, we additionally sampled the repulsive force strength $f_{\text{rep}}$ homogeneously from the interval given in Tab.~\ref{Tab:ModelParams}. Finally, to restore reciprocity of attraction and repulsion between disks $i$~and~$j$ for sampled parameters $\beta_i$ and $\beta_j$ (representing corresponding values of $F_{\text{st}}$ or $f_{\text{rep}}$), we symmetrized the linear coefficients of each pair-wise interaction in Eq.~(\ref{eq:FB}) as $F_{\text{st}},f_{\text{rep}}\rightarrow(\beta_i+\beta_j)/2$.\\

\textit{Phenomenological scaling of embryo spinning frequencies in large clusters:}
The emergent dynamics that our minimal model gives rise to can be tested by comparing spinning frequencies of embryos within clusters and whole-cluster rotation frequencies with experiments. The minimal model reveals that, as more disks join a cluster, the average nearest-neighbor distance is reduced, which leads to a slow-down of embryo spinning frequencies ($\omega_i$ in our model) within clusters due to torque exchanges (see experimental data in main text Fig.~2f). This causes a reduction in lateral force exchanges and, together with the increased drag experienced by larger clusters, slows down whole-cluster rotations (see experimental data in main text Fig.~2g). The interactions described in Eqs.~(\ref{eq:FB}) and (\ref{eq:TB}) with parameters from Tab.~\ref{Tab:ModelParams} quantitatively match the corresponding experimental observations up to cluster sizes of about 40 to 50 embryos. For larger clusters, we observed experimentally a more drastic reduction in the whole-cluster rotation speeds than captured by the local interactions described so far (main text Fig.~2g). The simplest way to take this into account without making assumptions about more complex spatial modulations of embryo interactions is to assume that the embryo's ability to generate a torque surface density is increasingly impeded in larger clusters. Formally, this amounts to a cluster-size-dependent reduction of $\omega_0$, where we heuristically found that a scaling $\omega_0\rightarrow \omega_0/[1+(N_{cl}/N_0)^2]$ in Eq.~(\ref{eq:TB}) ($N_0=80$ and $N_{cl}\ge2$ is the total number of disks in a given cluster) leads to whole-cluster rotation rates that fit the experimental data (main text Fig.~2g). A nontrivial prediction of the $\omega_0$-modulation are the emerging embryo spinning frequencies $\omega_i$ in larger clusters, which agree well with the experimental data (main text Fig. 2f).

\subsubsection{Details of numerical simulations}
The effective force and torque balance Eqs.~(\ref{eq:FB}) and (\ref{eq:TB}) with parameters shown in Tab.~\ref{Tab:ModelParams} were implemented in MATLAB and solved using the ordinary differential equation solver \verb+ode113+. Note that in practice Eq.~(\ref{eq:TB}) represents a separate linear system of equations for each connected component of the graph that is generated by connecting all disks within a distance of~$|\mathbf{r}_i-\mathbf{r}_j|<d_c+2R$. Because the weights of this linear system given by Eq.~(\ref{eq:Tinter}) depend only on positions~$\mathbf{r}_i$, it can be explicitly solved for the angular spinning frequencies $\omega_i$ at arbitrary intermediate time steps. The resulting values are directly used to evaluate the transverse interaction $\sim\hat{\mathbf{r}}_{ij}^{\top}$ in Eq.~(\ref{eq:FB}). 

To simulate the cluster formation shown in Fig.~2e (main text), we initiated 700 disks with individual parameters given in Tab.~\ref{Tab:ModelParams} and positioned them homogeneously on a circular domain with an approximate radius of 8\,cm.

To determine the data shown in Fig.~2f (main text) (``In small clusters"), we ran 30 simulations of pairs, triplets and groups of four disks corresponding to one, two, and three direct neighbors, respectively. 10 simulations of clusters with 100 disks were run to determine spinning frequencies ``In large clusters" Fig.~2f (main text). Mean and standard deviations of the spinning frequencies extracted from these simulations are depicted by the symbols and error bars in~Fig.~2f. 

To determine the size-dependent whole-cluster rotation rates shown in Fig.~2g (main text), we ran 5 simulations for each total numbers of disks $N_{cl}\in\{2,3,4,5,6,7,8,9,10,12,17,25,$ $37,54,78,114,165,220,320\}$ and extracted rotation frequencies of the final clusters that had formed. The standard variation of these whole-cluster rotation frequencies in the minimal model is smaller than the symbol size used in Fig.~2g (main text).

\subsubsection{Discussion}
In the following, we discuss the key assumptions and simplifications made in the minimal model introduced above in more detail.

To simplify the fluid dynamics, we have assumed hydrodynamic interactions are dominated by the Stokeslet properties of single bound embryos. The corresponding singularity description was used to describe a fixed attraction strength among embryos and therefore neglects potential changes of the ciliary activity and more complex aspects of hydrodynamic interactions when embryos come closer together. Similarly, we have considered an isotropic potential-like steric repulsion in the $xy$-plane that implies a circular  apparent shape of bound embryos. Taking more details of embryo morphology~(main text Fig.~1b and Fig.~\ref{fig:Morphology}) and associated hydrodynamic effects in tightly packed clusters faithfully into account will most likely require fully resolved, three-dimensional hydrodynamic simulations~\cite{2019Shen,2020Yan}, as well additional knowledge about the ciliary activity of embryos within clusters as compared to freely spinning~ones.

We did not explicitly include contributions of cilia-cilia interactions between embryos that may become relevant when embryos within clusters get very close to each other. In particular, the latter effect could lead to additional terms that complement the hydrodynamically motivated transverse near-field force and torques given in Eqs.~(\ref{eq:Finter}) and (\ref{eq:Tinter}). Using the phenomenological coefficients $f_0$ and $\tau_0$ and determining them from experimental measurements (Sec.~\ref{sec:ModPara}) partially accounts for this uncertainty. More generally, we restricted the force and torque balance equations to local interactions but note that hydrodynamic effects may lead, even among embryos within clusters, to longer-ranged interactions. 

The above mentioned factors -- cilia-cilia interactions and spatial three-dimensional details of the inter-embryonic fluid streaming -- but also potential feedback of the latter on ciliary activity might play a particularly important role in large clusters. While difficult to detect experimentally, such feedbacks could underlie the phenomenological $\omega_0$-scaling (see Sec.~\ref{sec:ModPara}) that is in the present model required to quantitatively explain embryo and whole-cluster rotation rates for large clusters.

For simplicity, we have considered a scenario with a constant number of interacting disks. In experiments, embryos at the clusters boundaries can depart from clusters and incoming embryos may fill up holes that arise, for example, when clusters merge. This additional level of effective noise in experiments smoothens ``sharp" edges of clusters, eliminates some of the vacancies or fills up larger holes that have formed, but is otherwise not expected to change the essential characteristics of cluster formation.

\subsection{Effective material properties of living chiral crystals}\label{sec:MappingDetails} 
Based on the minimal model described in Sec.~\ref{sec:HydrModel}, we derive in this section effective macroscopic material properties that arise for a collection of disks interacting within a hexagonal lattice. More complex details of interactions with the surrounding fluid, which may lead to effectively viscoelastic material properties of living chiral crystals, were for simplicity not included in the microscopic minimal model. Consequently, we focus here on the purely elastic response of such a system. To provide the required background, we first introduce the continuum description of isotropic elastic materials, closely following recent work on odd elasticity theory~\cite{Scheibner:2020gm,braverman2020topological} (Sec.~\ref{sec:OddModuli}) and proceed with the linearization (Sec.~\ref{Sec:ModelMapping}) and coarse-graining (Sec.~\ref{Sec:CGmodel}) of the interactions given in minimal model Eqs.~(\ref{eq:FB}) and (\ref{eq:TB}).

\subsubsection{Linear elasticity and strain cycles in isotropic solids}\label{sec:OddModuli}
The pair-wise interactions on the right-hand side of Eq.~(\ref{eq:FB}) describe a response to small displacements $\mathbf{u}(x,y,t)=(u_x,u_y)^\top$ from an equilibrium distance that is elastic and purely distance-dependent (translationally invariant). Hence, to linear order, a mean-field theory of effective stresses $\sigma_{ij}$ associated with crystal deformations will be of the form
\begin{equation}\label{eq:StrStrind}
\sigma_{ij}=\sigma^0_{ij}+C_{ijkl}\partial_ku_l,
\end{equation}
where $\sigma^0_{ij}$ is a pre-stress, $C_{ijkl}$ denotes the elastic moduli tensor, $\partial_ku_l$ is the displacement gradient tensor (Einstein summation convention $i,j,k,l=x,y$). Using the matrix basis~\cite{Scheibner:2020gm}
\begin{equation}
s^0=\begin{pmatrix}
1 & 0\\
0 & 1\\
\end{pmatrix}\hspace{1cm}
s^1=\begin{pmatrix}
0 & -1\\
1 & 0\\
\end{pmatrix}\hspace{1cm}
s^2=\begin{pmatrix}
1 & 0\\
0 & -1\\
\end{pmatrix}\hspace{1cm}
s^3=\begin{pmatrix}
0 & 1\\ 
1 & 0\\
\end{pmatrix},
\end{equation}
with $s_{ij}^{\alpha}s_{ij}^{\beta}=2\delta^{\alpha\beta}$ ($\alpha,\beta=0,1,2,3$), Eq.~(\ref{eq:StrStrind}) can be conveniently expressed as 
\begin{equation}\label{eq:StrStrproj}
\sigma^{\alpha}=\sigma^{\alpha}_0+C^{\alpha\beta}u^{\beta},
\end{equation}
which highlights the moduli $C^{\alpha\beta}=\frac{1}{2}C_{ijkl}s_{ij}^{\alpha}s_{kl}^{\beta}$ that couple strain components $u^{\alpha}=s^{\alpha}_{ij}\partial_iu_j$ ($u^0$: Compression/Expansion; $u^1$: Rotation; $u^2$, $u^3$: Shear components 1 and 2) to different stress components $\sigma^{\alpha}=s^{\alpha}_{ij}\sigma_{ij}$ ($\sigma^0$: Compressive/Expansive stress; $\sigma^1$: Torque, $\sigma^2$, $\sigma^3$: Shear stress components). Based on the nature of the interactions and the hexagonal lattice geometry in our system, we anticipate isotropic material properties to emerge at larger scales. Isotropy restricts potential pre-stresses to take the form of a pressure $\sigma^0_{ij}\sim\delta_{ij}$ and/or a torque density $\sigma^0_{ij}\sim\epsilon_{ij}$, and the most general isotropic moduli tensor is given by~\cite{Scheibner:2020gm,braverman2020topological}
\begin{equation}\label{eq:Cab}
C^{\alpha\beta}=\begin{pmatrix}
B & \Lambda & 0 & 0\\
A & \Gamma & 0 & 0\\
0 & 0 & \mu & K^o\\
0 & 0 & -K^o & \mu\\
\end{pmatrix}.
\end{equation}
Besides the bulk and shear moduli $B$ and $\mu$ commonly found in equilibrium solids, the moduli tensor Eq.~(\ref{eq:Cab}) contains additionally odd bulk and shear moduli $A$ and $K^o$ that can exist in isotropic nonequilibrium materials~\cite{Scheibner:2020gm}. While $B,\mu>0$ is required for mechanical stability, there are in principle no restrictions on the signs of $A$ and $K^o$. The moduli $\Gamma$ and $\Lambda$ respectively represent equilibrium and nonequilibrium moduli that couple rotations to isotropic stresses and torques, and therefore require interactions with a surrounding medium~\cite{braverman2020topological}. In general, all these moduli are compatible with the properties of living chiral crystals, which are inherently out-of-equilibrium, exhibit nonreciprocal transverse interactions and are embedded in and interacting with a surrounding medium.

Anticipating the experimental observation of closed cycles in the strain component spaces ($u^1,u^0$) and ($u^2,u^3$) (see Sec.~\ref{sec:StrCycAnal}, Fig.~\ref{fig:Entropy} and main text Fig.~4f,g), it is interesting to note that the \textbf{work that can be extracted from} a material with stress-strain relation Eqs.~(\ref{eq:StrStrproj}) and (\ref{eq:Cab}) during a strain cycle, $W=-\oint\sigma_{ij}du_{ij}$, can be expressed as~\cite{Scheibner:2020gm,Sousl2021}
\begin{align}
W&=\left(A-\Lambda\right)\times\left\{\text{Area enclosed by counter-clockwise cycle in $(u^1,u^0)$-space}\right\}\nonumber\\
&+\hspace{0.34cm}2K^o\hspace{0.34cm}\times\left\{\text{Area enclosed by counter-clockwise cycle in $(u^2,u^3)$-space}\right\}.\label{eq:work}
\end{align}

\subsubsection{Linearization of pair-interactions in the minimal model}\label{Sec:ModelMapping}
We now show that pair-interactions in the two-dimensional disk dynamics described by Eqs.~(\ref{eq:FB}) and (\ref{eq:TB}) can be mapped to odd elastic spring-like interactions~\cite{Scheibner:2020gm} when considering small displacements from an equilibrium distance. For convenience, we normalize forces and two-dimensional stresses (tension) throughout by $\eta R$, such that forces have units [Length/Time] and tensions have units~[1/Time].

We consider a pair of disks $i=1,2$ with centroid distance $r_{12}=|\mathbf{r}_1-\mathbf{r}_2|$ interacting according to Eqs.~(\ref{eq:FB}) and (\ref{eq:TB}) and focus for simplicity on the scenario of equal disk radii $R$ and free spinning frequency $\omega_0$. Interactions in Eq.~(\ref{eq:FB}) coming from Stokeslet attraction $\bar{\mathbf{v}}_{\text{st}}$ and repulsion $\mathbf{F}_{\text{rep}}$ are both central forces acting along the vector \hbox{$\hat{\mathbf{r}}_{12}=(\mathbf{r}_1-\mathbf{r}_2)/r_{12}$}. To linear order around some fixed relative distance $\tilde{x}=x/R$, the effective force associated with these interactions reads
\begin{equation}\label{eq:linattrrep}
F^{\parallel}(r_{12})=\frac{1}{\eta R}\left[12\frac{f_{\text{rep}}}{\tilde{x}^{13}}-\frac{F_{\text{st}}\tilde{x}}{4\pi(\tilde{x}^2+4)^{3/2}}\right]-\frac{1}{\eta R^2}\left[\frac{156f_{\text{rep}}}{\tilde{x}^{14}}-\frac{F_{\text{st}}(\tilde{x}^2-2)}{2\pi(\tilde{x}^2+4)^{5/2}}\right](r_{12}-x)+\mathcal{O}(\delta\tilde{r}^2),
\end{equation}
where $\delta\tilde{r}=(r_{12}-x)/R$. The first term in square brackets vanishes at the equilibrium distance 
\begin{equation}\label{eq:r0eq}
r_0:=x^*\approx2.2R,
\end{equation}
where we used the values for $F_{\text{st}}$ and $f_{\text{rep}}$ from Table.~\ref{Tab:ModelParams}. We recall that the Stokeslet strength~$F_{\text{st}}$ (relative to the viscosity) was determined from experimental measurements (Sec.~\ref{sec:PIVfit}) and $f_{\text{rep}}$ was chosen such the equilibrium distance between a pair of disks was equal to the experimentally measured mean distance of an orbiting pair of embryos (Sec.~\ref{sec:ModPara}). The latter is therefore self-consistently recovered in Eq.~(\ref{eq:r0eq}). At this rest-length $r_0$, the linear coefficient in Eq.~(\ref{eq:linattrrep}) defines an effective spring constant
\begin{equation}\label{eq:SprConst}
k:=\frac{1}{\eta R^2}\left[\frac{156f_{\text{rep}}}{\tilde{x}^{14}}-\frac{F_{\text{st}}(\tilde{x}^2-2)}{2\pi(\tilde{x}^2+4)^{5/2}}\right]\approx 8.1\,\text{s}^{-1}.
\end{equation}

To approximate the transverse interaction $\sim\hat{\mathbf{r}}_{ij}^{\perp}$ in Eq.~(\ref{eq:FB}), we have to determine the prefactor $\omega_1+\omega_2$ from the torque balance Eq.~(\ref{eq:TB}). By symmetry, the angular rotation frequencies of a pair of interacting disks must be equal,~$\omega_1=\omega_2=\omega$. Equation~(\ref{eq:TB}) then implies
\begin{equation}
    \omega=\frac{\omega_0}{1+2\tau_0\ln\frac{d_c}{d_{12}}},
\end{equation}
where $d_{12}=r_{12}-2R$ is the shortest distance between the two disk boundaries. The amplitude of transverse interaction in Eq.~(\ref{eq:FB}) for a pair of interacting disks therefore becomes
\begin{equation}
2\omega RF_{\text{nf}}(r_{12})=2R\omega_0f_0\frac{\ln\frac{d_{12}}{d_c}}{2\tau_0\ln\frac{d_{12}}{d_c}-1}=F_0^{\perp}-k_a\left(r_{12}-r_0\right)+\mathcal{O}(\delta r^2),
\end{equation}
where the zeroth order transverse interaction $F_0^{\perp}$ and an effective elastic constant $k_a$ are given~by\vspace{-0.4cm}
\begin{subequations}\label{eq:oddcoeffmap}
\begin{align}
F_0^{\perp}&:=2R\omega_0f_0\ln\frac{d_0}{d_c}\left(2\tau_0\ln\frac{d_0}{d_c}-1\right)^{-1}\approx0.04\,\text{mm/s}\\
k_a&:=\frac{2R\omega_0f_0}{d_0}\left(2\tau_0\ln\frac{d_0}{d_c}-1\right)^{-2}\approx2.0\,\text{s}^{-1}.
\end{align}
\end{subequations}
In these expressions, $d_0=r_0-2R=0.2R$ [see Eq.~(\ref{eq:r0eq})] denotes the disk surface distance at equilibrium, $d_c=0.5R$ the cut-off length of lubrication-induced transverse interactions and we used parameters from the experimentally calibrated disk model to determine concrete values for $F_0^{\perp}$ and $k_a$ (see Tab.~\ref{Tab:ModelParams}). All together, pairwise interactions in Eq.~(\ref{eq:FB}) imply in this limit the centroid dynamics
\begin{equation}\label{eq:odddynmap}
\frac{d\mathbf{r}_1}{dt}=-\frac{d\mathbf{r}_2}{dt}=-k\left(r_{12}-r_0\right)\hat{\mathbf{r}}_{12}+\left[F_0^{\perp}-k_a\left(r_{12}-r_0\right)\right]\hat{\mathbf{r}}_{12}^{\perp},
\end{equation}
where $\hat{\mathbf{r}}_{12}^{\perp}=[(\hat{\mathbf{r}}_{12})_y,-(\hat{\mathbf{r}}_{12})_x]^\top$ and the effective parameters $k$, $k_a$ and $F^\perp_0$ are given in Eqs.~(\ref{eq:SprConst}) and (\ref{eq:oddcoeffmap}).

\subsubsection{Coarse-grained odd material parameters of living chiral crystals}\label{Sec:CGmodel}
The linearized pair-wise dynamics Eq.~(\ref{eq:odddynmap}) corresponds to a spring-like interaction with linear spring constant $k$ and odd spring constant $k_a$ as introduced in~\cite{Scheibner:2020gm} and analyzed further in~\cite{braverman2020topological}. While the exact values of $r_0$, $k$ and $k_a$ differ from the values given in Eqs.~(\ref{eq:SprConst}) and (\ref{eq:oddcoeffmap}) when more than two disks interact within a hexagonal lattice, linear and transverse interactions will be qualitatively similar to Eq.~(\ref{eq:odddynmap}). Hence, we can refer to the coarse-graining results derived in ref.~\cite{braverman2020topological} for hexagonal lattices of particles interacting with nearest neighbors. The parameters given in Eqs.~(\ref{eq:oddcoeffmap}) then provide estimates for effective material moduli that emerge from transverse lubrication interactions and are given by~\cite{braverman2020topological}
\begin{equation}\label{eq:OddModuliCG}
A=\frac{\sqrt{3}}{2}\left(k_a+\frac{F_0^{\perp}}{r_0}\right)\approx1.9\,\text{s}^{-1}\hspace{1cm}K^o=\frac{\sqrt{3}}{4}\left(k_a-\frac{F_0^{\perp}}{r_0}\right)\approx0.8\,\text{s}^{-1}.
\end{equation}
These moduli are comparable to the standard bulk and shear moduli
\begin{equation}\label{eq:PassModuliCG}
B=\frac{\sqrt{3}}{2}k\approx7.0\,\text{s}^{-1}\hspace{1cm}\mu=\frac{\sqrt{3}}{4}k\approx3.5\,\text{s}^{-1}
\end{equation}
of a coarse-grained hexagonal lattice connected by linear springs with spring constant given in Eq.~(\ref{eq:SprConst}). Additionally, for a hexagonal lattice with nearest neighbor interactions the contribution $F_0^{\perp}$ in Eq.~(\ref{eq:odddynmap}) yields in the long-wavelength limit an anti-symmetric pre-stress $\sigma^0_{ij}=\sigma_0\epsilon_{ij}$ with~\cite{braverman2020topological}
\begin{equation}\label{eq:td}
\sigma_0=\frac{\sqrt{3}F_0^{\perp}}{r_0}\approx0.3\,\text{s}^{-1}.
\end{equation}
To express the parameter values in Eqs.~(\ref{eq:OddModuliCG})--(\ref{eq:td}) in units of tension, they have to be multiplied by the effective friction parameter $R\eta$.

Based on our minimal model, distance-dependent transverse lubrication interactions are therefore expected to give rise to effective odd material properties with $A, K^o>0$ and broken Maxwell-Betti reciprocity~\cite{braverman2020topological}. From this derivation, we see explicitly that signs of the odd moduli are determined by the sign of $\omega_0$ and thus by the intrinsic handedness of active embryo rotations. In agreement with this coarse-graining result, $A>0$ and $K^o>0$ is also found from the experimental analysis of strains surrounding topological lattice defects (Fig.~4a--c in the main text, Sec.~\ref{sec:StrAnalysis}). Finally, the sign of the anti-symmetric pre-stress $\sigma^0_{ij}=\sigma_0\epsilon_{ij}$ with $\sigma_0>0$ indicates an intrinsic tendency of the material to spin clockwise, consistent with the clockwise rotations exhibited by living chiral crystals.

\section{Data Analysis}
This final section details the different data analysis approaches used to quantitatively living chiral crystals.

\subsection{Starfish embryo centroid localization and rotation correction}\label{sec:EmbrProc}
To identify starfish embryos, we first performed a circular Hough transform using the MATLAB (2016b, MathWorks) function \verb+imfindcircles+ on inverted raw intensities of microscopy images. To calculate the rotation frequency of clusters, we first identified all embryos within a cluster. This was done by recursively finding all nearest neighbors within a 1.5 embryo diameter from a predetermined seed position (for large clusters) or by using intensity thresholding to identify clusters and identifying the embryo centroids that lie within a given cluster region (for small clusters). Particle tracking was performed using the Hungarian linker algorithm~\cite{Kuhn1955} (for large clusters) or Daniel Blair and Eric Dufresne's MATLAB adaptation of the IDL Particle Tracking software (for small clusters)~\cite{blair_dufresne}. The rotation angle between any two consecutive time points was calculated by finding a rigid body transformation that maximizes the overlap of the embryo centroid positions between two frames. We then applied the inverse transformation on subsequent time frames to obtain embryo centroid positions in the cluster's co-rotating frame for analysis, which we refer to as \textit{rotation-corrected} centroid data. Measured rotation frequencies of clusters as a function of the number of embryos are shown in Fig.~2g (main text).

\subsection{Analysis of top-view flow fields surrounding bound embryos, pairs and triplets}\label{sec:FFanalsis}
In the following, we describe a quantitative analysis of the stationary flow fields parallel to the surface ($xy$-plane) as they can be seen around isolated bound embryos, as well as around pairs and triplets rotating near the surface. To this end, we use experimental measurements of tracer particle velocities (see Sec.~\ref{sec:PIVexp} for experimental details) and hydrodynamic arguments to fit the resulting flow fields.

\subsubsection{Flow field measurements}\label{sec:FFM}

\textit{Embryo centroid tracking:} Embryos bound near the surface were segmented using Fiji's \cite{schindelin2012fiji} built-in function ``Threshold". Subsequently, incomplete or touching edges of embryos were delineated with the ``Pencil Tool" and the operation ``Fill Holes" was applied. Positions of embryo centroids were then extracted from the resulting shapes using the ``Analyze Particles" feature.\\

\textit{Tracer particle tracking:} For all measurements, tracer particles were segmented using Fiji's ``Threshold" ($i$)~with and ($ii$)~without applying a prior ``FFT Bandpass Filter" step (filtering structures over 100\,pixel $\approx200\,\mu$m). Segmented particles from $(i)$ and $(ii)$ were then tracked using the~``TrackMate" plugin (Simple LAP tracker)~\cite{tiv17}. Combining the results from $(i)$ and $(ii)$ enabled segmentation and tracking of particles with high fidelity both near and far away from the embryo surface.\\

\textit{Averaging tracer particle velocities:} Before averaging the velocities found from particle tracking, we removed measurements that indicated particle movements away from single embryos or cluster centers by more than~$45^\circ$ to narrow the depth of field. Furthermore, to determine meaningful average flow fields surrounding embryo pairs and triplets, their rotation dynamics had to be taken into account. To this end, the location of a given velocity measurement was additionally registered with respect to the current orientation of the pair or triplet and duplicated with respect to the corresponding cluster symmetry~(See Code Availability). Thereby, the final average field corresponds to the instantaneous flow field in the lab frame registered with respect some fixed orientation of the given pair or triplet.

\subsubsection{Flow field fitting}\label{sec:PIVfit}
To connect average flow fields to the ciliary activity of the embryos, we approximated the flow fields around isolated embryos, pairs and triplets using suitable sets of Stokes flow singularities~\cite{Drescher2009,Drescher2010}:\\

\textit{Single embryos:} The flow surrounding an isolated embryo was approximated by a Stokeslet with force $\mathbf{F}=-F_{\text{st}}\mathbf{e}_z$ at distance $h$ below a free surface, which we write for purpose of defining a general fit function in this data analysis section as
\begin{equation}
 \bar{\mathbf{v}}_{\text{st}}(\mathbf{r};\mathbf{F},\mathbf{r}_0)=\mathbf{v}_{\text{st}}(\mathbf{r};\mathbf{F},\mathbf{r}_0)+\mathbf{v}_{\text{st}}(\mathbf{r};\mathbf{F}',\mathbf{r}_0'),\label{eq:stok_forfit}
\end{equation}
where $\mathbf{F}'=(F_x,F_y,-F_z)^\top$, $\mathbf{r}'_0=(x_0,y_0,h)^\top$ (fluid surface at $z=0$) and
\begin{equation}
 \mathbf{v}_{\text{st}}(\mathbf{r};\mathbf{F},\mathbf{r}_0)=\frac{1}{8\pi\eta}\left(\frac{\mathbf{F}}{|\mathbf{r}-\mathbf{r}_0|}+\frac{\mathbf{F}\cdot\left(\mathbf{r}-\mathbf{r}_0\right)\left(\mathbf{r}-\mathbf{r}_0\right)}{|\mathbf{r}-\mathbf{r}_0|^3}\right).\label{eq:stok_forfit2}
\end{equation}
The Stokeslet position $\mathbf{r}_0=(x_0,y_0,-h)^\top$ is centered at the $(x,y)$-coordinate of a given embryo centroid. The parameters $h$, $F_{\text{st}}$, as well as the $z$-coordinate~$z_v$ of the apparent $xy$-plane in which the flow field is observed are determined by the fitting procedure described below.\\

\begin{figure}[!t]
\centering
\includegraphics[width=0.99\textwidth]{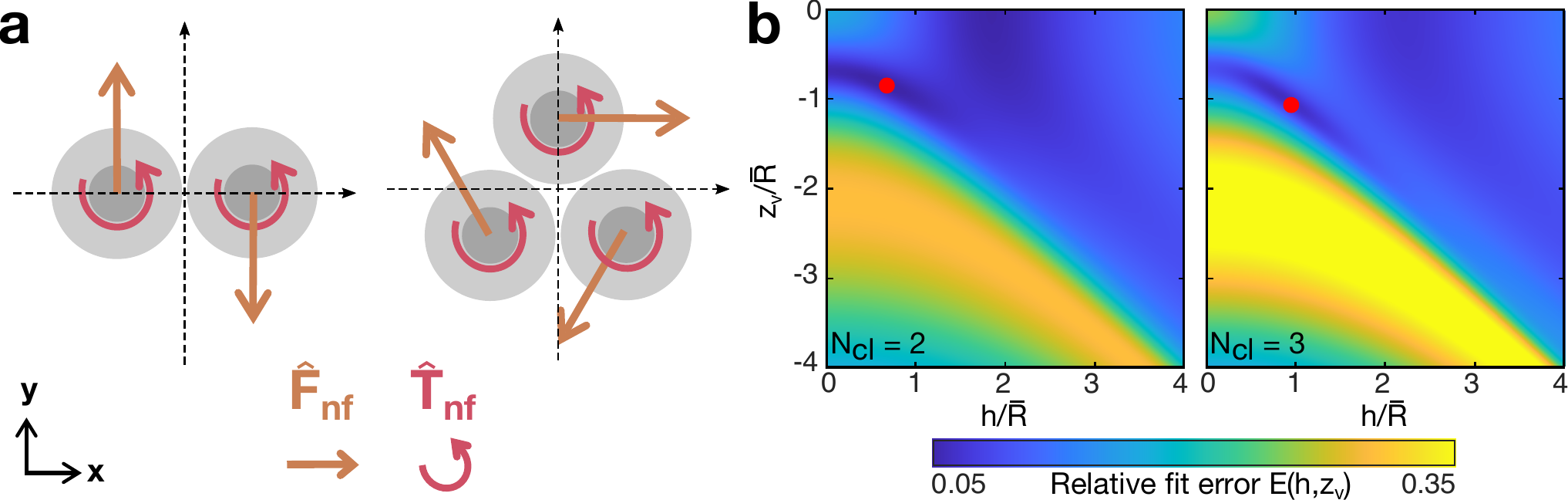}
\caption{\textbf{Fitting of experimental in-plane flow fields.} \textbf{a,}~Schematic of the locations~\smash{$\mathbf{r}_0^{(i)}$} of near-field Stokeslet forces~\smash{ $\mathbf{F}=\hat{F}_{\text{nf}}\mathbf{e}^{(i)}$} (brown arrows, see Tab.~\ref{Tab:FlowFieldFit}) used to fit flow fields around rotating pairs and triplets. Red arrows indicate corresponding rotlet-associated torques $\mathbf{T}=\hat{T}_{\text{nf}}\mathbf{e}_z$. \textbf{b,}~Relative least squares errors $E(h,z_v)=\langle|\mathbf{v}_{\text{fit}}-\mathbf{v}_{\text{exp}}|\rangle_{x,y}/\max_{x,y}|\mathbf{v}_{\text{exp}}|$ of fits to $xy$-plane flows around the rotating pair (left, $\bar{R}=108\,\mu$m) and triplet (right, $\bar{R}=111\,\mu$m) shown in Fig.~2c (main text) and in Fig.~\ref{fig:PIVfittingtrip}, respectively. Red dots indicate fit parameters that globally minimize $E(h,z_v)$.}
 \label{fig:PIVfitting}
\end{figure}

\textit{Additional contributions for rotating pairs and triplets:} Rotations of two closely nearby embryos give rise to additional hydrodynamic interactions (Fig.~2d, main text) that can in turn affect the surrounding flow field~\cite{Drescher2009}. Similar to the Stokeslet Eq.~(\ref{eq:stok_forfit2}), the generic flow around a torque $\mathbf{T}=(T_x,T_y,T_z)^\top$ located at $\mathbf{r}_0=(x_0,y_0,-h)^\top$ in an unbound fluid (``rotlet") is given~by
\begin{equation}
 \mathbf{v}_{\text{rot}}(\mathbf{r};\mathbf{T},\mathbf{r}_0)=\frac{1}{8\pi\eta}\frac{\mathbf{T}\times\left(\mathbf{r}-\mathbf{r}_0\right)}{|\mathbf{r}-\mathbf{r}_0|^3}.\label{eq:rot_uf}
\end{equation}
To meet the free-surface boundary conditions at $\mathbf{r}=(x,y,0)^\top$ this flow has to be complemented by a rotlet image. The rotlet below a free surface is given by
\begin{equation}
 \bar{\mathbf{v}}_{\text{rot}}(\mathbf{r},\mathbf{T},\mathbf{r}_0)=\mathbf{v}_{\text{rot}}(\mathbf{r};\mathbf{T},\mathbf{r}_0)+\mathbf{v}_{\text{rot}}(\mathbf{r};\mathbf{T}',\mathbf{r}_0'),\label{eq:rot}
\end{equation}
where $\mathbf{T}'=(-T_x,-T_y,T_z)^\top$ and $\mathbf{r}_0'=(x_0,y_0,h)$. To fit flow fields around pairs and triplets, we assume each embryo centroid is the source of an ``upright" Stokeslet flow with a common strength $F_{\text{st}}$. Furthermore, we assume that the exchange of near-field forces is homogeneous and pair-wise symmetric among embryos, such that it can be described by only one additional Stokeslet and rotlet strength, $\hat{F}_{\text{nf}}$ and $\hat{T}_{\text{nf}}$, respectively~(Fig.~\ref{fig:PIVfitting}a).\\

\textit{General fit function:} The final flow field that is fitted to measured tracer particle velocity fields around single embryos~($N_{cl}=1$), pairs~($N_{cl}=2$) and triplets~($N_{cl}=3$) is given by a fit function
\begin{equation}
\mathbf{v}_{\text{fit}}(\mathbf{r})=\sum_{i=1}^{N_{cl}}\left[\bar{\mathbf{v}}_{\text{st}}\left(\mathbf{r};-F_{\text{st}}\mathbf{e}_z,\mathbf{r}_0^{(i)}\right)+\bar{\mathbf{v}}_{\text{st}}\left(\mathbf{r};\hat{F}_{\text{nf}}\mathbf{e}^{(i)},\mathbf{r}_0^{(i)}\right)+\bar{\mathbf{v}}_{\text{rot}}\left(\mathbf{r};\hat{T}_{\text{nf}}\mathbf{e}_z,\mathbf{r}_0^{(i)}\right)\right],\label{eq:ufit}
\end{equation}
where $\bar{\mathbf{v}}_{\text{st}}$ and $\bar{\mathbf{v}}_{\text{rot}}$ are given in Eqs.~(\ref{eq:stok_forfit}) and (\ref{eq:rot}) and $\hat{F}_{\text{nf}}=\hat{T}_{\text{nf}}=0$ for isolated single embryos. For a comparison with measurements, Eq.~(\ref{eq:ufit}) is evaluated at $\mathbf{r}=(x,y,z_v)^\top$ and the parameters $F_{\text{st}},\hat{F}_{\text{nf}},\hat{T}_{\text{nf}}, h$ and $z_v$ are determined from a suitable fit procedure (see below). The positions~\smash{$\mathbf{r}_0^{(i)}$} of all flow singularities and the orientations~\smash{$\mathbf{e}^{(i)}$} of the near-field forces used for these fits are listed in Tab.~\ref{Tab:FlowFieldFit} and depicted in~Fig.~\ref{fig:PIVfitting}a.

\begin{table}[H]
\centering
\begin{tabular}{c|c|c|c}
    $N_{cl}$ & $\mathbf{r}_0^{(i)}$ & $\mathbf{e}^{(i)}$ & Fit parameters\\
    \toprule
    1 & $-h\mathbf{e}_z$ & $-$ & $\{h,z_v,F_{\text{st}}\}$\\
    \cmidrule{1-4}
    \multirow{2}{*}{2} & $-\frac{l}{2}\mathbf{e}_x-h\mathbf{e}_z$ & $\mathbf{e}_y$ & \multirow{2}{*}{$\{h,z_v,F_{\text{st}},\hat{F}_{\text{nf}},\hat{T}_{\text{nf}}\}$}\\
      & $\frac{l}{2}\mathbf{e}_x-h\mathbf{e}_z$ & $-\mathbf{e}_y$\\
    \cmidrule{1-4}
    \multirow{3}{*}{3} & $\frac{l}{2}\left(\mathbf{e}_x-\frac{1}{\sqrt{3}}\mathbf{e}_y\right)-h\mathbf{e}_z$ &  $-\frac{1}{2}\left(\mathbf{e}_x+\sqrt{3}\mathbf{e}_y\right)$ & \multirow{3}{*}{$\{h,z_v,F_{\text{st}},\hat{F}_{\text{nf}},\hat{T}_{\text{nf}}\}$}\\
     & $-\frac{l}{2}\left(\mathbf{e}_x+\frac{1}{\sqrt{3}}\mathbf{e}_y\right)-h\mathbf{e}_z$ & $\frac{1}{2}\left(-\mathbf{e}_x+\sqrt{3}\mathbf{e}_y\right)$ & \\
     & $\frac{l}{\sqrt{3}}\mathbf{e}_y-h\mathbf{e}_z$ & $\mathbf{e}_x$ & 
\end{tabular}
\caption{Positions $\mathbf{r}_0^{(i)}$ and near-field force orientations~$\mathbf{e}^{(i)}$ of flow singularities used to fit $\mathbf{v}_{fit}(\mathbf{r})$ given in Eq.~(\ref{eq:ufit}) to experimentally measured flow fields. $l$ denotes average nearest neighbor centroid distances. Fit parameters are given by the distance of the flow singularities to the surface ($h$), the apparent $z$-coordinate of the $xy$-plane in which the flow is observed ($z_v$), as well as the strengths $F_{\text{st}}$, $\hat{F}_{\text{nf}}$ and $\hat{T}_{\text{nf}}$ of the different flow singularities used in $\mathbf{v}_{\text{fit}}(\mathbf{r})$ given in Eq.~(\ref{eq:ufit}). (see Sec.~\ref{sec:PIVfit} for details). \label{Tab:FlowFieldFit}}
\end{table}

\textit{Least squares fit and nonlinear optimization:} 
Fitting parameters that need to be determined are in all cases given by the upright Stokeslet strength $F_{\text{st}}$, the distance of the flow singularities to the surface $h$, and the effective $z$-coordinate~$z_v$ of the plane in which the $xy$-plane flow is observed. Fits of flow fields surrounding pairs and triplets additionally include a near-field Stokeslet and rotlet strength, $\hat{F}_{\text{nf}}$ and $\hat{T}_{\text{nf}}$, respectively~(Fig.~\ref{fig:PIVfitting}a). To determine these fitting parameters from flow fields $\mathbf{v}_{\text{exp}}$ measured in the $xy$-plane, we proceeded in two steps. First, for a given observation plane~$z_v$ and distance of flow singularities below the surface~$h$, we note that the fit function $\mathbf{v}_{\text{fit}}$ given in Eq.~(\ref{eq:ufit}) depends linearly on the Stokes flow parameters $F_{\text{st}},\hat{F}_{\text{nf}}$ and~$\hat{T}_{\text{nf}}$. Hence, for given $h$ and $z_v$, these singularity strengths can be determined from an exact least squares minimization of the error $E(h,z_v)=\langle|\mathbf{v}_{\text{fit}}-\mathbf{v}_{\text{exp}}|\rangle_{x,y}/\max_{x,y}|\mathbf{v}_{\text{exp}}|$, where only the in-plane $x$- and $y$-components of $\mathbf{v}_{\text{fit}}$ contribute to this error. For an embryo (or pairs and triplets) of apparent (average) radius $\bar{R}$, the error $E(h,z_v)$ was determined on a $100\times100$ parameter grid on which $h/\bar{R}$ and $z_v/\bar{R}$ are varied in intervals $[0,4]$ and $[-4,0]$, respectively, and $F_{\text{st}},\hat{F}_{\text{nf}}$ and~$\hat{T}_{\text{nf}}$ are determined by least squares fitting $\mathbf{v}_{\text{fit}}$ to measured flow fields $\mathbf{v}_{\text{exp}}$ at each point of this ($h,z_v$)-parameter grid~(Fig.~\ref{fig:PIVfitting}b). In a second step, the final fit parameters are chosen from the fit that globally minimizes the error $E(h,z_v)$. Exemplary error maps $E(h,z_v)$ of pair and triplet fits are shown in Fig.~\ref{fig:PIVfitting}b.\\

\begin{figure}[!t]
\centering
\includegraphics[width=0.99\textwidth]{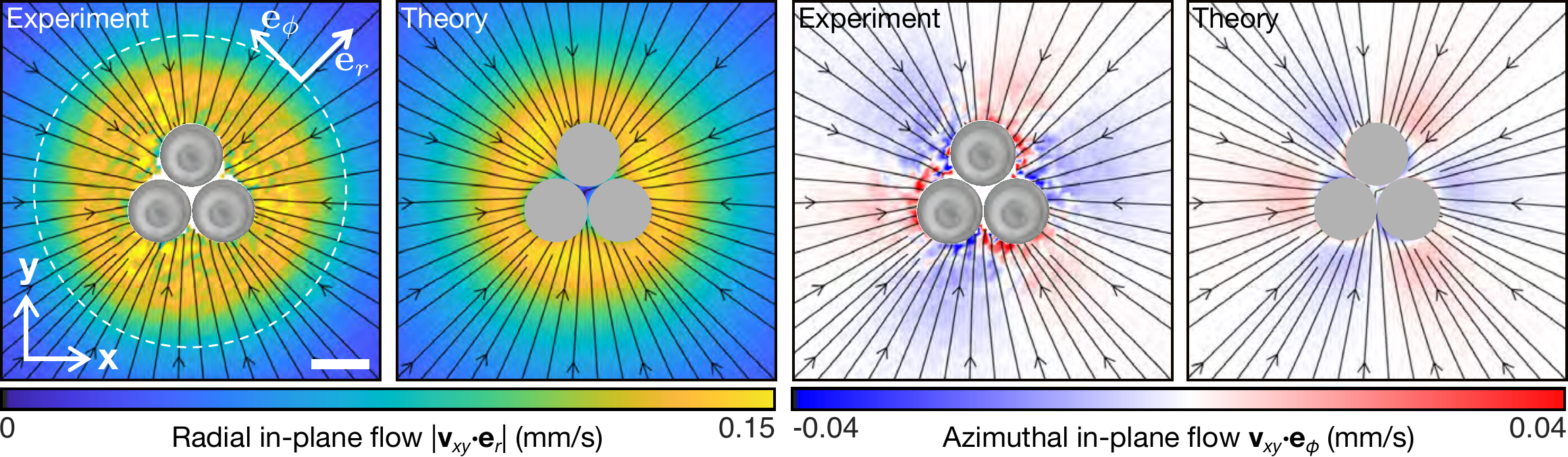}
\caption{\textbf{Fitting of top-view flow fields around triplet bound near the fluid surface.} Measured radial and azimuthal flow field components surrounding a rotating triplets (``Experiment") can be quantitatively described with suitable solutions of the Stokes equation~(``Theory", see Sec.~\ref{sec:PIVfit}) by invoking hydrodynamic interactions illustrated in Fig.~2d (main text). Scale bar, 200\,$\mu$m.}
 \label{fig:PIVfittingtrip}
\end{figure}

\textit{Fit results:} The single-embryo Stokeslet strengths $F_{\text{st}}$ found by the fitting procedure described above are shown in Fig.~\ref{fig:StokesletHydrod}a and an exemplary fit represented as a radial projection is shown in main text Fig.~2b. For the flow field around the rotating pair shown in Fig.~2c (main text), we find \hbox{$F_{\text{st}}=1.1\,$nN}, $\hat{F}_{\text{nf}}=25.4\,$pN and~$\hat{T}_{\text{nf}}=11\,$pN$\cdot\bar{R}$ ($\bar{R}=108\,\mu$m) assuming a viscosity of $\eta=1\,$mPa$\cdot$s to estimate the involved forces and torques. For the flow field around the rotating triplet shown in Fig.~\ref{fig:PIVfittingtrip}, we find \hbox{$F_{\text{st}}=1.1\,$nN}, $\hat{F}_{\text{nf}}=19.8\,$pN and~$\hat{T}_{\text{nf}}=23.3\,$pN$\cdot\bar{R}$ ($\bar{R}=111\,\mu$m). Compared to fitting results for single embryos (Fig.~\ref{fig:StokesletHydrod}a), the Stokeslet strength $F_{\text{st}}$ is reduced, which is most likely due to a lack of flow screening in a pure singularity description of multiple swimmers. Importantly, the signs ($\mathrel{\hat=}$\,orientations) of the fitted near-field contributions are consistent with the orientations of forces and torques expected from the hydrodynamic interactions (main text Fig.~2d, Sec.~\ref{sec:HydrModel}): Consistent with $\hat{F}_{\text{nf}}>0$ (see Fig.~\ref{fig:PIVfitting}a), clockwise spinning of bound embryos indeed leads to a clockwise rotation of pairs and triplets, and consistent with $T_z=\hat{T}_{\text{nf}}>0$, excess torques that arises from the reduced embryo spinning frequencies are expected to point along the positive $z$-direction. Furthermore, we note that the ratio of near-field forces and torques from these fits,  \hbox{$\bar{R}\hat{F}_{\text{nf}}/\hat{T}_{\text{nf}}\approx2.3$} (pair) and $\bar{R}\hat{F}_{\text{nf}}/\hat{T}_{\text{nf}}\approx0.8$ (triplet), are similar to the ratio $f_0/\tau_0=0.5$ of near-field force and torque strengths used in the minimal model (Eqs.~(\ref{eq:Finter}),(\ref{eq:Tinter}), see Tab.~\ref{Tab:ModelParams}), where the latter had been determined by matching rotation frequencies of pairs and triplets between theory and experiment (Sec.~\ref{sec:ModPara}, Fig.~\ref{fig:EmbryoProps}c).

\subsection{Analysis of side-view flow fields surrounding confined embryos}\label{sec:FFanalsisSide}
Here, we describe the experimental procedures and fitting methods used to spatially resolved the fluid flows generated along the lateral embryo surface. While the analysis described in Sec.~\ref{sec:FFanalsis} captures flows from a top-view and provides a direct quantification of the Stokeslet strength and effective hydrodynamic attraction, the analysis described here spatially resolves flow properties along the embryo's AP~body axis and provides the quantitative basis for the developmental stability analysis in Secs.~\ref{sec:StabAnalFull}--\ref{sec:stabtimefinal}.

\subsubsection{Flow field measurements}\label{sec:FFMSide}
\begin{figure}[!t]
\centering
\includegraphics[width=0.99\textwidth]{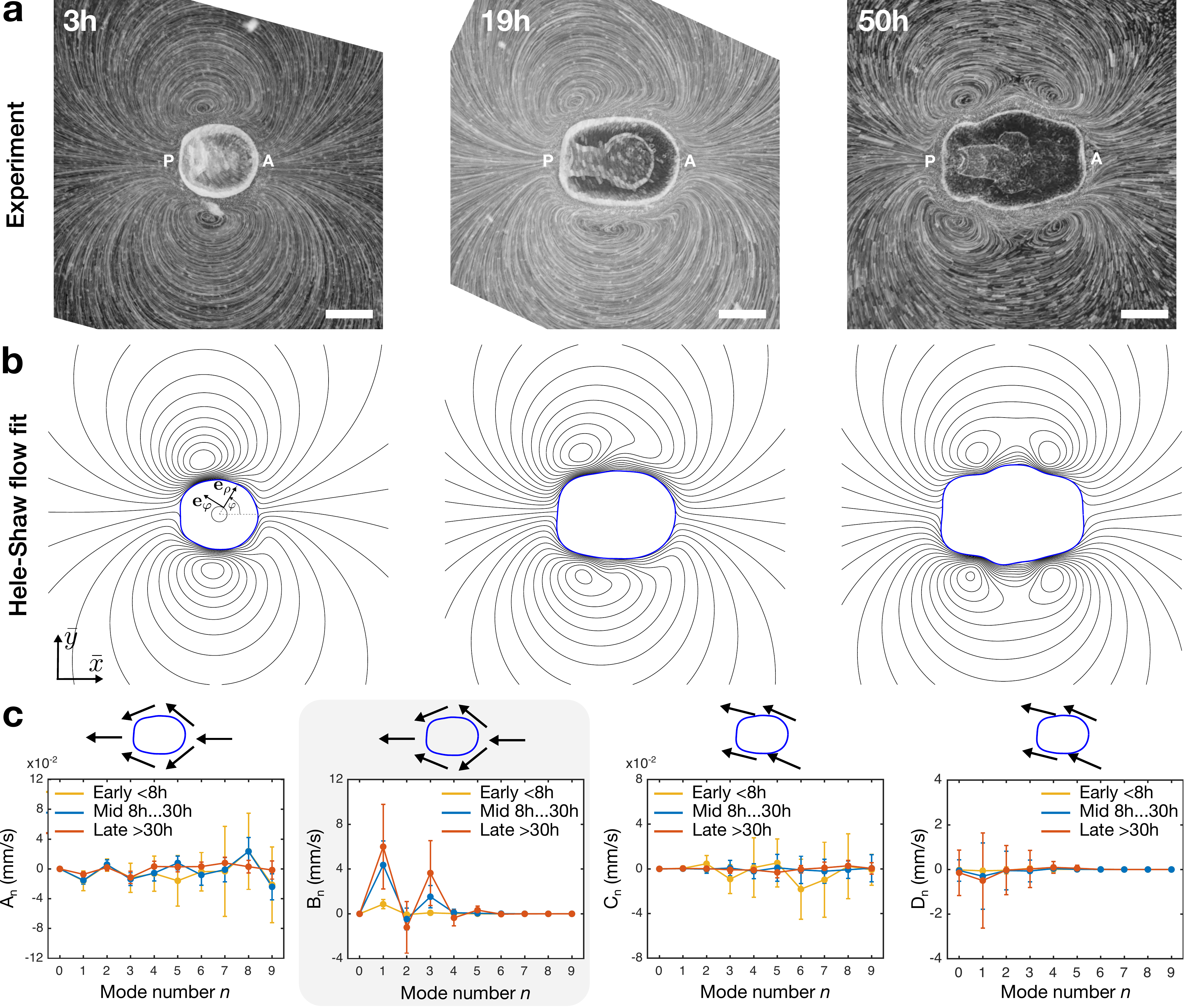}
\caption{\textbf{Fitting of experimental side-view flow fields.} \textbf{a,}~Flow field streamlines surrounding embryos confined between glass slide and cover slip at different developmental stages (maximum projections, see Sec.~\ref{sec:FFMSide}). Over development, an initially symmetric vortex pair moves posteriorly (P) and a second vortex pair appears near the anterior end (A). Scale bar, 100\,$\mu$m. \textbf{b,}~Streamlines of Hele-Shaw flow solutions Eqs.~(\ref{eq:HSsol}) fitted to experimental data in~\textbf{a} (see Sec.~\ref{sec:PIVfitSide}). \textbf{c,}~Mean and standard deviations of fitted mode amplitudes in Eq.~(\ref{eq:HSsol}) for embryos during early ($n=7$), mid ($n=14$) and late ($n=16$) development. These coefficients suggest a minimal parametrization of fluid flows near the embryo surface. Indeed, using only the first three dominant modes $B_1,\,B_2$ and $B_3$, is sufficient to capture the key properties of measured surrounding flows~(see Fig.~\ref{fig:SideViewFitResultsfortheo}).}
 \label{fig:SideViewFitResults}
\end{figure}

Embryos were confined between a glass slide and a cover slip (separation distance $\approx\,100\,\mu$m) such that their long axes were parallel to the imaging plane. Surrounding flow fields were characterized by analyzing the motion of tracer beads though particle image velocimetry (PIV) using the PIVlab plugin for MATLAB \cite{thielicke2014pivlab,thielicke2021particle,thielicke2014flapping}. Default PIVlab setting were used, with the exception that the ``Pass 1 interrogation area" and ``Step" were set to 128 and 64 pixels, respectively, (corresponding to $\approx$\,164\,$\mu$m and $\approx$\,82\,$\mu$m) and the ``Pass 2 interrogation area" was set to 64 pixels ($\approx$\,82\,$\mu$m). The resulting velocity fields were temporally averaged over either 100 or 500 frames (frame rate 20\,fps). Exemplary maximum projections over 100 frames that reveal the streamlines of these flow fields are shown in Fig.~\ref{fig:SideViewFitResults}a. Additionally, the outer contour of each embryo at a given time was found by first generating a binary mask using an intensity threshold to identify pixels corresponding to the embryo. Holes in this mask were then filled, and the mask was smoothed and spurious pixels removed through repeated erosion and dilation. The largest connected region provides a binary mask of the embryo, and the gradient of this mask approximates the embryo's outer contour. A smooth representation of the latter was generated by Fourier-transforming the radial-azimuthal representation $r(\varphi)$ of the approximate contour and using only the first 10 Fourier-modes for further analysis~(blue outlines in Fig.~\ref{fig:SideViewFitResults}b).

\subsubsection{Flow field fitting}\label{sec:PIVfitSide}
In the following, we describe how the mode decomposition of embryo surface flows used in Sec.~\ref{sec:DevModePara} was extracted from  side-view~PIV data. We denote the flows in this analysis by $\mathbf{V}(\bar{x},\bar{y})$ to emphasize that the Cartesian coordinate plane $\{\bar{x},\bar{y}\}$ considered in these measurements is distinct from the ``top-view" $xy$-plane, where the latter is part of the three-dimensional domain in which clusters were analyzed throughout most of this work. We define cylindrical coordinates $\{\rho,\varphi\}$ in this plane by $\bar{x}=\rho\cos\varphi$ and $\bar{y}=\rho\sin\varphi$ (see Fig.~\ref{fig:SideViewFitResults}b). To take into account the confinement under which flow data was acquired (see Sec.~\ref{sec:FFMSide}), we aim to describe measured flows in terms of solutions of the Hele-Shaw equation~\cite{Jeckel2019}
\begin{equation}\label{eq:HS}
\eta\Delta\mathbf{V}-\nabla P=\kappa'\mathbf{V},    
\end{equation}
where $P(\bar{x},\bar{y})$ is the pressure enforcing incompressibility $\nabla\cdot\mathbf{V}=0$ and $\kappa'=12\eta/H^2$ is an effective friction parameter resulting from interactions of the fluid flow with top and bottom surface of the channel of height $H$ ($\approx\,100\,\mu$m in experiments). The most general solution of Eq.~(\ref{eq:HS}) for which $\mathbf{V},P\rightarrow0$ at infinity can be written in cylindrical coordinates as \hbox{$\mathbf{V}=V_{\rho}\mathbf{e}_{\rho}+V_{\varphi}\mathbf{e}_{\varphi}$} with
\begin{subequations}\label{eq:HSsol}
\begin{align}
V_{\rho}&=\sum_{n=1}^{\infty}\left(A_n\cos n\varphi-C_n\sin n\varphi\right)\left(\frac{\rho}{L}\right)^{-n-1}+\sum_{n=1}^{\infty}\left(B_n\cos n\varphi-D_n\sin n\varphi\right)\frac{K_{n+1}-K_{n-1}}{2}\label{eq:ur2D}\\
V_{\varphi}&=\sum_{n=1}^{\infty}\left(A_n\sin n\varphi+C_n\cos n\varphi\right)\left(\frac{\rho}{L}\right)^{-n-1}+\sum_{n=0}^{\infty}\left(B_n\sin n\varphi+D_n\cos n\varphi\right)\frac{K_{n+1}+K_{n-1}}{2},
\end{align}
\end{subequations}
where $\kappa=\sqrt{\kappa'/\eta}=\sqrt{12}H^{-1}$, $L=(\ell_{\text{min}}+\ell_{\text{maj}})/2$ is the characteristic embryo size, $K_{n}(\kappa\rho)$ denote modified Bessel functions of the second kind and the coefficients $\{A_n,B_n,C_n,D_n\}$ represent integration constants that will be determined by fitting the measurement data described in Sec.~\ref{sec:FFMSide}. Qualitatively, the coefficients $A_n$ and $B_n$ represent flows that are mirror-symmetric with respect to the $x$-axis, the coefficients $B_n$ and $D_n$ describe flows whose mirror-symmetry with respect to the $x$-axis is broken. To define the least-square optimization underlying these fits, we additionally introduce the stream function
\begin{equation}
\Psi(\rho,\varphi)=L\sum_{n=1}^{\infty}\left(\frac{A_n}{n}\sin n\varphi+\frac{C_n}{n}\cos n\varphi\right)\left(\frac{\rho}{L}\right)^{-n}+\kappa^{-1}\sum_{n=0}^{\infty}\left(B_n\sin n\varphi+D_n\cos n\varphi\right)K_n(\kappa\rho)
\end{equation}
that generates the flow field components in Eqs.~(\ref{eq:HSsol}) via $V_{\rho}=\rho^{-1}\partial_{\varphi}\Psi$ and $V_{\varphi}=-\partial_{\rho}\Psi$. We then solve the least-square problem
\begin{equation}\label{eq:LSopt}
\min_{A_n,B_n,C_n,D_n}\Vert\mathbf{V}-\mathbf{v}_{\text{exp}}\Vert_2+\Vert\mathbf{e}_s\cdot\nabla\Psi|_{\mathcal{C}}\Vert_2,
\end{equation}
where $\mathbf{v}_{\text{exp}}$ denotes data from flow measurements~(Sec.~\ref{sec:FFMSide}), $\mathbf{e}_s$ denotes the unit tangent vector along the curve $\mathcal{C}$ prescribed by the embryo boundary~(blue outlines in Fig.~\ref{fig:SideViewFitResults}b). The norm~$\Vert\cdot\Vert_2$ indicates the square of the Euclidean norm at which experimental velocities~$\mathbf{v}_{\text{exp}}$ were measured. The second term in Eq.~(\ref{eq:LSopt}) penalizes flow contributions from~$\mathbf{V}$ that are not tangential at the embryo surface. The prefactor of this term corresponds to a weight, which is set here to $1$, while we note that the final fits and conclusions drawn from them do not dependent on this particular choice. Exemplary fits resulting from this approach for embryos are shown in~Fig.~\ref{fig:SideViewFitResults}b. The corresponding mode amplitudes, pooled at three different developmental stages, are depicted in~Fig.~\ref{fig:SideViewFitResults}c. Mode coefficients $B_n$ (gray box in Fig.~\ref{fig:SideViewFitResults}c) that describe mirror-symmetric flows along the embryo's AP axes are the dominating contribution to these fits and are used in Sec.~\ref{sec:DevModePara} to parametrize changes of near-surface flow fields surrounding developing embryos.

\subsection{Embryo spinning frequencies}
\textit{Analysis of embryo spinning frequencies in small clusters:} 
Average embryo spinning frequencies were determined from the manually measured duration of 10 embryo rotations~(data ``In small clusters" in Fig.~2f, main text).\ \\

\textit{Automated analysis of embryo spinning frequencies in large clusters:} 
To measure embryo spinning frequencies within clusters containing $\approx100$ embryos, we took advantage of the inhomogeneous intensity profile within the embryo body that results from the uneven positioning of developing internal organs. We first processed raw videos using Fiji~\cite{schindelin2012fiji} by inverting the pixel intensity and performing a background subtraction. For each embryo, we shifted the origin of the coordinate system to the embryo centroid, and considered pixels within a circular neighborhood with radius equal to the apparent embryo radius. From these pixels, we constructed an angular intensity profile $I(\theta,t)$ at each time $t$ by averaging the pixel intensity within sectors of angular width~$\pi/90$. We then smoothed the angular intensity profile using adjacent averaging of 10 points in $\theta$ and 500 points in $t$ to reduce noise and suppress slow global intensity variations, respectively. We then computed the cross-correlation between angular intensity profiles $I(\theta+\Delta\theta,t+\Delta t)$ and $I(\theta,t)$ at two time points, and determined the angular lag $\Delta\theta=\Delta\theta_{max}$ that maximizes this cross-correlation. While the analysis can be performed between successive frames, performing a running average over $\Delta t$ frames yields a higher signal to noise ratio and hence more accurate results. In practice, we chose $\Delta t = 3$ frames and subsequently determined the measured dynamics $\bar{\theta}(t)$ from a cumulative sum over $\Delta\theta_{max}$. Finally, we fitted $\theta(t)=\omega t^2/2$ to the cumulative sum of $\bar{\theta}(t)$ to determine a noise-robust estimate of the embryo's spinning frequency $\omega/(2\pi)$~(data ``In larger clusters" in Fig.~2f, main text).\ \\ 

\textit{Analysis of embryo spinning frequencies in the oscillating cluster:} Embryo spinning frequencies inside oscillating clusters (main text Fig.~4d) were not amenable to automated tracking. Therefore, we manually determined the spinning rates of 40 embryos within the cluster and found a value of $(0.33\pm0.13)\,$min$^{-1}$ (mean$\,\pm\,$standard deviation,~$n=40$).

\subsection{Quantifying sources of effective noise contributing to cluster dissolution}
As embryos develop, they exhibit notable changes in their hydrodynamic properties (see Sec.~\ref{sec:FFMSide} and Fig.~\ref{fig:SideViewFitResults}) and in their morphology (see main text Fig.~1b, Fig.~\ref{fig:SideViewFitResultsfortheo}d,e for views in the plane of the embryo's anterior-posterior axis). These changes introduce sources of an effective noise in embryo interactions that leads to a gradual loss of positional order (Fig.~3a,b) and eventually facilities cluster dissolution. We quantified two features associated with this effective noise that can be directly observed in top-view images of clusters: Body shape anisotropies perpendicular to the anterior-posterior axis that make embryos ``bump" into each other when closely packed and spinning within a cluster, as well as the tilt angle of embryos at cluster boundaries that increases their tendency to swim away from or to be scattered off the cluster boundary. The corresponding data is shown in Fig.~2h,i (main text).

\subsubsection{Shape anisotropy of embryo cross-sectional area}\label{sec:shapanis}
Of particular importance for interactions in clusters is the morphological symmetry breaking in the dorsal/ventral-left/right (DV-LR) plane perpendicular to the embryo's AP axis. When embryos are bound below the fluid surface, this plane is parallel to the ``top-view" $xy$-plane defined in Fig.~1e. To quantify the anisotropy of embryo shape in the DV-LR plane, we have fitted ellipses to outlines of embryos in the bulk of the crystal (red outlines in main text Fig.~2h). To find outlines from brightfield images, we first identified the white spaces between the embryo center and boundary using the pixel classification functionality of the software ilastik~\cite{berg2019}. The labeling of training data is done at different time points to improve the classification quality. We then fit ellipses onto the identified region using MATLAB's \verb+regionprops+ function. To eliminate spuriously defined regions, we only consider regions with areas within 9000 and 26000\,$\mu$m$^2$ (120 and 350 pixels squared). From fitted ellipses, we obtain the minor and major axis lengths $\ell_{>}$ and $\ell_{<}$, respectively, and compute the ellipticity of the embryo's top-view cross section as $1-\ell_{>}/\ell_{<}$ (data shown in main text Fig.~2h).

\subsubsection{Tilt angle of embryos at cluster boundaries}
Focusing now on embryos at cluster boundaries, we want to determine tilt angles of the embryo's AP axis away from the $z$-axis (see main text Fig.~1e and Fig.~2j inset for definition of $z$-axis and tilt angle). With basic shape information given, tilt angles can be inferred from projected embryo outlines in the $xy$-plane. Similar to the approach in Sec.~\ref{sec:shapanis}, outlines were characterized by the principal axis lengths $\bar{\ell}_{>}$ and $\bar{\ell}_{<}$ of ellipses that we placed manually over embryo outlines. An automated segmentation could not be realized for this analysis due to the increasingly complex embryo morphology at later developmental stages.

While 3D embryo shapes are approximately given by ellipsoids with possibly three distinct principal axes (see Fig.~\ref{fig:SideViewFitResultsfortheo} and main text Fig.~2h for shape anisotropies in the AP-LR- and in the DV-LR-plane, respectively), we found that the projected minor axis $\bar{\ell}_{<}$ of tilted embryos hardly changed across different measurement time points. In this case, tilt angles can be systematically determined from the projected major axis $2\bar{\ell}_{>}$, if the embryo's actual AP axis length $2\ell_{\text{maj}}$ (see Fig.~\ref{fig:SideViewFitResultsfortheo}c) is known. We estimated the latter independently from horizontally swimming embryos ($\ell_{\text{maj}}=[155, 186, 212]\,\mu$m for the 3 time points $t_d=[11,28,33]\,$hours in Fig.~2i). We then determined a ``look-up table" $\ell_{xy}(\theta)$ by projecting a correspondingly rotated ellipsoid with minor and major axis lengths $2\bar{\ell}_{<}$ and $2\ell_{\text{maj}}$, respectively, onto the $xy$-plane and compared the values $\ell_{xy}(\theta)$ to measurements of~$\bar{\ell}_{>}$ to infer the tilt angle~$\theta$ shown in Fig.~2i (main text).

\subsection{Orientational order parameter}\label{sec:OrientOrd}
To compute the local bond orientational order for each embryo $i$, the nearest neighbor embryo is first determined by a delaunay triangulation of embryos within the cluster (using the MATLAB function \verb+delaunay+). A threshold of $1.5a$ with lattice constant $\bar{a}\approx207\,\mu$m was applied to exclude cases where nearest neighbors are anomalously determined at the cluster boundary due to irregular cluster shapes. The local bond orientational order parameter~\cite{nelson1979dislocation} is then defined as
\begin{equation}\label{eq:hexorder}
\psi_6(\mathbf{r}_i)=\frac{1}{N_i}\sum_{j=1}^{N_i}e^{i6\phi_{ij}}\equiv |\psi_6|_ie^{i\phi_i},
\end{equation}
where the sum is over the $N_i$ nearest neighbors of embryo~$i$, and~$\phi_{ij}$ is the angle between the $x$-axis of the co-rotated cluster frame and the bond connecting embryos $i$ and $j$. $|\psi_6|_i$ quantifies the magnitude of local hexagonal order and $\phi_i=\arg{\psi_6}$ measures the angle of the local bond orientational order parameter.

\subsection{Pair distribution function}
\begin{figure}[!t]
\centering
\includegraphics[width=0.97\textwidth]{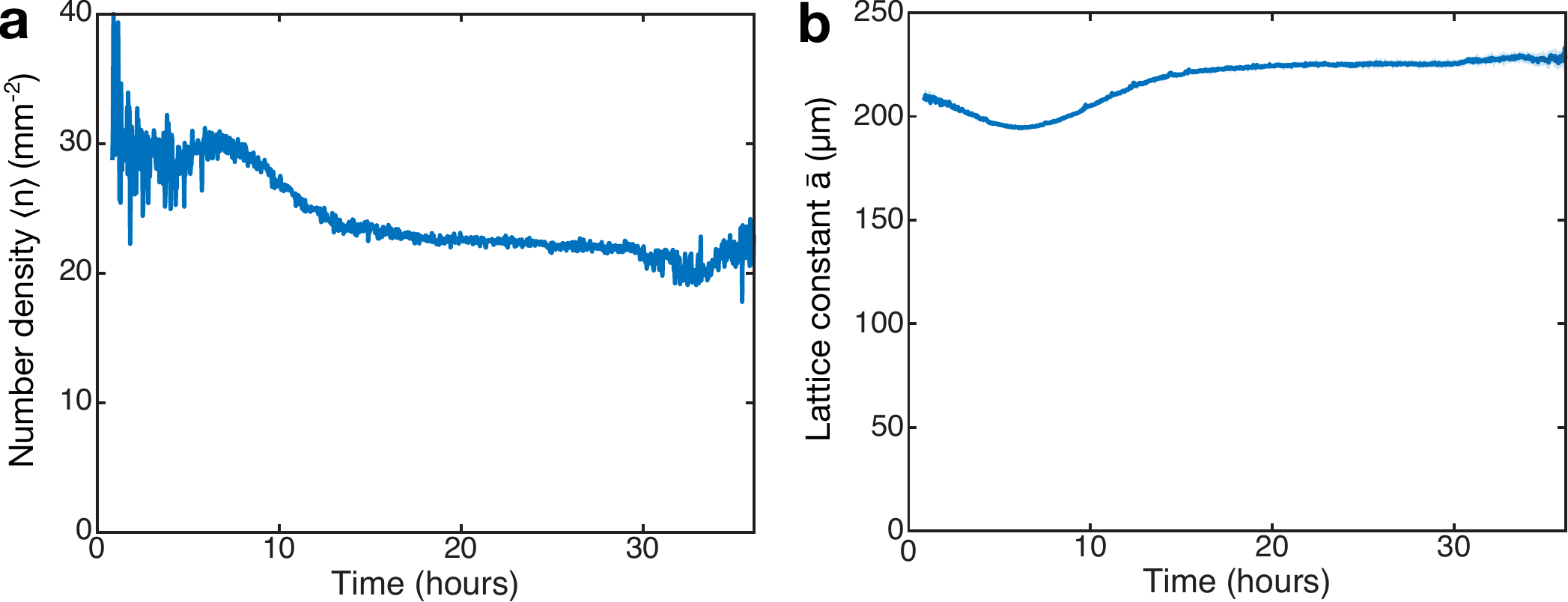}
\caption{\textbf{Time course of average lattice properties.}
\textbf{a,}~Number density $\langle n\rangle$ of embryos in a crystal over the time course of the experiment. This information was used to compute the radial pair distribution function $g(r)$ given in Eq.~(\ref{eq:gr_discr}) and shown as an inset in Fig.~3f (main text). \textbf{b,}~Lattice constant~$\bar{a}$ as a function of time from fitting the first, nearest-neighbors peak of $g(r)$ (see also main text Fig.~3f, inset).}
 \label{fig:lattice_props}
\end{figure}

Following~\cite{chaikin_lubensky_1995}, we define the radial pair distribution function~$g(r)$ as
\begin{equation}
    g(r)=\frac{1}{2\pi rN_{cl}\langle n\rangle}\left\langle\sum_{i\neq j}\delta(r-|\mathbf{r}_i -\mathbf{r}_j|)\right\rangle,\label{eq:gr}
\end{equation}
where $N_{cl}$ is the total number of centroids in a cluster and $\langle\cdot\rangle$ denotes the average over all centroids at positions~$\mathbf{r}_i$ and $\mathbf{r}_j$. The number density $\langle n\rangle$ in Eq.~(\ref{eq:gr}) was estimated from the rotation-corrected centroid data~(see Sec.~\ref{sec:EmbrProc}) by identifying boundary centroids using the MATLAB function \verb+boundary+, finding the area defined by these boundary centroids using \verb+polyarea+, and dividing the total number of centroids by this area. A time course of the number density $\langle n\rangle$ is shown in Fig.~\ref{fig:lattice_props}a. The pair distribution $g(r)$, e.g. as shown in the inset of Fig~3f (main text), was then computed at~100\,s intervals by approximating  Eq.~(\ref{eq:gr}) at discrete points $r_k=k\Delta r$ ($k=1,2,3,...$) as
\begin{equation}\label{eq:gr_discr}
    g(r_k)= \frac{1}{2\pi kN_{cl}\langle n\rangle\left(\Delta r\right)^2}\sum_{i\neq j}\mathbbm{1}_{r_k\leq|\mathbf{r}_i -\mathbf{r}_j|<r_{k+1}},
\end{equation}
which uses an indicator function
\begin{equation}  
\mathbbm{1}_{r_k\leq|\mathbf{r}_i -\mathbf{r}_j|<r_{k+1}}=\begin{cases} 
      1 & \text{if } r_k\leq|\mathbf{r}_i -\mathbf{r}_j|<r_{k+1} \\
      0 & \text{else}.
   \end{cases}.\label{eq:indicfunc}
\end{equation}   
The bin width was $\Delta r\approx8.6\,\mu$m. Finally, the nearest neighbor peak of $g(r_k)$ was fit to a Gaussian function $G(r)=C\exp\left[-(r-r_{\mu})^2/\left(2\sigma^2\right)\right]$ (Fig.~3f, inset) with fitting parameters describing the amplitude~$C$, mean~$r_{\mu}$, and width~$\sigma$ of the peak. That latter was used as the ``First Peak Width" shown in Fig.~3f (main text), and $r_{\mu}$ was taken as the lattice constant~$\bar{a}$ (Fig.~\ref{fig:lattice_props}b).

\subsection{Dynamic Lindemann parameter}\label{sec:LM} 
Following~\cite{Zahn1999} and using rotation-corrected centroid positions~(see Sec.~\ref{sec:EmbrProc}), we define the dynamic Lindemann parameter as 
\begin{equation}
\gamma_{L}(\tau) = \frac{1}{2\bar{a}^2}\sum_{j,j+1}\left\langle\left[\Delta\mathbf{r}_j(\tau,t) - \Delta\mathbf{r}_{j+1}(\tau,t)\right]^2 \right\rangle_t,\label{eq:LM}
\end{equation}
where $\Delta\mathbf{r}_i(\tau,t)=\mathbf{r}_i(t+\tau)-\mathbf{r}_i(t)$ denotes the displacement of embryo $i$ between two time points of duration $\tau$ apart. Index pairs $j$ and $j+1$ in Eq.~(\ref{eq:LM}) correspond to nearest neighbor pairs, and $a$ is the lattice constant. The analysis takes into account variation of $a$ with developmental time (Fig.~\ref{fig:lattice_props}b). Two embryos were considered to be nearest neighbors if their initial positions were separated by a distance smaller than $1.2a$. 

\begin{figure}[!t]
\centering
\includegraphics[width=0.55\textwidth]{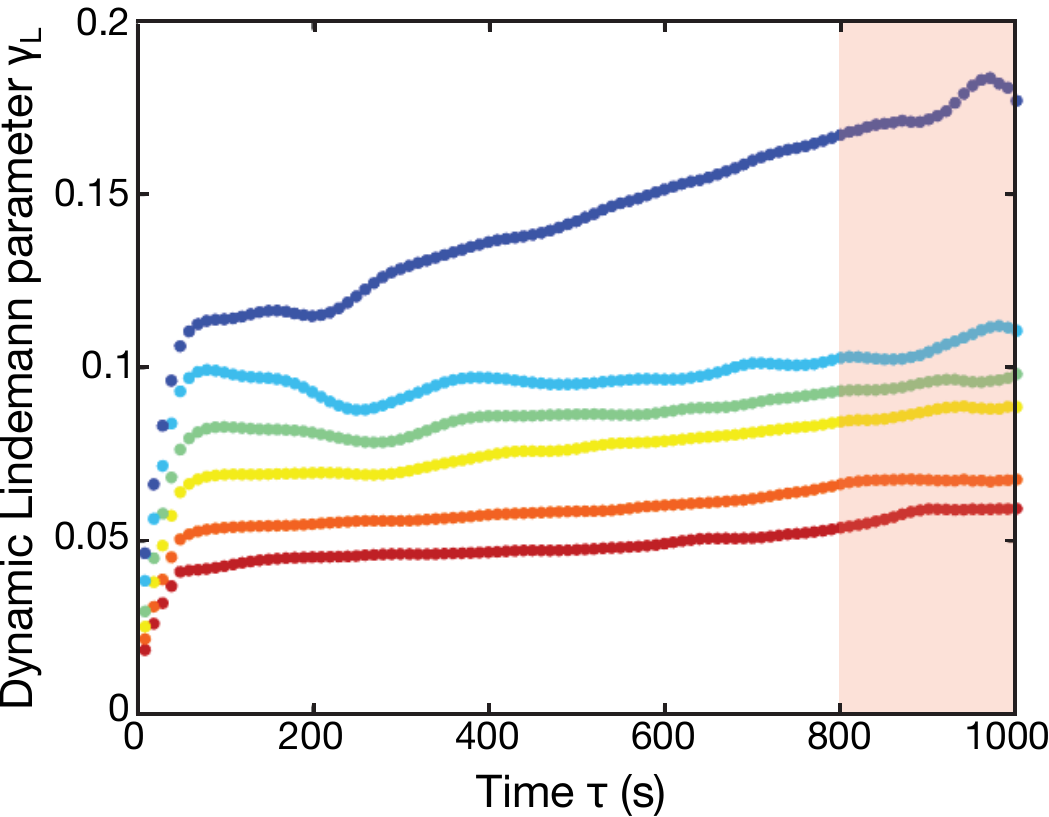}
\caption{\textbf{Dynamic Lindemann parameter.} Representative plots for the dynamic Lindemann parameter $\gamma_L(\tau)$ (see Sec.~\ref{sec:LM}) for developmental times $t_d$ of 8\,h (red), 12\,h (orange), 16\,h (yellow), 20\,h (green), 24\,h~(light blue) and 28\,h (dark blue). The average values and variance of $\gamma_L(\tau)$ within the shaded region for different developmental times $t_d$ are shown in Fig.~3g (main text).}
 \label{fig:LMexample}
\end{figure}

To generate Fig.~3g in the main text, we considered a subset of consecutive developmental time points $t_d$ that are $100$\,s apart from each other. The Lindemann parameter $\gamma_{L}(\tau)$ is then computed for each $t_{d}$ within a centered $1000$\,s interval, i.e. the average in Eq.~(\ref{eq:LM}) was performed over times $t\in[t_{d}-500$\,s, $t_{d}+500$\,s$]$. Fig.~\ref{fig:LMexample} shows representative examples of~$\gamma_{L}(\tau)$ computed at different developmental time points~$t_d$. For each $t_d$, the mean and standard deviation of $\gamma_L(\tau)$ over the range \hbox{$800$\,s $< \tau\le$ $1000$\,s} (20 consecutive time points in total) are plotted as data points and error bars in Fig.~3g (main text), respectively. At very late times $>30\,$h, due to dissolution of the cluster, embryo trajectories do not last over $1000$\,s, and hence, the dynamic Lindemann parameter cannot be calculated in this regime. 

\subsection{Displacement field and strain components}\label{sec:SFcalc}
To determine displacement and strain fields of oscillating clusters (Fig.~4), rotation-corrected centroid trajectories~$\mathbf{r}_i(t)$ (see Sec.~\ref{sec:EmbrProc}) of embryos $i=1,2,...$ were first smoothed with a 5-frame (50\,s) moving average to remove noise. We then determined long-time averaged embryo positions $\mathbf{R}_i(t)=\langle \mathbf{r}_i(t) \rangle_{[t-25,t+25]}$ by using a 50-frame (500\,s) moving average. The displacement of each embryo from its average position is given by $\mathbf{u}(\mathbf{r}_i,t):=\mathbf{u}_i(t)=\mathbf{r}_i(t)-\mathbf{R}_i(t)$ and an exemplary displacement time series is shown in the inset of Fig.~4a (main text).
A Fourier-analysis of $\mathbf{u}(\mathbf{r}_i,t)$ for all embryos that were continuously tracked throughout a time window of 4000\,s revealed an average displacement oscillation frequency of $(0.26\pm0.04)\,$min$^{-1}$ (mean\,$\pm$\,standard deviation, $n=389$).

A~continuous displacement field $\mathbf{u}(\mathbf{r},t)$ was approximated from all embryo displacements $\mathbf{u}_i(t)$ by applying a 2D Gaussian filter of radius $\bar{a}\approx 216\,\mu$m (approximately one lattice constant). From $\mathbf{u}(\mathbf{r},t)$, we computed the displacement gradient tensor $u_{ij}=\partial_iu_j$ in the $xy$-plane \hbox{($i,j\in\{x,y\}$)}. As described in Sec.~\ref{sec:OddModuli} and following the convention in~\cite{Scheibner:2020gm}, the displacement gradient tensor can be decomposed into four independent strain components given by (i)~the divergence \hbox{$u^0(\mathbf{r},t)=u_{xx}+u_{yy}$}, (ii) the curl $u^1(\mathbf{r},t)=u_{yx}-u_{xy}$, as well as by the two shear components (iii)~$u^2(\mathbf{r},t)=u_{xx}-u_{yy}$ (shear~1), and (iv) $u^3(\mathbf{r},t)=u_{yx}+u_{xy}$ (shear~2). 

\begin{figure}[!t]
\centering
\includegraphics[width=0.97\textwidth]{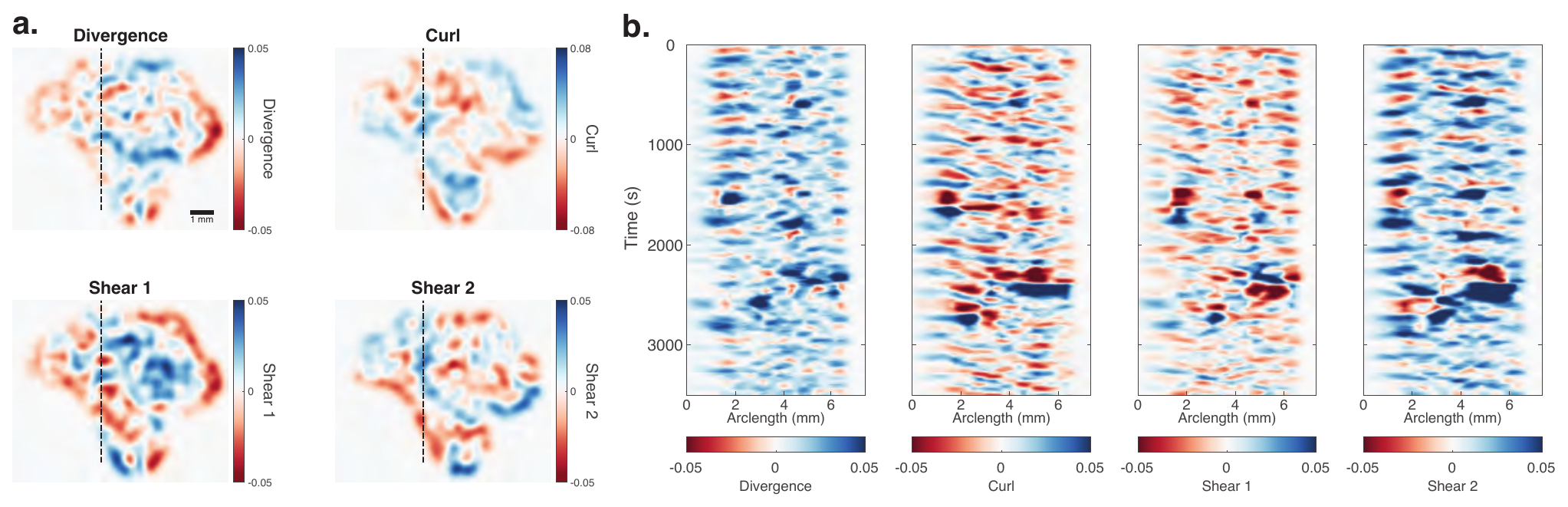}
\caption{\textbf{Strain waves in bulk.}~\textbf{a,}~Exemplary snapshots of the four strain components: divergence, curl, shear 1 and shear 2 during cluster oscillations. \textbf{b,}~Space-time kymographs of the four principal strain components measured along the black dashed lines in \textbf{a}. Details of the analysis are described in Sec.~\ref{sec:SFcalc}.}
 \label{fig:StrainInBulk}
\end{figure}

To determine space-time kymographs of the strain component dynamics along the boundary shown in maint text Fig.~4e,g, we first determined a parametrization~$\mathbf{r}_s$ of the cluster boundary (MATLAB function \verb+bwboundaries+) in terms of the boundary arc length $s$. Finally, components of the displacement gradient tensor $u_{ij}$ were projected onto a local basis composed of the boundary tangent~$\partial_s\mathbf{r}_s$ and the boundary normal pointing away from the cluster to compute transformed strain components $u^{\alpha}$ ($\alpha=0,1,2,3$) analog to the definitions above. This transformation leaves the divergence and curl components $u^0$ and $u^1$, respectively, invariant and corresponds to a rotation of the strain component vector $(u^2,u^3)^\top$ that conserves the total shear strain amplitude $u_s=\sqrt{[u^2]^2+[u^3]^2}$ ($[u^{\alpha}]^2$ with $\alpha=2,3$ denotes squared shear strain components). Strain components computed in the co-rotating Cartesian basis along a section through the bulk (dashed line in Fig.~\ref{fig:StrainInBulk}a) are shown in Fig.~\ref{fig:StrainInBulk}b and exhibit similar amplitudes and wavelengths, indicating that the corresponding excitations are present at the boundary and in the bulk of the cluster.

\subsection{Mode chirality analysis of displacement waves}\label{sec:ModChir}
To characterize the bulk dynamics of cluster oscillations discussed in Fig.~4d--h (main text) with respect to their chiral symmetry, we consider a complex representation of the displacement field given by
\begin{equation}
U(\mathbf{r},t)=u_x(\mathbf{r},t)+iu_y(\mathbf{r},t).\label{eq:complrepr}
\end{equation}
Here, $i$ denotes the imaginary unit, and $u_x(\mathbf{r},t)$ and $u_y(\mathbf{r},t)$ represent the Cartesian components of the smoothed, dynamic displacement field $\mathbf{u}(\mathbf{r},t)$ described in Sec.~\ref{sec:SFcalc}. Using the spatio-temporal Fourier transform of $U(\mathbf{r},t)$ given by
\begin{equation}
    \tilde{U}(\mathbf{q},\omega)=\int d\mathbf{r}\int dt\,U(\mathbf{r},t)\exp\left[-i\left(\mathbf{q}\cdot\mathbf{r}+\omega t\right)\right],\label{eq:FT}
\end{equation}
a mode chirality parameter can be defined as
\begin{equation}
C(\omega)=\frac{\langle|\tilde{U}(\mathbf{q},\omega)|-|\tilde{U}(-\mathbf{q},-\omega)|\rangle_{\text{BZ}}}{\langle|\tilde{U}(\mathbf{q},\omega)|+|\tilde{U}(-\mathbf{q},-\omega)|\rangle_{\text{BZ}}}.\label{eq:ChirMeas}
\end{equation}
Averages $\langle\cdot\rangle_{\text{BZ}}$ over wave vectors $\mathbf{q}$ are taken within the first Brillouin zone defined by the lattice constant $\bar{a}\approx220\,\mu$m of the hexagonal embryo cluster. The quantity $|C(\omega)|$ characterizes the symmetry of the displacement field's Fourier spectrum $\tilde{U}(\mathbf{q},\omega)$ with respect to point-reflections at the Fourier space origin. It vanishes if the spectrum is perfectly point-symmetric and becomes unity if the point-symmetry of a given mode is maximally broken. In practice, $C(\omega)$ given in Eq.~(\ref{eq:ChirMeas}) can be used to detect and quantify signatures of chirality in oscillating displacement fields. To see this explicitly, it is instructive to consider a minimal displacement wave of the form
\begin{subequations}
\begin{align}
u_x(\mathbf{r},t)&=u^{(0)}_x\cos\left(\Omega t-\lambda x\right)\label{eq:wv1}\\
u_y(\mathbf{r},t)&=u^{(0)}_y\sin\left(\Omega t-\lambda x\right)\label{eq:wv2},
\end{align}\label{eq:testwave}%
\end{subequations}
for some frequency $\Omega$ and wavelength $\lambda$. The wave described by Eqs.~(\ref{eq:testwave}) represents a pure longitudinal (transverse) wave if $u^{(0)}_x\ne0$ and $u^{(0)}_y=0$ ($u^{(0)}_x=0$ and $u^{(0)}_y\ne0$). In either case, the Fourier amplitude of the complex representation Eq.~(\ref{eq:complrepr}) is point-symmetric, i.e. $|\tilde{U}(\mathbf{q},\omega)|=|\tilde{U}(-\mathbf{q},-\omega)|$ and consequently $C(\omega)=0$ [see Eq.~(\ref{eq:ChirMeas})] for purely longitudinal or transverse waves. However, if both amplitudes of the minimal wave in Eq.~(\ref{eq:testwave}) are finite, $u^{(0)}_x\ne0$ and $u^{(0)}_y\ne0$, the wave acquires a chiral character, as seen by the well-defined rotation sense of displacement vectors at every point $\mathbf{r}$. In this case, the point-symmetry in Fourier-space is lost and $|C(\omega=\Omega)|\ne0$. Interestingly, for $u^{(0)}_x=u^{(0)}_y$ the displacement vector Eq.~(\ref{eq:testwave}) draws out perfect cirlces at every point $\mathbf{r}$, a characteristic of emergent displacement waves that can appear in purely odd elastic materials~\cite{Scheibner:2020gm}. In this case, the chirality measure defined in Eq.~(\ref{eq:ChirMeas}) becomes maximal at the wave frequency, i.e. $|C(\omega=\Omega)|=1$.

\begin{figure}[!t]
\centering
\includegraphics[width=0.97\textwidth]{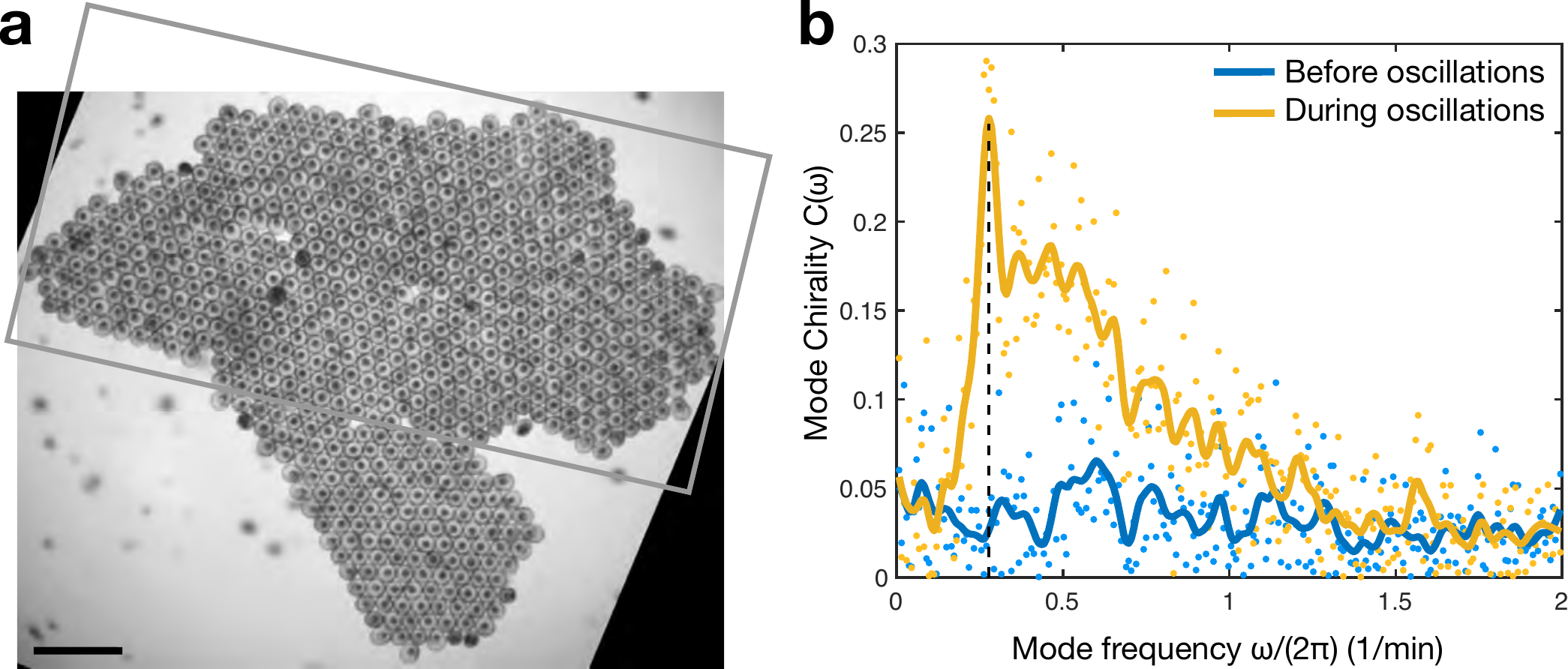}
\caption{\textbf{Mode chirality analysis of displacement waves.}~\textbf{a,}~The mode chirality analysis described in Sec.~\ref{sec:ModChir} was performed on a $7.8\times 3.9\,$mm region of interest indicated by the gray box over a time window of 135\,min. Scale bar, 1\,mm. \textbf{b,}~Mode chirality parameter $C(\omega)$ given in Eq.~(\ref{eq:ChirMeas}) before the onset of (blue dots) and during (yellow dots) visible oscillations. A smoothed representation of this mode data is depicted by solid lines and serves as guide to the eye. The black dashed line indicates most prominent chiral oscillations at a frequency of approximately $0.28\,$min$^{-1}$.}
 \label{fig:ModeChirality}
\end{figure}

We have determined $C(\omega)$ from displacement field information $\mathbf{u}(\mathbf{r},t)$ located in the domain shown in Fig.~\ref{fig:ModeChirality}a (gray box, $900\times 450$\,pixel, corresponding to $7.8\times 3.9\,$mm). Points of this domain outside the cluster were zero-padded. We then considered temporal sections of 800 consecutive time points ($\approx135\,$min in total) before the onset of and during cluster oscillations. For each section, we computed a fast Fourier transform (MATLAB function~\verb+fftn+ \cite{fft}) of the resulting $900\times450\times800$ data-cubes to approximate the Fourier transform Eq.~(\ref{eq:FT}) and determine $C(\omega)$ given in Eq.~(\ref{eq:ChirMeas}). During oscillations, a broad spectrum of frequencies shows chiral signatures~(yellow curve in Fig.~\ref{fig:ModeChirality}b), including a distinct peak with frequency \hbox{$\approx0.28\,$min$^{-1}$}~(black dashed line). An almost identical frequency (\hbox{$0.26\,$min$^{-1}$}) is found from a direct analysis of the space-time kymographs that characterize strain component oscillation along the cluster boundary~(see main text Fig.~4e,g and Sec.~\ref{sec:SFcalc}). In the absence of visible cluster oscillations $|C(\omega)|$ flattens substantially~(blue curve in Fig.~\ref{fig:ModeChirality}b).

\subsection{Extracting elastic moduli from strain fields near crystal lattice defects}\label{sec:StrAnalysis}
Interactions between embryos within clusters give rise to emergent mechanical properties\linebreak(Sec.~\ref{Sec:CGmodel}). To analyze these properties experimentally, we use the fact that living chiral crystals typically contain topological lattice defects, such as edge dislocations, that locally displace embryos away from a regular hexagonal arrangement. The associated displacement and strain fields contain information about the effective mechanical properties of a living chiral crystal (Fig.~4a--c, main text). In this section, we characterize these properties in terms of suitable elastic moduli. To this end, we compare experimental measurements with recently derived predictions about strain fields around edge dislocations in the most general linearly elastic isotropic material~\cite{braverman2020topological} that can also be out of equilibrium~\cite{Scheibner:2020gm}. As suggested by the minimal model of the microscopic embryo interactions (Sec.~\ref{Sec:ModelMapping}), and consistent with experimental observations of overdamped chiral displacement waves (Sec.~\ref{sec:ModChir}) and strain cycles~(Fig.~4d--h, main text), we also allow for odd elastic moduli~(see Sec.~\ref{sec:OddModuli}) when fitting experimental displacement fields. Throughout this analysis, we use rotation-corrected centroid data (see~Sec~\ref{sec:EmbrProc}) and analyze embryo crystals at time points in which oscillations are absent.

\subsubsection{Determining strain fields around lattice defects}\label{sec:SFD}
To properly resolve strain fields near defects, we follow the procedure suggested in~\cite{braverman2020topological} and briefly described below. Away from any lattice defects, this approach is consistent with the analysis in Sec.~\ref{sec:SFcalc}, while it does not require spatio-temporal smoothing and therefore provides a more detailed characterization of strain fields in the vicinity of defects. We consider a Delauny triangulation of embryo centroids to identify nearest neighbors (Fig.~\ref{fig:DefectStrainSI1}) and define the ``local displacement field" around embryo $i$ as 
\begin{equation}\label{eq:locDisp}
    \textbf{u}_i(j_n)=\textbf{r}_{j_n(i)}-\textbf{r}^\text{Tem}_{j_n(i)},
\end{equation}
which corresponds to the set of six displacement vectors $\textbf{u}_i(j_n)$ associated with the six nearest neighbors $j_1(i),...,j_6(i)$ of embryo $i$, located at positions $\textbf{r}_{j_n(i)}$. The positions $\textbf{r}^\text{Tem}_{j_n(i)}$ correspond to the corners of a template hexagon that is defined as follows. First, we determine from the Delaunay triangulation global averages of the lattice constant $\bar{a}^\text{Tem}$ and of the hexagonal bond orientation $\phi^\text{Tem}=\arg\langle\psi_6\rangle/6$ (mod~$\pi/3$) with $\psi_6$ defined in Eq.~(\ref{eq:hexorder}), where edges connecting to embryos without 6 neighbors or near the cluster boundary (closer than 650\,$\mu$m) are excluded. Template positions surrounding each embryo $i$ are then given by the corners of a hexagon located~at
\begin{equation}\label{eq:Def_Tem}
    \mathbf{r}^\text{Tem}_{j_n(i)}=\mathbf{r}_{i}+\bar{a}^\text{Tem}\begin{bmatrix}
\cos\left(\frac{(n-1)\pi}{3}+\phi^\text{Tem}\right)\vspace{0.05cm}\\
\sin\left(\frac{(n-1)\pi}{3}+\phi^\text{Tem}\right)
\end{bmatrix}.
\end{equation}
Neighbor positions $\mathbf{r}_{j_n(i)}$ in Eq.~(\ref{eq:locDisp}) are indexed in counter-clockwise order and such that the bond angle between embryos $\mathbf{r}_{i}$ and $\mathbf{r}_{j_1(i)}$ is the one closest to $\phi^\text{Tem}$. Finally, we perform a linear regression (using MATLAB's~\verb+polyfitn+ function) and fit the 6 local displacement vectors~by
\begin{equation}\label{eq:Du}
\textbf{u}_{i,\text{fit}}(j_n)=\textbf{u}_0+\mathbf{S}(\mathbf{r}_i)^\top\cdot\textbf{r}_{j_n(i)}^\text{Tem},
\end{equation}
to determine a vector $\textbf{u}_0$ that captures any residual translations, while the fitted matrix
\begin{equation}\label{eq:Duf}
\mathbf{S}(\mathbf{r}_i)=\begin{pmatrix}
S_{11} & S_{12}\\
S_{21} & S_{22}
\end{pmatrix}
\end{equation}
approximates the local displacement gradient tensor $S_{ij}\approx\partial_iu_j$ in the basis in which the components the of $\textbf{u}_{i,\text{fit}}(j_n)$ and \smash{$\textbf{r}_{j_n(i)}^\text{Tem}$} are provided. From this matrix, we can therefore extract strain components at position $\mathbf{r}_i$ in analogy to the definitions in Sec.~(\ref{sec:SFcalc}): \hbox{$u^0_{\text{fit}}=S_{11}+S_{22}$} (compression/expansion), $u^1_{\text{fit}}=S_{21}-S_{12}$ (rotation), $u^2_{\text{fit}}=S_{11}-S_{22}$ and $u^3_{\text{fit}}=S_{12}+S_{21}$ (shear components).

\subsubsection{Identification, tracking and characterization of edge dislocations}\label{sec:EdgeDisl}
\begin{figure}[!t]
\centering
\includegraphics[width=0.7\textwidth]{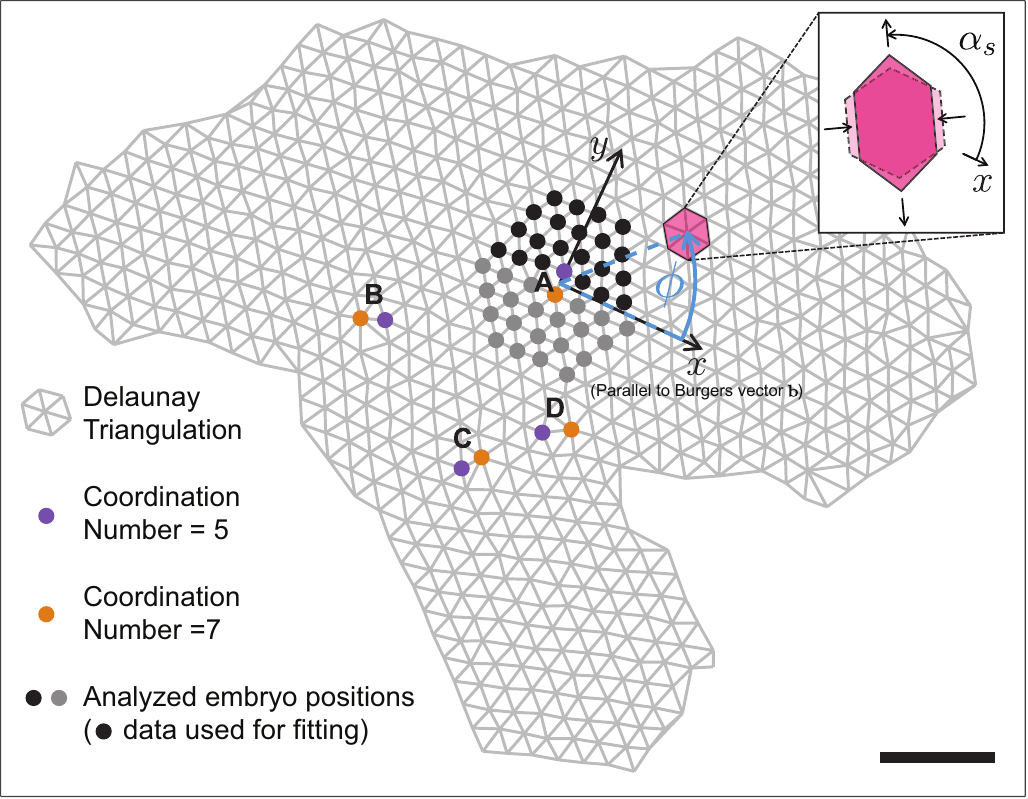}
\caption{\textbf{Strain analysis around edge dislocations.} From a Delaunay triangulation (grey lines) four edge dislocations -- 5,7-coordinated embryo pairs -- labeled $A$,\,$B$,\,$C$, and $D$ can be identified that move little and are present throughout most of the time window of the strain analysis. The Burgers vector $\mathbf{b}$ associated with a dislocation (see Sec.~\ref{sec:EdgeDisl}) defines the $x$-axis direction of a local coordinate system with azimuthal angle $\phi$. Shear elongation axis orientations $\alpha_s$ of embryos colored in gray and black seen along~$\phi$ are shown in Fig.~\ref{fig:DefectStrainSI3}. To minimize spurious effects from crystal boundaries and nearby dislocations, we focus the fitting in Secs.~\ref{sec:axisElong} and \ref{sec:ObservedStrain} on dislocation~$A$, specifically on the 21~embryos colored in black. Scale bar, 1\,mm.}
 \label{fig:DefectStrainSI1}
\end{figure}

The defects we are interested in for this analysis are edge dislocations. In a hexagonal lattice, they can be identified as a bound pair of 5- and 7-fold coordinated embryos (orange and purple dots in Fig.~\ref{fig:DefectStrainSI1}). The coordination number of an embryo corresponds to the number of its edges in the Delaunay triangulation. To minimize the influence of other dislocations or from cluster boundaries, we restrict the strain analysis to defects that are isolated and located in the crystal bulk (dislocation $A$ in Fig.~\ref{fig:DefectStrainSI1}). Other than the ``ideal" edge dislocation~$A$, a second fairly isolated defect ($B$ in Fig.~\ref{fig:DefectStrainSI1}) was the result of an abnormally positioned embryo and therefore not considered for further analysis. Due to their close proximity to each other, edge dislocations $C$ and $D$ also had to be excluded from the analysis.

Focusing on the edge dislocation $A$, we track its center -- defined as the average position of the two constituent 5- and 7-coordinated embryos -- over 500 frames ($\approx80\,$min) using MATLAB's \verb+simpletracker+ function~\cite{Jean-Yves2021}. For each frame, the dislocation can be characterized by a Burgers vector ~\cite{chaikin_lubensky_1995,braverman2020topological}
\begin{equation}\label{eq:Burgers}
    \textbf{b}:=\oint_{\mathcal{C}}\mathbf{S}^\top\cdot\mathrm{d}\textbf{r},
\end{equation}
where the contour $\mathcal{C}$ represents a counter-clockwise loop enclosing the defect, and the fitted strain matrix~$\mathbf{S}$ given at embryo positions $\mathbf{r}_i$ was introduced in Sec.~\ref{sec:SFD}. In practice, we take~$\mathcal{C}$ to be a circular contour of radius $\approx500\,\mu$m centered at the edge dislocation, interpolate $\mathbf{S}(\mathbf{r}_i)$ on 1000 equidistant points along $\mathcal{C}$ using MATLAB's \verb+scatteredInterpolant+ function (method setting ``natural") and approximate the integral Eq.~(\ref{eq:Burgers}) numerically using a Riemann sum with midpoint rule.

\subsubsection{Registration and averaging of strain measurements around edge dislocations}\label{sec:Collapse_strain}
Due to positional fluctuations of dislocations and embryos around it even in the rotation-corrected data and in the absence of whole-cluster oscillations, the amplitude $|\mathbf{b}|$ and orientation~$\phi_{\mathbf{b}}$ of the Burgers vector $\mathbf{b}=|\mathbf{b}|(\cos\phi_{\mathbf{b}},\sin\phi_{\mathbf{b}})^\top$ defined in Eq.~(\ref{eq:Burgers}) changes between frames. However, to define meaningful averages of strain profiles around a dislocation they have to be measured in a common reference frame with respect to the dislocation. Specifically, we want to analyze and average strain fields in a frame in which the Burgers vector defines at all times the $x$-axis of a local coordinate system (see Fig.~\ref{fig:DefectStrainSI1}). The correspondingly rotated strain matrix $\mathbf{S}'$ is given by \smash{$\mathbf{S}'=\mathbf{R}(\phi_{\mathbf{b}})\cdot\mathbf{S}\cdot\mathbf{R}(\phi_{\mathbf{b}})^\top$}, where we use the rotation matrix
\begin{equation}
\mathbf{R}(\phi_{\mathbf{b}})=\begin{pmatrix}
\cos\phi_\textbf{b} & -\sin\phi_\textbf{b}\\
\sin\phi_\textbf{b} & \cos\phi_\textbf{b}
\end{pmatrix}.
\end{equation}

This transformation leaves the fitted strain components related to compression and rotation defined below Eq.~(\ref{eq:Duf}) unchanged, $u_{\text{fit}}^{\prime\,0}=u_{\text{fit}}^{0}$ and  $u_{\text{fit}}^{\prime\,1}=u_{\text{fit}}^1$, and rotates the strain component vector according to $(u_{\text{fit}}^{\prime\,2},u_{\text{fit}}^{\prime\,3})^\top=\mathbf{R}(2\phi_{\mathbf{b}})\cdot(u_{\text{fit}}^2,u_{\text{fit}}^3)^\top$. Accordingly rotated embryo positions~$\mathbf{r}_i$ surrounding the edge dislocation fall nicely onto a common hexagonal lattice (black dots in Fig.~4a, main text), demonstrating that this registration and the following calculation of averages are meaningful. From here on, we work with the registered strain components $u_{\text{fit}}^{\prime\,\alpha}$ ($\alpha=0,1,2,3$), but drop primes~$'$ and ``$\text{fit}$"-labels again to simplify the notation.

After these pre-processing steps, we arrive at a time series of well-defined strain components given at each 6-fold coordinated embryo positions close to the edge dislocation $A$ (black and gray dots in Fig.~\ref{fig:DefectStrainSI1}). In the following, we will use these strain components to estimate effective elastic material properties of living chiral crystals.

\subsubsection{Estimating effective material parameters from shear orientations}\label{sec:axisElong}
\begin{figure}[!t]
\centering
\includegraphics[width=0.7\textwidth]{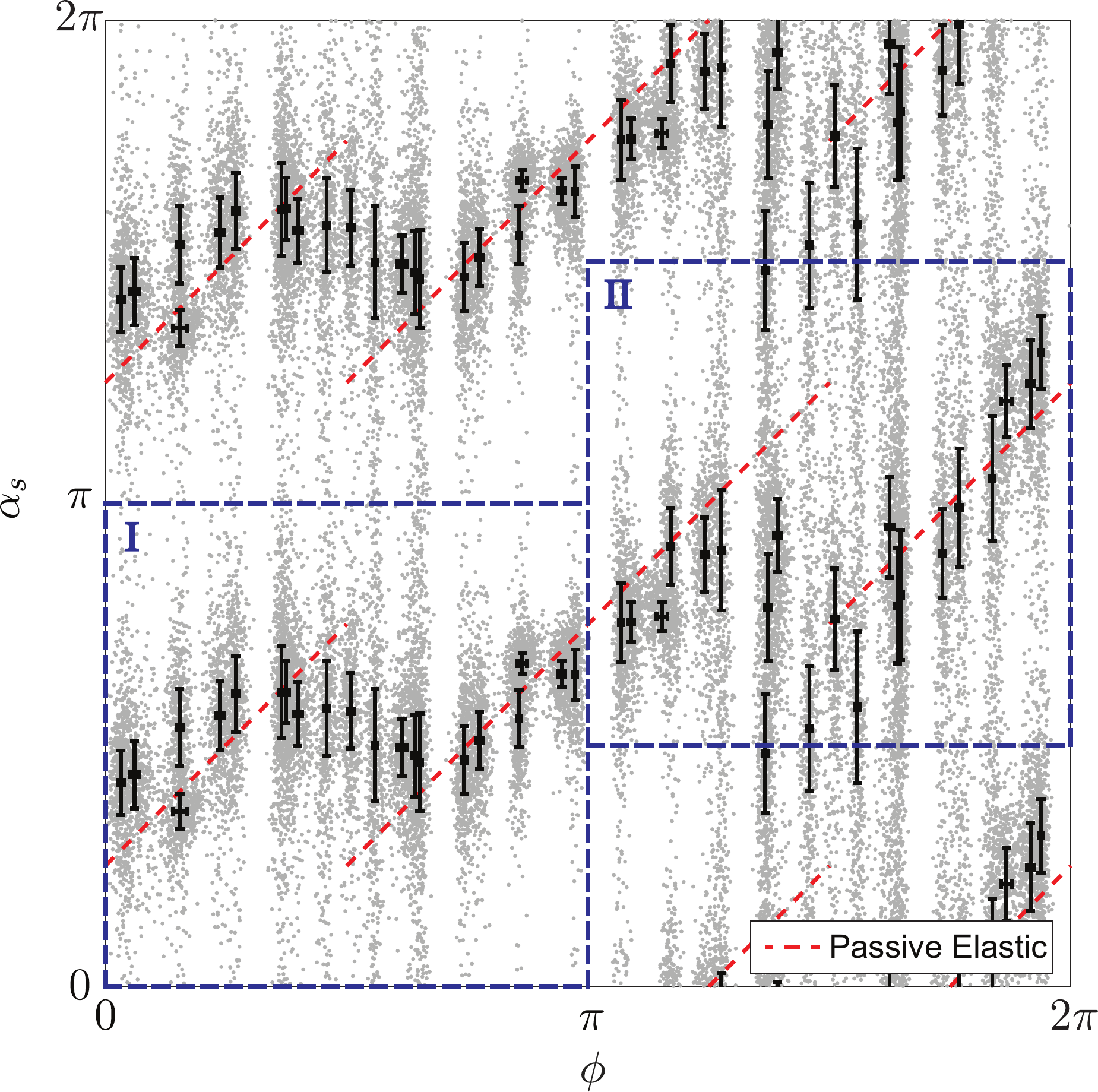}
\caption{\textbf{Orientation of $\boldsymbol{\alpha_s}$ the shear elongation axis around a dislocation.} Each gray dot represents a ``measurement" of the shear elongation axis orientation $\alpha_s$ (see Sec.~\ref{sec:axisElong}) in an azimuthal direction $\phi$ relative to the Burgers vector $\mathbf{b}$ of the dislocation (Fig.~\ref{fig:DefectStrainSI1}). Black dots and errorbars denote means and standard deviation of data pooled at each of the 42 nearby lattice sites (depicted in gray and black in Fig.~\ref{fig:DefectStrainSI1}). For reference, red dashed lines depict orientations $\alpha_s$ expected in a passive isotropic elastic solid (Eq.~(\ref{eq:alphaphi}) with $\delta\alpha_s=0$). To test the theoretically expected symmetry Eq.~(\ref{eq:alphasim}), we periodically extent the shear elongation axis orientation, which is uniquely defined in an interval~$[0,\pi)$. Equation~(\ref{eq:alphasim}) then predicts the data in box~II are a copy of those in box~I. The fact that this symmetry is not clearly present is most likely due to the presence of other nearby dislocations ($B-D$ in Fig.~\ref{fig:DefectStrainSI1}) in the direction \hbox{$\pi<\phi<2\pi$} of the local coordinate system associated with dislocation $A$. Fits are therefore restricted to data in the interval $0\le\phi\le\pi$ (data shown in Fig.~4b, main text).}
 \label{fig:DefectStrainSI2}
\end{figure}

We first analyze the local axis of shear elongation near the dislocation. In particular, we are interested in the angle $\alpha_s$ of local shear elongations, which is defined by\linebreak $(u^2,u^3)^\top=u_s(\cos2\alpha_s,\sin2\alpha_s)^\top$, where $u_s=\sqrt{[u^2]^2+[u^3]^2}$ ($[u^{\alpha}]^2$ with $\alpha=2,3$ denotes squared shear strain components) is the total shear strain amplitude. Note that the shear axis of elongation has nematic symmetry and therefore uniquely defined in the interval  $\alpha_s\in[0,\pi)$. Rotations described in the previous section transformed measured strain components into a coordinate system in which the Burgers vector is parallel to the $x$-axis, i.e. $\mathbf{b}=|\mathbf{b}|\mathbf{e}_x$, and we use in the following the azimuthal angle $\phi\in[0,2\pi)$ defined in this coordinate system (see Fig.~\ref{fig:DefectStrainSI1}). We then plot the values of $\alpha_s$ as determined from $u^2$ and $u^3$ for each embryo and for each frame as a function of this azimuthal angle $\phi$, where it is instructive to periodically extent the axis orientation data up to $\alpha_s=2\pi$ (Fig.~\ref{fig:DefectStrainSI2}; gray dots: individual measurements, black dots/error bars: circular means and standard deviation of strain orientations averaged over times).

For the most general linearly elastic isotropic solid the profile of shear elongation axis orientations~$\alpha_s$ surrounding a dislocation is predicted to be independent of the distance to the dislocation. The general profile is given by~\cite{braverman2020topological}
\begin{equation}\label{eq:alphaphi}
\alpha_s(\phi)=\phi+\frac{\pi}{4}\text{sgn}(\cos\phi)+\delta\alpha_s(\phi),   
\end{equation}
with
\begin{equation}\label{eq:dalphaphi}
\delta\alpha_s(\phi)=\frac{1}{2}\arg\left\{(1+\nu+2i\nu^o )\cos\phi+[2\gamma_1+i(\gamma_2-1)]\sin\phi\right\}-\frac{1}{2}\arg(\cos\phi).
\end{equation}
This expression is a function of the (effective) material parameters
\allowdisplaybreaks
\begin{subequations}\label{eq:strain_param}
    \begin{align}
        \nu&=\frac{(B-\mu)(\mu+\Gamma)+(A-K^o)(K^o-\Lambda)}{(B+\mu)(\mu+\Gamma)+(A+K^o)(K^o-\Lambda)}\\
        \nu^o&=\frac{BK^o-A\mu}{(B+\mu)(\mu+\Gamma)+(A+K^o)(K^o-\Lambda)}\\
        \gamma_1&=\frac{K^o\Gamma+\Lambda\mu}{(B+\mu)(\mu+\Gamma)+(A+K^o)(K^o-\Lambda)}\\
        \gamma_2&=\frac{(B+\mu)(\mu-\Gamma)+(A+K^o)(K^o+\Lambda)}{(B+\mu)(\mu+\Gamma)+(A+K^o)(K^o-\Lambda)},\label{eq:g2}
    \end{align}
\end{subequations}
which are related to the 6 independent elastic moduli $B,\,\mu,\,A,\,K^o,\,\Gamma,\,\Lambda$ introduced in Sec.~\ref{sec:OddModuli}. Hence, the profile of elongation axis orientations can encode information about material properties. For standard passive elastic solids ($B,\mu>0$, and $\,A,\,K^o,\,\Gamma,\,\Lambda = 0$) one finds $\delta\alpha_s=0$; in this case $\alpha_s(\phi)$ is independent of material parameters (red dashed lines in Fig.~\ref{fig:DefectStrainSI2} and in Fig.~4b, main text). More generally, the symmetry of an edge dislocation implies for any value of the 6 elastic moduli that 
\begin{equation}\label{eq:alphasim}
    \alpha_s(\phi+\pi)=\alpha_s(\phi)+\frac{\pi}{2}.
\end{equation}
Graphically, we therefore expect from the theory Eq.~(\ref{eq:alphaphi}) that the values in the blue box~II in Fig.~\ref{fig:DefectStrainSI2} are merely a copy of those in box~I. The fact that this symmetry is not clearly present in the data and standard deviations in box~II are larger than in box~I is most likely due to the presence of other nearby dislocations ($B-D$ in Fig.~\ref{fig:DefectStrainSI1}) that face the dislocation~$A$ approximately in the direction~$\pi<\phi<2\pi$ of the local coordinate system. We thus restrict the fitting in this and in the following Sec.~\ref{sec:ObservedStrain} to data from the interval $0\le\phi\le\pi$ (see also Fig.~4b of the main text) that contains the azimuthal position $\phi$ and shear axis orientations $\alpha_s$ from 21 lattice sites surrounding the defect (black dots in Fig.~\ref{fig:DefectStrainSI1}).

\begin{table}[b!]
\centering
\begin{tabular}{c|c|c}
    Fit Parameter & Fit value & Standard Error\\
    \toprule
    $p_1:=2\nu^o/(1+\nu)$ & 0.048 & 0.098 \\
    \cmidrule{1-3}
    $p_2:=2\gamma_1/(1+\nu)$ & 0.11 & 0.094 \\
    \cmidrule{1-3}
    $p_3:=(\gamma_2-1)/(1+\nu)$ & -0.16 & 0.085 \\
    \bottomrule
\end{tabular}
\caption{Best fit parameters of the shear elongation axis orientation $\alpha(\phi)$ (see Sec.~\ref{sec:axisElong}) around a dislocation (Figs.~\ref{fig:DefectStrainSI1} and \ref{fig:DefectStrainSI2}). The resulting fit  $\alpha_\text{fit}(\phi)=\phi+\frac{\pi}{4}\text{sgn}(\cos\phi)+\delta\alpha_\text{fit}(\phi)$ with $\delta\alpha_{\text{fit}}(\phi)$ given in Eq.~(\ref{eq:alphafit}) is shown as a solid red curve in main text Fig.~4b.}\label{Tab:alphafit}
\end{table}

Finally, we fit the theoretical prediction Eq.~(\ref{eq:alphaphi}) to the measured shear elongation orientations. To this end, define in accordance with Eqs.~(\ref{eq:alphaphi})--(\ref{eq:dalphaphi}) a fit function $\alpha_\text{fit}(\phi)=\phi+\frac{\pi}{4}\text{sgn}(\cos\phi)+\delta\alpha_\text{fit}(\phi)$ with
\begin{equation}\label{eq:alphafit}
\delta\alpha_\text{fit}(\phi)=\frac{1}{2}\arg\left[(1+ip_1)\cos\phi+(p_2+ip_3)\sin\phi\right]-\frac{1}{2}\arg(\cos\phi).
\end{equation}
Weighted estimates and standard errors of the three dimensionless fitting parameters\linebreak\hbox{$p_1=2\nu^o/(1+\nu)$}, $p_2=2\gamma_1/(1+\nu)$ and $p_3=(\gamma_2-1)/(1+\nu)$ (assuming $\nu>-1$) are determined using MATLAB's~\verb+nlmfit+ function~(Tab.~\ref{Tab:alphafit}). The curve $\alpha_\text{fit}(\phi)$ corresponding to the best fit is shown as a red solid line in Fig.~4b of the main text.

In a standard passive linearly elastic solid with $B,\mu>0$ and all other moduli zero, one expects \hbox{$p_1=p_2=p_3=0$}, which seems -- despite a fairly large standard error -- not sufficient to fit strain elongation axis orientation (main text Fig.~4b, Tab.~\ref{Tab:alphafit}). The fact that $p_3$ is notably different from zero suggests that stress-strain couplings mediated through interactions with the surrounding fluid [$\Gamma,\Lambda$, see Eq.~(\ref{eq:g2})] are contributing to the effective mechanical response of the crystal. In addition, we note that for a coarse-grained hexagonal network with ideal linearly elastic odd interactions one finds $p_1\sim\nu^o=0$~\cite{Scheibner:2020gm}. It is therefore not possible to directly infer unambiguous information about odd bulk and shear moduli $A$ and $K^o$, respectively, from fits to the shear elongation axis alone. A complementary fitting approach that overcomes this limitation is described in the following section.

\subsubsection{Estimating elastic moduli from shear strain fields}\label{sec:ObservedStrain}
The fits from the previous section are consistent with the presence of odd moduli in the elastic response of a living chiral crystal, but do not provide unambiguous information about the relative magnitude and signs of such moduli. To determine the latter, we now explicitly fit theoretical predictions of the spatial dependence of strain components to measurements. Importantly, these fits specify magnitude and signs of odd moduli, which in turn can be compared against predictions based on embryo-embryo interactions within the crystal (see~Sec.~\ref{Sec:CGmodel}).

Using the coordinate system in which $\textbf{b}=|\mathbf{b}|\mathbf{e}_x$ and centering the dislocation at the coordinate origin as before, the spatial dependence of strain components in a general isotropic linearly elastic solid is given by~\cite{braverman2020topological} 
\begin{equation}\label{eq:uapred}
u^{\alpha}=\frac{|\mathbf{b}|}{2\pi r^2}\tilde{u}^{\alpha}    
\end{equation}
with 
\begin{subequations}\label{eq:Strain_theory}
\begin{align}
        \tilde{u}^0(x,y)&=-2\gamma_1x-(1-\nu)y\label{eq:u0pred}\\
        \tilde{u}^1(x,y)&=(1+\gamma_2)x-2\nu^oy\label{eq:u1pred}\\
        \tilde{u}^2(x,y)&=-[2\nu^ox+(\gamma_2-1)y]\frac{x^2-y^2}{x^2+y^2}-[(1+\nu)x+2\gamma_1y]\frac{2xy}{x^2+y^2}\label{eq:u2pred}\\
        \tilde{u}^3(x,y)&=-[2\nu^ox+(\gamma_2-1)y]\frac{2xy}{x^2+y^2}-[(1+\nu)x+2\gamma_1y]\frac{y^2-x^2}{x^2+y^2}\label{eq:u3pred}.
\end{align}
\end{subequations}

\begin{table}[b!]
\centering
\begin{tabular}{c|c|c}
    Fit Parameter & Fit value & Standard Error\\
    \toprule
    $\nu$ & 0.049 & 0.0097 \\
    \cmidrule{1-3}
    $\nu^o$ & 0.053 & 0.0039 \\
    \cmidrule{1-3}
    $\gamma_1$ & 0.010 & 0.0046 \\
    \cmidrule{1-3}
    $\gamma_2$ & 0.74 & 0.0088 \\
    \bottomrule
\end{tabular}
\caption{Mean and standard errors for fits of the shear strain components $u^2$ and $u^3$ using theoretical predictions~\cite{braverman2020topological} Eq.~(\ref{eq:uapred}) with $\tilde{u}^2$ and $\tilde{u}^3$ given in Eqs.~(\ref{eq:u2pred})~and~(\ref{eq:u3pred}). The corresponding spatial shear strain profiles together with experimental measurements are depicted in Fig.~\ref{fig:DefectStrainSI3}a. Predictions for divergent and rotary strain components using these fitting parameters in Eq.~(\ref{eq:uapred}) with $\tilde{u}^0$ and $\tilde{u}^1$ are shown in Fig.~\ref{fig:DefectStrainSI3}b (same data as shown in main text Fig.~4c). \label{Tab:strainfit1}}
\end{table}

\begin{figure}[!t]
\centering
\includegraphics[width=1\textwidth]{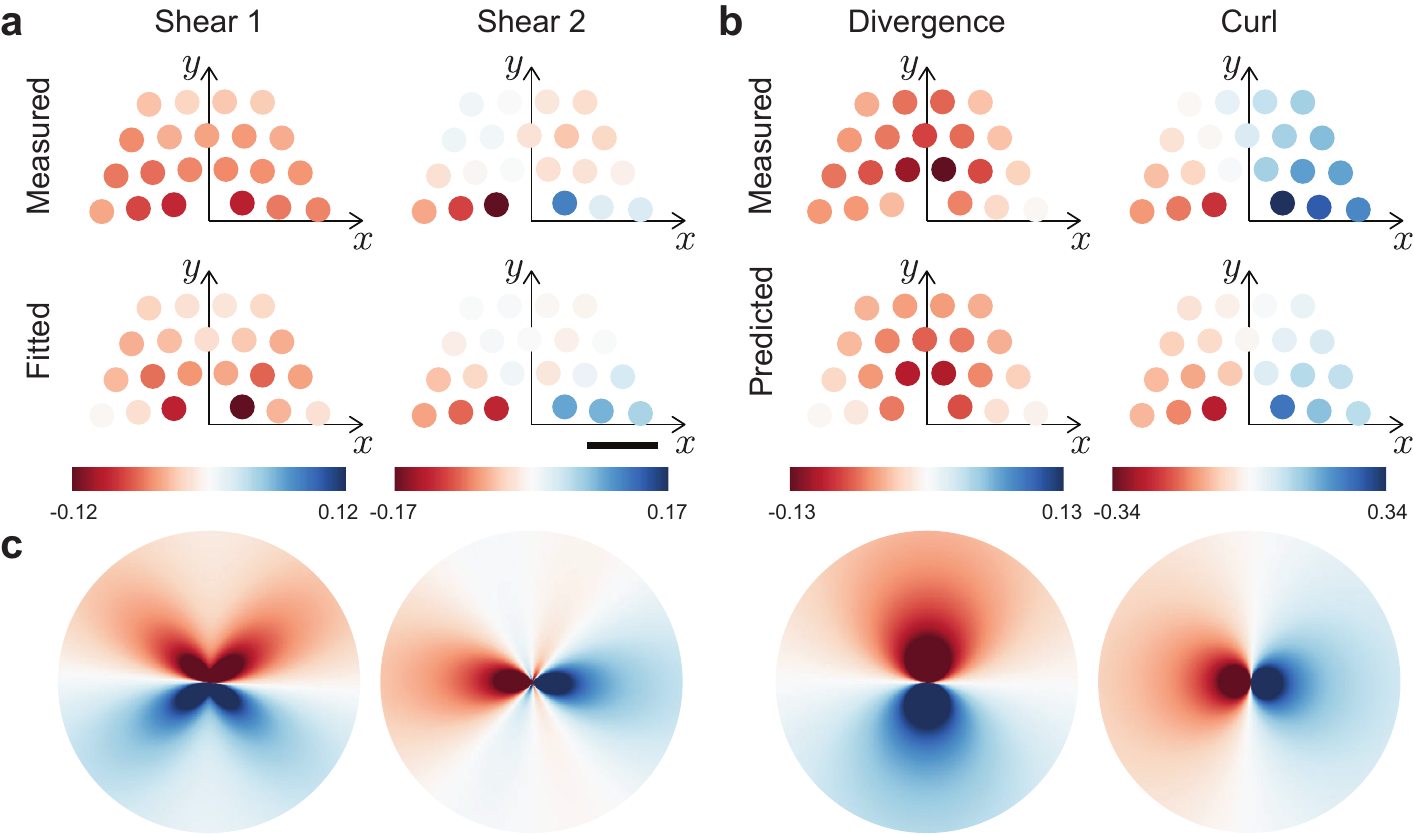}
\caption{\textbf{Spatial strain component analysis near a dislocation.} \textbf{a,}~Shear components measured (top, see Sec.~\ref{sec:SFD}--\ref{sec:Collapse_strain}) and fitted (bottom, see Sec.~\ref{sec:ObservedStrain}) using Eq.~(\ref{eq:uapred}) with $\tilde{u}^2$ (Shear~1) and $\tilde{u}^3$ (Shear~2). Each disk represents an average strain measurement or value of the fitting function centered at the average position of lattice sites near the dislocation (depicted black in Fig.~\ref{fig:DefectStrainSI1}). The resulting fit parameters are given in Tab.~\ref{Tab:strainfit1}. \textbf{b,}~Comparison of measured divergent and rotary strain components (top) with predictions (bottom) from Eq.~(\ref{eq:uapred}) with $\tilde{u}^0$ (Divergence) and $\tilde{u}^1$ (Curl) using the fitting paramers determined in \textbf{a} (same data as shown in main text Fig.~4c). Scale bar,~0.4\,mm. \textbf{c,}~Spatial profiles of strain components on a continuous domain as computed from Eq.~(\ref{eq:uapred}) with $\tilde{u}^{\alpha}$ given in Eqs.~(\ref{eq:u0pred})--(\ref{eq:u3pred}) using parameters given in Tab.~\ref{Tab:strainfit1} found from fitting data in \textbf{a} (top). Scales and color bars same as in panels \textbf{a} and \textbf{b}.}
 \label{fig:DefectStrainSI3}
\end{figure}

The goal is then to determine $\nu,\,\nu^o,\,\gamma_1$ and $\gamma_2$ by fitting measured spatial profiles of strain around the dislocation. To make this analysis comparable to the fitting of shear orientations described the previous section, we determine $\nu,\,\nu^o,\,\gamma_1$ and $\gamma_2$ by fitting spatial profiles of the shear strain components $u^2$ and $u^3$ (Eq.~(\ref{eq:uapred}) with $\tilde{u}^2$ and $\tilde{u}^3$). As before, we restrict the fitting to a specific subset of embryo positions around the dislocation (black dots in Fig.~\ref{fig:DefectStrainSI1}) to mitigate spurious effects from other nearby dislocations and from domain boundaries (see Sec.~\ref{sec:EdgeDisl} for details). The shear component fits are implemented using MATLAB's \verb+lsqcurvefit+ function. The mean and the standard error of each fit parameter, computed from the collection of single-frame fits, are depicted in Tab.~\ref{Tab:strainfit1}. The strain components $u^2$ and $u^3$ corresponding to this best fit are plotted for comparison with measurements in Fig.~\ref{fig:DefectStrainSI3}a.

As a consistency check for this alternative fitting approach, we compute the parameters $p_1\,,p_2$ and $p_3$ found from strain elongation orientation fits (see Sec.~\ref{sec:axisElong}) using the effective material parameters from Tab.~\ref{Tab:strainfit1}, where we find $p_1\approx0.1$, $p_2\approx0.02$ and $p_3\approx-0.25$. These values deviate, but are still within the error margins of the fit values listed Tab.~\ref{Tab:alphafit}.
 
The fit results in Tab.~\ref{Tab:strainfit1} lead to nontrivial predictions for divergent and rotary strain components surrounding a dislocation (Eq.~(\ref{eq:uapred}) with $\tilde{u}^0$ and $\tilde{u}^1$), which so far had not been included into our analysis. Importantly, these predictions can be compared against independent experimental measurements of the corresponding strain components in the crystal, where we find good quantitative agreement~(see main text Fig.~4c, reproduced for convenience in Fig.~\ref{fig:DefectStrainSI3}b).

\begin{table}[b!]
\centering
\begin{tabular}{c|c|c}
    Modulus & Estimate & Standard Error\\
    \toprule
    $A/\mu$ & 7.7 & 0.61 \\
    \cmidrule{1-3}
    $K^o/\mu$ & 7.1 & 0.59 \\
    \cmidrule{1-3}
    $\Gamma/\mu$ & 0.32 & 0.082 \\
    \cmidrule{1-3}
    $\Lambda/\mu$ & -1.0 & 0.097 \\
    \bottomrule
\end{tabular}
\caption{Relative values of elastic moduli determined from an inversion of Eqs.~\eqref{eq:strain_param} for given fit parameters $\nu,\,\nu^o,\,\gamma_1$ and $\gamma_2$ from Tab.~\ref{Tab:strainfit1}. The signs of the odd bulk and shear moduli $A>0$ and $K^o>0$ ($\mu>0$ required for stability) are consistent with predictions based on the handedness of transverse embryo interactions (see Sec.~\ref{Sec:CGmodel}) and indicate broken Maxwell-Betti reciprocity~\cite{braverman2020topological}.\label{Tab:strainfit2}}
\end{table}

Finally, we explicitly determine elastic moduli by inverting Eqs.~(\ref{eq:strain_param}) for given fit parameters $\nu$, $\nu^o$, $\gamma_1$ and $\gamma_2$ from Tab.~\ref{Tab:strainfit1}. From dimensionless strain measurements, we can only expect to find relative values of elastic moduli, where we choose the passive shear modulus $\mu$ to make all other moduli dimensionless. To make the inversion result of Eqs.~(\ref{eq:strain_param}) unique, we additionally impose $B=2\mu$, which holds for a linearly elastic hexagonal spring network with nearest neighbor interactions~\cite{Scheibner:2020gm,braverman2020topological}. Using MATLAB's \verb+solve+ function, we can then determine from the values Tab.~\ref{Tab:strainfit1} and Eqs.~(\ref{eq:strain_param}) a unique solution for the parameters $A/\mu$ and $K^o/\mu$ (odd bulk and shear moduli), as well as for $\Gamma/\mu$ and $\Lambda/\mu$ (equilibrium and nonequilibrium moduli coupling rotations to torque and pressure). The final parameter values and standard errors of these moduli are given in Tab.~\ref{Tab:strainfit2}.

According to the fit results Tab.~\ref{Tab:strainfit2}, strain measurements around a dislocation suggest that odd bulk and shear moduli $A$ and $K^o$, respectively, contribute to the effective material properties of a living chiral crystal. Their relative values compared to the shear modulus $\mu$ are larger than estimated from the coarse-graining of the microscopic model [see Sec.~\ref{Sec:CGmodel}, Eqs.~(\ref{eq:OddModuliCG}) and (\ref{eq:PassModuliCG})]. However, the latter was based on interactions among an isolated pair of embryos, which neglects interaction modulations that are most likely present within clusters. Importantly, the signs determined for $A$ and $K^o$ agree between the coarse-graining and the strain measurement approaches. Ultimately, these signs result from the handedness of the microscopic transverse interactions and therefore represent a macroscopic signature of the well-defined handedness of embryo rotations within the cystal. Furthermore, $A,K^o>0$ (together with $\Lambda<0$) implies broken Maxwell-Betti reciprocity~\cite{braverman2020topological} and suggests that strain cycles observed during the emergence of chiral displacement waves (main text Fig.~4d--h, Sec.~\ref{sec:Hand}, Fig.~\ref{fig:Handedness}a,b) do work on the surrounding [see Eq.~(\ref{eq:work})], highlighting once again the nonequilibrium nature of living chiral crystals. In addition, we find the equilibrium ($\Gamma$) and nonequlibrium ($\Lambda$) moduli coupling rotations to stress (see Sec.~\ref{sec:OddModuli}) are nonzero, which is most likely due to interactions of the crystal with the surrounding fluid.

\subsection{Phase space analysis of strain cycles}\label{sec:StrCycAnal}
In this section, we describe the systematic statistical analysis of the living chiral crystal dynamics in strain space. In particular, we find closed cycles in strain space that provide information about contributions to entropy production~\cite{li2019quantifying} and, in the context of effective odd elastic properties, can be related to mechanical work that is done on the environment~\cite{Scheibner:2020gm}.

\subsubsection{Phase space currents and partial entropy production}\label{sec:EntrProd}
The entropy production rate $\dot{S}$ of an overdamped stochastic system that follows Langevin dynamics can be determined from~\cite{li2019quantifying} 
\begin{equation}
    \dot{S}=k_{\text{B}}\int d\textbf{x}\,\frac{\mathbf{j}(\textbf{x})\cdot \mathbf{D}^{-1}(\textbf{x})\cdot \mathbf{j}(\textbf{x})}{\rho(\textbf{x})},
    \label{eq:entropyproduction}
\end{equation}
where $\rho(\textbf{x})$ and $\mathbf{j}(\textbf{x})$ denote the probability density and corresponding probability current of a particular system configuration $\textbf{x}$, respectively. $\mathbf{D}(\textbf{x})$ is an effective diffusion matrix with inverse~$\mathbf{D}^{-1}(\textbf{x})$. In the following, we describe how $\rho(\textbf{x})$, $\mathbf{j}(\textbf{x})$ and $\mathbf{D}(\textbf{x})$ can be approximated from experimental data of the strain component dynamics to estimate the system's partial entropy production rate shown in main text Fig.~4f,h.

\begin{figure}[!t]
\centering
\includegraphics[width=0.97\textwidth]{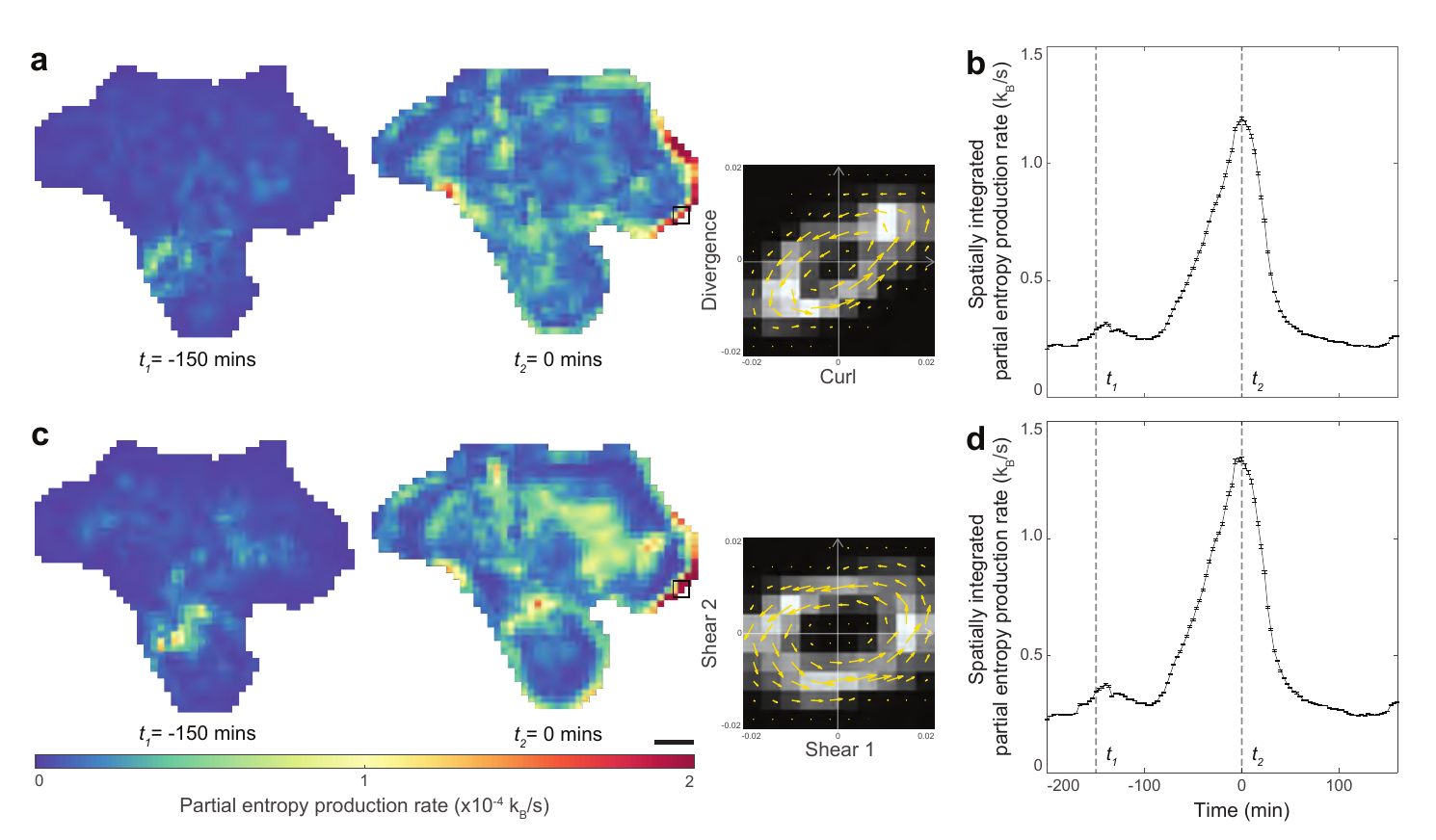}
\caption{\textbf{Partial entropy production rate during cluster oscillations.}
\textbf{a,}~Spatial map the local partial entropy production rate in the curl-divergence space before the emergence of strain waves (left) and in the presence of strain waves (right). Details of the analysis are described in Sec.~\ref{sec:EntrProd}.
\textbf{b},~Spatially integrated entropy production rate in the curl-divergence space. Error bars indicate standard deviation over 100 bootstraps (see Sec.~\ref{sec:EntrProd}). The time point of maximum entropy production rate $t_2$ is defined as 0\,min.
\textbf{c,\,d,}~Same analysis as \textbf{a} and \textbf{b} in the shear strain component space. Scale bar, 1\,mm.}
 \label{fig:Entropy}
\end{figure}

For the spatially resolved analysis shown in Fig.~\ref{fig:Entropy}b,d, we tiled the cluster domain into square regions of $200\,\mu$m (approximately one embryo diameter). The center of each tile is located at some position~$\hat{\mathbf{r}}$. In each of these squares, we spatially average the strain components $u^{\alpha}(\mathbf{r},t)$ ($\alpha=0,1,2,3$, see Sec.~\ref{sec:SFcalc}) to determine local strain component values $\hat{u}^{\alpha}(\hat{\mathbf{r}},t)$. We then introduce a strain component pair vector $\hat{\bm{u}}(\hat{\mathbf{r}},t) = [\hat{u}^{\alpha}(\hat{\mathbf{r}},t),\hat{u}^{\beta}(\hat{\mathbf{r}},t)]^\top$. Inspired by the curl-divergence and shear~1-shear 2 cycles that were suggested as signatures of odd elastic oscillations~\cite{Scheibner:2020gm}, we considered for our analysis the corresponding pairs $\alpha=1,\beta=0$ (Fig.~4f in the main text) and $\alpha=2,\beta=3$ (Fig.~4h in the main text). 

To estimate partial entropy production rates, we used time series of strain component pairs \hbox{$\{\hat{\bm{u}}_1,\hat{\bm{u}}_2,...,\hat{\bm{u}}_N\}$}, where $\hat{\bm{u}}_k:=\hat{\bm{u}}(\hat{\mathbf{r}},k\Delta t)$ with $k=1,2,...,N$ denotes pairs of strain components at $N=200$ successive time points in $\Delta t=10$\,s intervals. The continuous probability  density~$\rho(\textbf{x})$ and the associated current~$\mathbf{j}(\textbf{x})$ were then estimated as~\cite{just2003nonequilibrium}:
\begin{subequations}\label{eq:EPdefs}
\begin{align}
    \hat{\rho}(\textbf{x})&=\frac{1}{N} \sum_{i=1}^{N} K(\textbf{x},\hat{\bm{u}}_i,\Sigma),
    \label{eq:density}\\
    \hat{\mathbf{j}}(\textbf{x})&=\frac{\hat{\rho}(\textbf{x})}{2\Delta t}
    \frac{\sum_{i=2}^{N-1} K(\textbf{x},\hat{\bm{u}}_i,\Sigma)(\hat{\bm{u}}_{i+1}-\hat{\bm{u}}_{i-1})}{\sum_{i=2}^{N-1} K(\textbf{x},\hat{\bm{u}}_i,\Sigma)},
    \label{eq:current}
\end{align}    
\end{subequations}
where system configurations $\textbf{x}=(u^{\alpha},u^{\beta})^\top$ are defined in the space of a strain component pair. In Eqs.~(\ref{eq:EPdefs}), $K(\bm{x},\bm{\mu},\Sigma )=\exp[-(\bm{x}-\bm{\mu})^\top\Sigma^{-1}(\bm{x}-\bm{\mu})/2] /[2\pi\det(\Sigma)]^{-1/2}$ is the bivariate Gaussian with a bandwidth $\Sigma$ determined using the so-called ``rule of thumb"~\cite{bowman1997applied,li2019quantifying}. The latter aims to heuristically minimize the asymptotic mean squared error of the estimated fields $\hat{\rho}(\textbf{x})$ and $\hat{\mathbf{j}}(\textbf{x})$ under the assumption of a standard normal distribution as reference distribution. In two dimensions, it suggests the bandwidth $\Sigma$ is a diagonal $2\times2$ matrix with components \hbox{$\Sigma_{\alpha\beta}=\frac{1}{2}N^{-1/6}\sigma_{\alpha}\delta_{\alpha\beta}$} (no summation), where $N$ denotes the total number of time points and $\sigma_{\alpha}$ is the standard deviation of the local strain component values $\{\hat{u}_{1}^{\alpha},\hat{u}_{2}^{\alpha},\dots,\hat{u}_{N}^{\alpha}\}$ defined above.

Examples of the probability densities and fluxes estimated at some position in the cluster using Eqs.~(\ref{eq:EPdefs}) are shown in the insets of Fig.~4f and h (main text). Similarly, the effective diffusion matrix is estimated as~\cite{just2003nonequilibrium}
\begin{equation}
    \bar{\mathbf{D}}(\textbf{x})=
    \frac{\hat{\rho}(\bm u)}{\Delta t}\frac{\sum_{i=1}^{N-1} K(\textbf{x},\hat{\bm{u}}_i,\Sigma)(\hat{\bm{u}}_{i+1}-\hat{\bm{u}}_{i})\otimes(\hat{\bm{u}}_{i+1}-\hat{\bm{u}}_{i})}{\sum_{i=1}^{N-1} K(\textbf{x},\hat{\bm{u}}_i,\Sigma)},\label{eq:difftens1}
\end{equation}
where $\otimes$ denotes a dyadic product. To ensure an inverse diffusion matrix required in Eq.~(\ref{eq:entropyproduction}) is well-defined, we finally use $\bar{\mathbf{D}}(\textbf{x})$ given in Eq.~(\ref{eq:difftens1}) to define a constant, weighted mean diffusion matrix $\hat{\mathbf{D}}$ as
\begin{equation}
    \hat{\mathbf{D}}=\int d\textbf{x}\,\bar{\mathbf{D}}(\textbf{x})\hat{\rho}(\textbf{x}) \mathbbm{1}_{\hat{\rho}(\textbf{x}) > c}.
    \label{eq:diffusivetensor}
\end{equation}
Here, $\mathbbm{1}_{\hat{\rho}(\textbf{x}) > c}$ is the indicator function defined analog to Eq.~(\ref{eq:indicfunc}) and $c$ is chosen such that only sufficiently well populated regions of the phase space with $\hat{\rho}d\textbf{x}>0.01$ are included in the analysis. The estimates Eqs.~(\ref{eq:EPdefs}) and Eq.~(\ref{eq:diffusivetensor}) are then used in  Eq.~(\ref{eq:entropyproduction}) to approximate a local partial entropy production rate \smash{$\hat{\dot{S}}(\hat{\mathbf{r}})$}. 
Fig.~\ref{fig:Entropy}a and c each show two exemplary spatial maps of the partial entropy production rate before (left) and during (right) the presence of displacement~waves.

Finally, we consider the spatially integrated entropy production rate for the whole cluster
\begin{equation}
    \hat{\dot{S}}_{\text{tot}} = \sum_{\hat{\mathbf{r}}} \hat{\dot{S}}(\hat{\mathbf{r}}).
\end{equation}
 In Fig.~\ref{fig:Entropy}b,d shows that the integrated entropy production rate calculated in both strain component spaces peaks when the displacement waves are most prominent. 

To quantify the robustness of the spatially integrated entropy production rate, we performed the same analysis on bootstrapped~\cite{Efron1979,battle2016broken} time series of the strain component pairs. In particular, we randomly sampled $N-2$ elements from the time series of strain component pair vectors \hbox{$\{\hat{\bm{u}}_2,\hat{\bm{u}}_3,...,\hat{\bm{u}}_{N-1}\}$}. With each sampled vector $\hat{\bm{u}}_k$, we additionally stored $\hat{\bm{u}}_{k-1}$ and $\hat{\bm{u}}_{k+1}$ to be able to compute the flux and diffusion matrix given in Eqs.~(\ref{eq:current}) and (\ref{eq:diffusivetensor}). Entropy production rates were finally determined from this bootstrapped configuration space information. Repeating this procedure 100 times yields a standard deviation that is depicted by error bars in Fig.~\ref{fig:Entropy}b,d.

\begin{figure}[!t]
\centering
\includegraphics[width=0.5\textwidth]{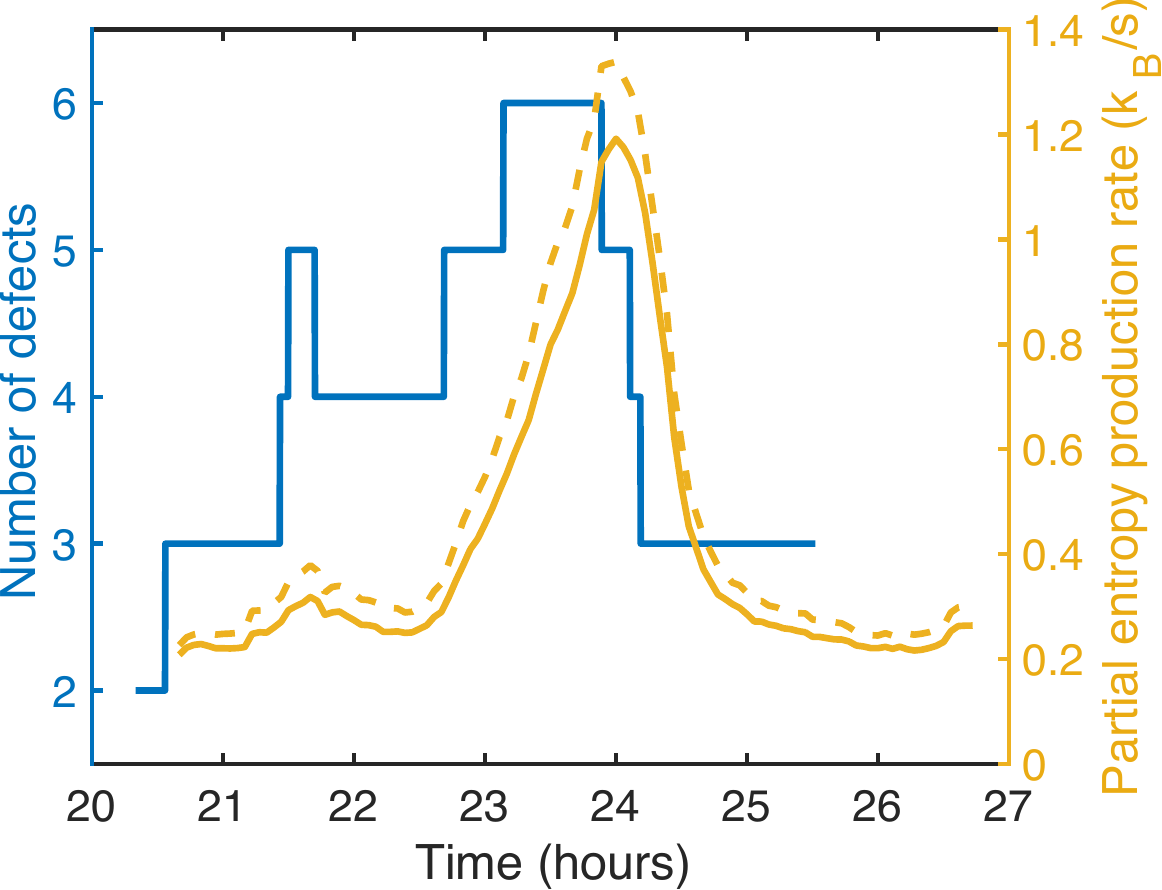}
\caption{\textbf{Dynamics of total defect number and partial entropy production rate.} The increase and decrease of the spatially integrated partial entropy production rate (see Sec.~\ref{sec:EntrProd} for details of the analysis) is accompanied by a corresponding increase and decrease of the total number of defects.}
 \label{fig:DefectNEnt}
\end{figure}

\ \\
Combining the dynamic analysis of lattice defects (Sec.~\ref{sec:EdgeDisl}) with the quantification of partial entropy production (Fig.~\ref{fig:Entropy} and Sec.~\ref{sec:EntrProd}), we can study the correlation of entropy production with the crystalline order of the cluster. Specifically, we plot the total number of defects over time together with the spatially integrated entropy production in Fig.~\ref{fig:DefectNEnt}, which reveals a striking correlation between increasing (and decreasing) defect numbers with the increase (and decrease) in the entropy production rate. Qualitatively, we suspect that the presence of defects provides additional space for embryos to fluctuate around their mean positions within the crystal and thereby facilitates the emergence of substantial strain waves. Once the defects are expelled through embryo rearrangements from the finite crystal domain (Fig.~\ref{fig:Entropy}), this extra space -- and consequently strain waves  -- cease to exists.

\subsubsection{Strain cycle handedness}\label{sec:Hand}
\begin{figure}[!t]
\begin{center}
\includegraphics[width=0.97\textwidth]{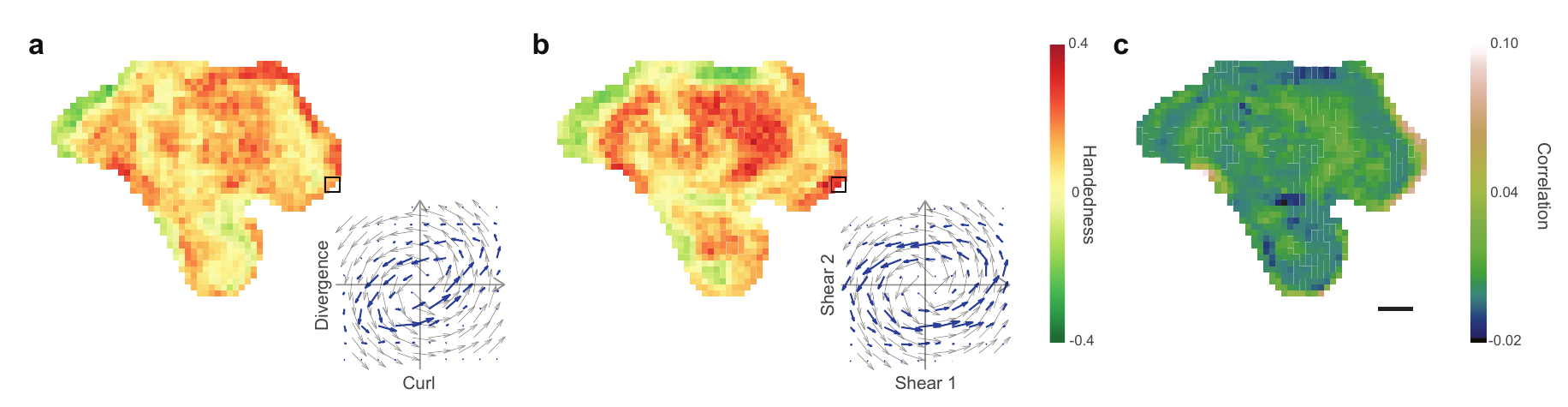}
\end{center}
\caption{\textbf{Handedness maps of strain cycles during cluster oscillations}.
\textbf{a}--\textbf{b,}~Handedness maps of strain cycles computed via Eq.~(\ref{eq:handedn}) during cluster oscillations at the time of maximum partial entropy production (denoted $t_2$ in Fig.~\ref{fig:Entropy}b,d). In both strain spaces $\mathbf{x}=(u^1,u^0)^\top$ (curl--divergence, \textbf{a}) and $\mathbf{x}=(u^2,u^3)^\top$ (shear 1--shear~2,~\textbf{b}), cycles are oriented mostly counter-clockwise.
\textbf{c,}~Correlation between the handedness maps in $\textbf{a}$ and $\textbf{b}$, confirming the visual impression that the cycle handedness in both strain spaces is strongly correlated. Scale bar, 1\,mm.}
 \label{fig:Handedness}
\end{figure}

The handedness of the phase space currents in strain space computed in Sec.~\ref{sec:EntrProd} corresponds to the handedness of strain cycles observed during cluster oscillations. To quantify cycle handedness, we define a counter-clockwise reference unit current field as $\mathbf{j}_{\text{ref}}(\mathbf{x}) =\{-\sin\phi_{\alpha\beta}, \cos\phi_{\alpha\beta}\}$, where $\phi_{\alpha\beta}\in[0,2\pi]$ represents the azimuthal angle in the strain spaces spanned by $\mathbf{x}=(u^1,u^0)^\top$ ($\alpha=1,\,\beta=0$) and $\mathbf{x}=(u^2,u^3)^\top$ ($\alpha=2,\,\beta=3$, see insets in Fig.~\ref{fig:Handedness}a,b). The strain cycle handedness is then at each point in the embryo cluster empirically defined as
\begin{equation}\label{eq:handedn}
    H_{\alpha\beta} = \frac{1}{\max(|\hat{\mathbf{j}}|)}\int\hat{\mathbf{j}}\cdot\mathbf{j}_{\text{ref}}\,d\mathbf{x},
\end{equation}
such that $H_{\alpha\beta}=1$ and $H_{\alpha\beta}=-1$ indicate perfectly counter-clockwise and clockwise cycles, respectively. Strain cycles handedness maps, computed at the time of maximum partial entropy production of strain wave oscillations ($t_2$ in Fig.~\ref{fig:Entropy}b,d), are shown in Fig.~\ref{fig:Handedness}a and b. These maps reveal a predominant counter-clockwise cycle handedness ($H_{\alpha\beta}>0$). Comparing the handedness distribution with the local partial entropy production rate map in Fig.~\ref{fig:Entropy}, we found that handedness amplitudes $|H_{\alpha\beta}|$ are larger in regions with high partial entropy production rate. Finally, we analyzed the map of local correlation of the handedness in either of the two strain spaces. Specifically, Fig.~\ref{fig:Handedness}c depicts the handedness product $H_{10}H_{23}$, which is largely positive and therefore indicates a strong correlation between counter-clockwise oriented strain cycles in both strain component spaces $\mathbf{x}=(u^1,u^0)^\top$ and $\mathbf{x}=(u^2,u^3)^\top$. 

\newpage
\section{Table of symbols}
{\centering\footnotesize
\begin{tabular}{|c|l|}
\toprule
$\mathbf{v}$ & Fluid velocity\\
$\eta$ & Viscosity\\
$v_s$ & Sedimentation velocity of inactive embryos\\
Re & Reynolds number \\
$\mathbf{r}=(x,y,z)^\top$ & Position vector with magnitude $r=|\mathbf{r}|$\\
$a$, $b$, $c$, $d$ & Stokeslet, force-dipole, force-quadrupole, source-dipole singularity amplitude\\
$\mathbf{p}$ & Embryo AP body-axis orientation\\
$\alpha$, $\beta$, $\gamma$ & Coefficients to describe flows near the surface of a microswimmer\\
$A_n$, $B_n$, $C_n$, $D_n$ & Mode coefficients in solution of the Hele-Shaw equation\\
$\mathbf{r}_i$ & Centroid position of a disk/particle $i$\\
$t_d$ & Developmental time, $t_d=0$: onset of cluster formation\\
$h$ & Distance of flow singularities from the fluid-air interface\\
$\mathbf{g}$/$\theta_g$ & Gravity/Orientation of gravity relative to fluid surface normal\\
$\mathbf{r}_{ij}=\mathbf{r}_i-\mathbf{r}_j$ & Relative coordinate between disk $i$ and $j$\\
$\hat{\mathbf{r}}_{ij}$, $\hat{\mathbf{r}}_{ij}^\perp$ & Normalized relative coordinate and orthogonal normalized relative coordinate\\
$\omega_0$ & Single embryo angular spinning frequency\\
$R$ & Apparent top-view radius of surface-bound embryos (in the disk model)\\
$\omega_i$ & Angular spinning frequency of disk/particle $i$\\
$\mathbf{F}_{\text{rep}}$, $f_{\text{rep}}$ & Effective repulsion force and effective repulsion force strength\\
$F_{\text{nf}}(|\mathbf{r}_i-\mathbf{r}_j|)$ & Effective lubrication forces in the disk model\\
$T_{\text{nf}}(|\mathbf{r}_i-\mathbf{r}_j|)$ & Effective lubrication torques in the disk model\\
$f_0$, $\tau_0$ & Strength of lubrication forces and torques in the disk model\\
$d_c$ & Cut-off length of lubrication interactions\\
$d_{ij}$ & Shortest disk surface distance between disks $i$ and $j$\\
$N_{cl}$ & Total number of embryos or disks bound in a cluster \\
$\mathbf{e}_x,\,\mathbf{e}_y,\,\mathbf{e}_z$ & 3D Cartesian basis\\
$\mathbf{e}_r,\,\mathbf{e}_{\theta},\,\mathbf{e}_{\phi}$ & 3D spherical coordinates basis, $x=r\sin\theta\cos\phi,y=r\sin\theta\sin\phi,z=r\cos\theta$\\
$\mathbf{e}_{\rho},\,\mathbf{e}_{\varphi}$ & 2D cylindrical coordinate basis, $\bar{x}=\rho\cos\varphi,\,\bar{y}=\rho\sin\varphi$.\\
$l$ & Average nearest neighbor centroid distance in bound pairs and triplets\\
$\ell_{\text{maj}}$, $\ell_{\text{min}}$ & Semi-major and semi-minor embryo body axis\\
$e=\ell_{\text{min}}/\ell_{\text{maj}}$ & Body axis ratio\\
$\chi=(1-e^2)/(1+e^2)$ & Microswimmer body shape anisotropy\\
$L=(\ell_{\text{maj}}+\ell_{\text{min}})/2$ & Characteristic embryo size, $L=150\,\mu$m\\
$F_{\text{st}}, F_g$ & Stokeslet force amplitude, Negatively buoyant weight force\\
$\mathbf{r}_0$, $\mathbf{r}_0'$ & Position of flow singularity and of image flow singularity\\
$\mathbf{E}$, $\mathbf{E}'$ & Symmetric strain rate tensor/Image above free fluid surface\\
$\mathbf{F}$, $\mathbf{F}'$ & Stokeslet force and image Stokeslet force/Image above free fluid surface\\
$\mathbf{T}$, $\mathbf{T}'$ & Rotlet torque and image rotlet torque/Image above free fluid surface\\
$\bar{\mathbf{v}}(\mathbf{r};\mathbf{p},\mathbf{r}_0)$ & Singularity construction below free, nondeforming fluid surfaces\\
$\mathbf{u}$, $u_{ij}=\partial_iu_j$ & Displacement field and displacement gradient tensor \\
$u^{\alpha}$ & 2D strain components: dilation ($\alpha=0$), rotation ($\alpha=1$), shear ($\alpha=2,3$)\\
$\sigma^{\alpha}$ & 2D stress components: pressure ($\alpha=0$), torque ($\alpha=1$), shear ($\alpha=2,3$)\\
$B$, $\mu$ & Bulk modulus, Shear modulus\\
$A$, $K^o$ & Odd bulk modulus, Odd shear modulus\\
$\Lambda$, $\Gamma$ & Elastic moduli coupling rotations to stresses\\
$\nu$, $\nu^o$ & Poisson's ratio, Odd ratio\\
$\gamma_1$, $\gamma_2$ & Effective material parameters of an isotropic elastic solid\\
$\hat{F}_{\text{nf}}$, $\hat{T}_{\text{nf}}$ & Stokeslet and rotlet fitting parameters for flows surrounding bound pairs and triplets\\
$\bar{a}$ & Living chiral crystal lattice constant ($\approx220\,\mu$m)\\
$\psi_6(\mathbf{r}_i)$ & Local hexagonal bond orientation parameter of embryo $i$\\
$g(r)$ & Radial pair distance distribution function \\
$\gamma_L(\tau)$ & Dynamic Lindemann parameter as a function of lag time $\tau$\\
$\mathbf{b}$ & Burgers vector\\
$\phi_{\mathbf{b}}$ & Azimuthal Burgers vector orientation in cluster frame\\
$u_s$, $\alpha_s$ & Magnitude and orientation of the shear elongation axis\\
$\mathbf{x}=(u^{\alpha},u^{\beta})^\top$ & Strain space coordinate\\
$\rho(\textbf{x})$, $\mathbf{j}(\textbf{x})$, $\mathbf{D}(\textbf{x})$ & Probability density, probability current and diffusion matrix in strain space\\
$\hat{\rho}(\textbf{x})$, $\hat{\mathbf{j}}(\textbf{x})$, $\hat{\mathbf{D}}(\textbf{x})$ & Corresponding functions estimated from data\\
$H_{\alpha\beta}\ $ & Strain cycle handedness averaged in the strain space $\mathbf{x}=(u^{\alpha},u^{\beta})^\top$\\
\smash{$\hat{\dot{S}}(\mathbf{r})$, $\hat{\dot{S}}_{\text{tot}}$} & Local and spatially integrated partial entropy production rate estimated from data\\
\bottomrule
\end{tabular}
}

\section{Extended Data Figures}
\renewcommand{\figurename}{Extended Data Figure}
\renewcommand{\thefigure}{\arabic{figure}}
\setcounter{figure}{0}
\begin{figure}[H]
\centering
\includegraphics[width=1\textwidth]{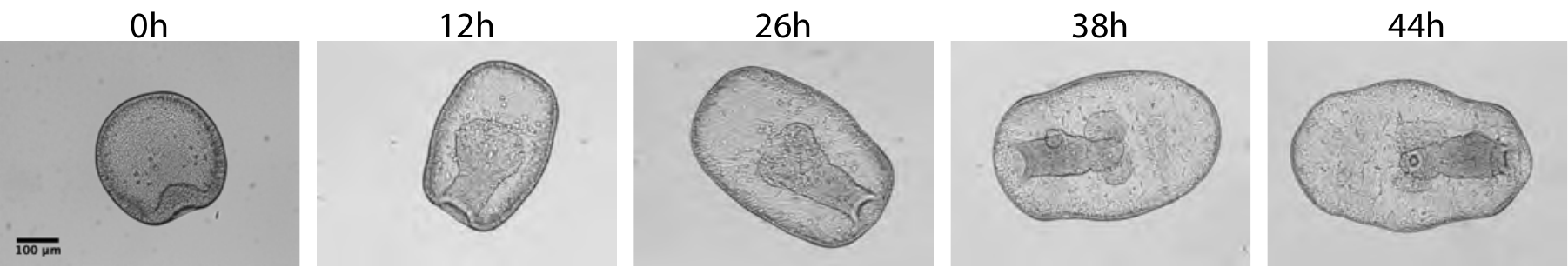}
\caption{Uncropped embryo morphology images shown in Fig.~1b in the main text.}
\end{figure}

\newpage
\bibliographystyle{unsrt}
\bibliography{references}